\documentclass[a4paper,11pt]{article}
\pdfoutput=1
\usepackage[table,xcdraw]{xcolor}
\usepackage{jcappub}
\usepackage[T1]{fontenc} 
\usepackage[utf8]{inputenc}
\usepackage[T1]{fontenc} 
\usepackage{natbib}
\usepackage{bm}
\usepackage{float}
\usepackage[varg]{txfonts}

\newcommand{\Hbb}{\mathbb{H}}
\newcommand{\Qbb}{\mathbb{Q}}
\newcommand{\Rbb}{\mathbb{R}}
\newcommand{\Jbb}{\mathbb{J}}
\newcommand{\Sbb}{\mathbb{S}}

\usepackage{appendix}

\usepackage{comment}
\usepackage{multirow}  

\usepackage[varg]{txfonts}

\hypersetup{
  bookmarksnumbered=true,
  linkcolor=red,
  citecolor=green,
  urlcolor=cyan
}

\newcommand{\simgt}{\lower.5ex\hbox{$\; \buildrel > \over \sim \;$}}
\newcommand{\simlt}{\lower.5ex\hbox{$\; \buildrel < \over \sim \;$}}

\newcommand{\roy}[1]{\textcolor{magenta}{#1}}

\title{The anisotropic expansion rate of the local Universe and its covariant  cosmographic interpretation}

\author[]{Basheer Kalbouneh$^{1}$,  Christian Marinoni$^{1}$,  Roy Maartens$^{2,3}$, Julien Bel$^1$, Jessica Santiago$^{4}$, Chris Clarkson$^{5,2}$, Maharshi Sarma$^{1}$ and Jean-Marc Virey$^{1}$}

\affiliation[1]{\small{Aix Marseille Univ, Universit\'e de Toulon, CNRS, CPT, Marseille, France}}
\affiliation[2]{\small{Department of Physics \& Astronomy, University of the Western Cape, Cape Town 7535, South Africa}}
\affiliation[3]{\small{National Institute for Theoretical \& Computational Sciences, Cape Town 7535, South Africa}}
\affiliation[4]{\small{Leung Center for Cosmology and Particle Astrophysics,
National Taiwan University, Taiwan}}
\affiliation[5]{\small{Department of Physics \& Astronomy, Queen Mary University of London, London E1 4NS, United Kingdom}}

\emailAdd{}

\abstract{
Without making any assumption on the underlying geometry and metric of the local Universe, we provide a
measurement of the  expansion rate fluctuation field using the Cosmicflows-4 and Pantheon+ samples in the redshift range $0.01 < z < 0.1$ ($30 \,h^{-1}\,\mathrm{Mpc} < R < 300\,h^{-1}\,\mathrm{Mpc}$).
The amplitude of the anisotropic fluctuations is found to be of order  a few percent relative to the monopole of the expansion rate.

We further decompose the expansion rate fluctuation field into spherical harmonic components and analyze their evolution with redshift across the studied redshift range. 
At low redshift, the dipole is clearly dominant, with an amplitude of  $\sim (2.2 \pm 0.15)\times 10^{-2}$, significantly larger than the higher--order 
modes. As redshift increases, the dipole amplitude steadily decreases, reaching  roughly half its value in the highest redshift bin investigated. The quadrupole 
is also significant, at about half the dipole amplitude, and persists across all  redshift bins, with no clear decreasing trend, although uncertainties grow at  higher redshift. A nonzero octupole is detected at low redshift with a 
signal-to-noise ratio of $\sim 3$, but it becomes unconstrained at higher  redshift.  The dipole, quadrupole, and octupole components are found to be aligned, exhibiting axial symmetry around a common axis ($l = 295^\circ,\, b = 5^\circ$).

We interpret the observed fluctuations in the expansion rate within the framework  of covariant cosmography. Our results indicate that the multipoles of the  expansion rate fluctuation field are primarily driven by a strong quadrupole in  the covariant Hubble parameter, together with dipole and octupole contributions  from the covariant deceleration parameter. These few parameters suffice to  reconstruct the luminosity distance with high precision out to $z \sim 0.1$, in  a manner that is model--independent, non--perturbative, and free from assumptions  about peculiar velocities.

Finally, we find that the CMB frame is not locally comoving with the matter fluid, and that a matter  fluid element, roughly a spherical region of size 
in the range $38\,  \lesssim {R\, (\mathrm{Mpc})} \lesssim 100$ 
centered on the observer position, moves relative to the CMB frame with a velocity of $188 \pm 22\,\mathrm{km/s}$,  along the axis of symmetry.
}
\begin{document}
\maketitle
\flushbottom

\section{Introduction}
Although observational programs have become increasingly powerful and precise, they have not yet fully converged on the fundamental parameters of the standard cosmological model \cite{DiValentino:2021izs, Perivolaropoulos:2021jda, Schoneberg:2021qvd, Abdalla:2022yfr, Capozziello:2024stm, Bengaly:2024ree}. These ongoing tensions may suggest the need to revisit certain underlying assumptions of cosmology, including aspects of the geometry of cosmic spacetime
 \cite{Schwarz:2007wf, Kashlinsky:2008ut, Antoniou:2010gw, Cai:2011xs, Marinoni:2012ba, Kalus:2012zu, Wang:2014vqa, Yoon:2014daa, Tiwari:2015tba, Javanmardi2015, Bengaly:2015nwa, Colin:2017juj, Rameez:2017euv, Migkas:2020fza, Migkas:2021zdo, Secrest:2020has, Siewert:2020krp, Luongo:2021nqh, Krishnan:2021jmh, Sorrenti:2022zat, Aluri:2022hzs, Cowell:2022ehf, Hu:2023eyf,Hu:2023jqc, Dainotti:2021pqg,CosmoVerseNetwork:2025alb,Mazurenko:2024gwj,Sah:2024csa,Hu:2024big,Franco:2024dvc,Adam:2024kgs,Lopes:2024vfz,Rameez:2024xsn,Celerier:2024dvs}.

The Friedmann–Lemaître–Robertson–Walker (FLRW) metric provides an effective description of the large–scale properties of the Universe, including its expansion and overall evolution. However, its underlying assumptions of exact homogeneity and isotropy limit its applicability as a fully realistic model. In particular, it does not capture the complexity of the local Universe on scales ($r \lesssim 150\,h^{-1}\,\mathrm{Mpc}$) where direct astrophysical measurements of the Hubble parameter are typically carried out (e.g., \cite{Riess2018}). Therefore, a more detailed, fully relativistic characterization of the cosmic expansion rate on local scales ($z < 0.1$) is essential, extending beyond the single-parameter description provided by the Hubble constant $H_0$ within the standard cosmological model.

This work is the fifth contribution in an ongoing series \cite{paper0,paper1,paper2,paper3}, through which we aim to open a new window on studies of the local expansion rate by developing a new method to characterize it in a model-independent and non-perturbative manner, without  assuming the FLRW metric, Einstein’s field equations, or invoking concepts such as peculiar velocities or density fluctuations.  Central to this  framework is the expansion rate fluctuation field, $\eta$ (introduced in \cite{paper0,paper2}), a scalar Gaussian observable designed to identify and classify deviations from isotropy in the redshift–distance relation.
An important caveat is in order. With this cosmological observable, we are not probing perturbations of the Hubble constant $H_0$, which would require assuming a fixed background and relying on FLRW modeling in the estimation of expansion-rate anisotropies. Although our fully model-independent approach does not provide a specific value of $H_0$, its key advantage is that it remains unaffected by distance-dependent selection biases, such as the Malmquist bias.

We used the expansion rate fluctuation $\eta$  to analyze the multipolar structure of the redshift-distance relation in the local Universe, facilitating clearer interpretation of the fluctuations present in the Hubble diagram. Previous analysis of the Cosmicflows-3 catalog (see \cite{paper0}) indicated that the expansion rate $\eta$ shows significant dipolar, quadrupolar, and octupolar components. Notably, the local Universe's expansion rate up to $z < 0.05$ exhibits an {\it axially symmetric pattern of anisotropy}, marked by the intriguing alignment  of the low-order multipole orientations with the bulk component of the Local Group velocity.

We also demonstrated how to interpret the expansion rate fluctuations signal $\eta$ to gain insight into the geometry of local spacetime, using invariant physical quantities: the covariant cosmographic parameters \cite{paper1}. Additionally, we showed \cite{paper2,paper3} that this formalism is applicable in the local Universe, allowing for unbiased estimation of the dominant $\ell \leq 4$ multipoles of the lowest cosmographic parameters (Hubble, deceleration, and jerk) -- even in the presence of large density fluctuations.

The purpose of this paper is twofold. 
\begin{itemize}
\item On the observational side, we introduce an enhanced analysis framework that exploits the most recent datasets, Cosmicflows-4 \cite{Tully_CF4:2022rbj} and Pantheon+ \cite{Scolnic:2021amr}, in order to extract new insights beyond what was accessible in our preliminary studies. Our goal is to achieve a higher-precision characterization of local expansion dynamics and to extend the analysis into previously underexplored spatial regimes, moving beyond the dominant influence of the Shapley supercluster $(z \approx 0.05)$ and probing scales out to $z = 0.1$.

\item On the theoretical front, we compute the lowest-order multipoles of the covariant cosmographic parameters: in particular, the quadrupole of the generalized Hubble parameter, together with the dipole and octupole components of the covariant deceleration parameter.  
Building on the demonstrated robustness of the CC formalism in local cosmological patches, established both analytically \cite{paper2} and through numerical simulations \cite{Adamek:2024hme,Macpherson:2025qec}, our aim is to characterize fluctuations in the expansion rate field over the redshift range $0.01 < z < 0.1$. This characterization is achieved in terms of a restricted set of coefficients which, being model-independent, enable direct comparison with the predictions of different metric-based cosmological models for the line element.
\end{itemize}

The paper is organized as follows. In Section 2, we briefly describe the samples of redshift-independent distances employed in our analysis. In Section 3, we introduce the expansion rate fluctuation field $\eta$ and detail the methodology used to estimate it from discrete datasets. The observed multipolar structure of $\eta$ and its scaling with redshift are presented and discussed in Section 4. In Sec. 5, the observational results are compared with the theoretical expectations from standard cosmology, while Section 6 provides their interpretation within the covariant cosmographic framework. Finally, Section 7 offers a summary and concluding remarks.

In the following, we present all formulas in natural units ($c=1$) and, when necessary, we adopt the standard $\Lambda$CDM model, defined as the spatially flat FLRW spacetime that provides the best fit to the Planck 2018 data \cite{Planck:2018nkj}. Observables measured in the CMB rest frame are denoted by $\tilde{x}$, while the same quantities in the matter (dust) rest frame are denoted by $x$ (see \cite{paper3}).

\section{Data}  \label{sec:data2}

\subsection{Cosmicflows-4 sample}

\begin{figure*}
\begin{center}
\includegraphics[scale=0.5]{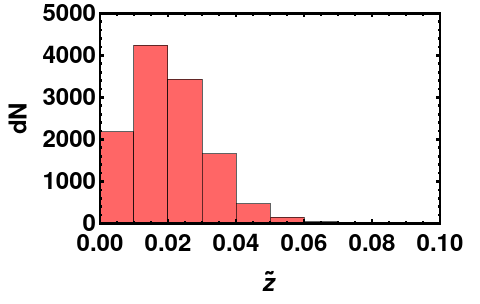}
\includegraphics[scale=0.5]{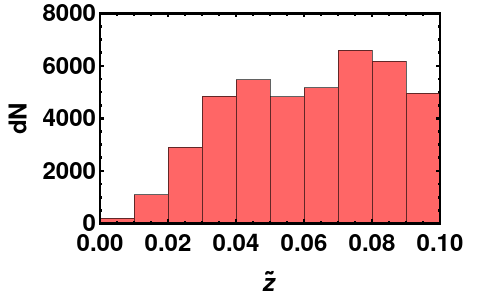}
\;\;\;\;
\includegraphics[scale=0.45]{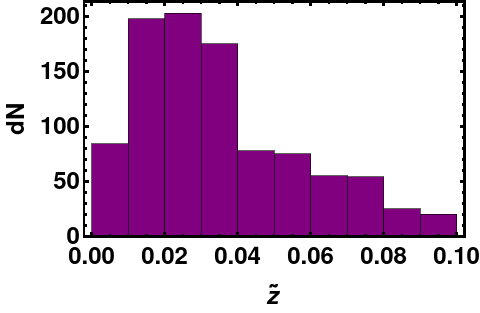}
\;\;\;\;
\includegraphics[scale=0.5]{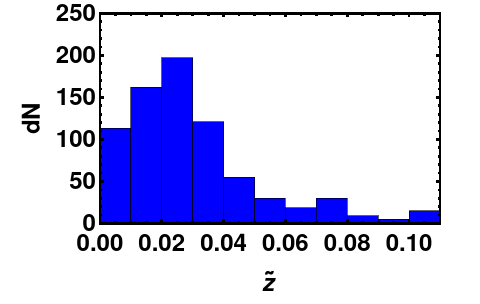}
\caption{Histograms showing number counts as a function of redshift. CF4TF and CF4FP are presented in the first row, while CF4SN and Pantheon+ are shown in the second row.}
	\label{Fig_1f}
		\centering
	\end{center}
\end{figure*}

The Cosmicflows-4 (CF4) dataset \cite{Tully_CF4:2022rbj} is a homogenized compilation of several independent distance catalogs, rather than a single uniform survey. It contains $55{,}874$ nearby galaxies ($\tilde z \leq 0.1$) with redshift-independent distance estimates derived from complementary methods.  
Distances for about three-quarters of the sample ($42{,}221$ early-type galaxies) are obtained using the Fundamental Plane (FP) relation. The dominant contributions come from the SDSS-PV catalog \cite{Howlett:2022len}, which extends to $\tilde z \approx 0.1$ and is confined to the SDSS footprint centered on the northern Galactic pole, and from the 6dFGRSv survey  \cite{Magoulas_2016_6dF}, which covers the southern hemisphere up to $\tilde z \approx 0.055$.  
Most of the remaining entries ($12{,}221$ late-type galaxies) are based on the Tully–Fisher (TF) relation, primarily derived from SDSS HI data \cite{Kourkchi_2022}, which provide near all-sky coverage with a redshift distribution peaking at $\tilde z \approx 0.02.$  
In addition, CF4 incorporates $\sim 1{,}000$ Type Ia supernova distances, mainly from the Pantheon+ and SH0ES compilations \cite{Scolnic:2021amr,Riess_2022}, concentrated at $\tilde z < 0.03$ but extending to $\tilde z \approx 0.1$.  
We hereafter refer to these subsamples as CF4FP, CF4TF, and CF4SN, respectively.  

Figure \ref{Fig_1f} illustrates the redshift dependence of morphological composition, with early-type galaxies becoming increasingly dominant at larger redshifts.  
The angular distribution of objects (top panel of Figure \ref{eta_l_map2}  shows an isotropic coverage within $\tilde z \sim 0.03$, moderately uniform sampling out to $\tilde z \sim 0.05$, and strong anisotropies beyond this scale and up to $\tilde z\sim 0.1$. The Galactic plane blocks observations within  $\sim 10^{\circ}$, producing the so-called Zone of Avoidance (ZoA)

The average uncertainty in the distance modulus  for the CF4FP, CF4TF, and CF4SN samples 
are  $\sim 0.4$ and $\sim 0.5$ and $\sim 0.15$ respectively (or relative error in the luminosity distance $\sim 18\%$, $\sim 23\%$ and $\sim 7\%$). The sample does not provide a covariance matrix, so no correlations between the measurements are assumed.
Although the distance errors for TF  and FP  estimates are approximately two and three times larger, respectively, than those for SNIa—making each measurement roughly four and nine times less valuable in a statistical weighting scheme—the large number of spiral and elliptical galaxies in the sample helps reduce the overall statistical uncertainty.
In order to further reduce the noise in each distance estimate, when additional distance measurements for the same object are available (such as those obtained using the surface brightness fluctuation or tip of the red-giant branch methods), the average of these measurements is calculated to determine the distance modulus, with each measurement weighted according to its associated uncertainty. In the same spirit, when possible, distances to CF4SN objects are computed as the average of estimates provided by multiple sources.

Given its density, we can divide the CF4 sample in various redshift bins. We impose a lower redshift cut at $\tilde z = 0.01$ because, as will be clarified in Section \ref{sec_cosmp}, the cosmographic covariant approach—used to interpret our results—relies on a fluid approximation that treats discrete data as a continuous field. This approximation breaks down below this redshift threshold.

In the next sections, we will use the following designations to avoid repetition:\\ CF4 ~~\, $0.01 < \tilde z < 0.1$,\\ CF4a ~ $0.01 < \tilde z < 0.03$,~ CF4b ~ $0.03 < \tilde z < 0.05$,~ CF4c ~ $0.05 < \tilde z < 0.075$,~   CF4d ~ $0.075 < \tilde z < 0.1$

Since a large number of high-$\sigma$ deviations is not statistically expected under Gaussian-distributed distance moduli, we apply Chauvenet's criterion \cite{taylor1997error} to exclude eight outliers (PGC 5037, 10302, 12384, 31586, 32512, 43423, 53805, 59808). These objects deviate by more than $5\sigma$ from predictions of both the standard cosmological model and the covariant cosmographic model (see Section \ref{sec_cosmp}).

The distribution of CF4 objects  over the sky in Galactic coordinates, for the four redshift ranges (CF4a, CF4b, CF4c, and CF4d), is shown below in Figure \ref{eta_l_map2} (the first row). The galaxies are relatively unevenly distributed across the sky, with some unsurveyed regions—particularly beyond the plane of the Milky Way—standing out conspicuously. The degree of inhomogeneity becomes more pronounced with increasing redshift, which also leads to greater sparsity in the distribution of objects. This complex pattern in the angular and radial distribution functions calls for a simulation-based assessment of the potential impact of selection biases on our results, as will be detailed in the following sections.

\subsection{Pantheon+ sample}

The Pantheon+ SNIa compilation \cite{Scolnic:2021amr} contains 1701 measurements in the range $0.001 < \tilde z < 2.26$. As done in the CF4 case, we exclude objects with $\tilde z < 0.01$ and limit the sample to $\tilde z < 0.1$, so as to allow for proper comparison with CF4. This results in a dataset of 695 measurements.

The uncertainties in the distance measurements are represented by a covariance matrix, with a typical relative uncertainty of 11\%.
Figure \ref{Fig_1f} (lower right panel) shows their redshift distribution. Compared to the older Pantheon catalog, this compilation features an increased number of objects, with around 80\% of the new measurements at low redshift ($\tilde z \lesssim 0.1$), which is in the range of this study. 

Since we focus on measuring fluctuations in the Hubble diagram via the observable $\eta$ field (defined in \eqref{defeta1}) and not on the monopole component of the expansion, our analysis is independent of the absolute calibration of distances. This frees us from distance-dependent effects such as selection biases and calibration errors. In particular, because our results do not rely on the zero-point calibration of absolute magnitudes, we can exclude Cepheid-based calibrations from the analysis, thereby simplifying the statistical treatment. Consequently, we may use either the raw SNIa magnitudes or the calibrated distance moduli from the Pantheon+ sample, as both yield equivalent results for our purposes.

\section{Expansion rate fluctuation field in the local Universe}\label{sec_eta}

The {\it expansion rate fluctuation field} $\eta$  \cite{paper0,paper2}, 
\begin{eqnarray}
    \eta(z,\boldsymbol{n}) &\equiv & \log\left(\frac{{z}}{{d}_{L}(z,\boldsymbol{n})}\right)-\frac{1}{4\pi}\int_S \log\left(\frac{{z}}{{d}_{L}(z,\boldsymbol{n})}\right)\mathrm{d}\Omega \,,
\label{defeta1}
\end{eqnarray}  
measures anisotropic angular fluctuations in the luminosity distance-redshift relation. The second term on the right is the monopole of the logarithmic ratio, which we will denote by $\mathcal{M}$. It guarantees that $\eta$ averages to zero on each spherical shell $S$ of radius $z$  and width $\delta z$, centered on the observer, where $\boldsymbol{n}$ is a unit vector specifying the observer's line of sight.
Redshifts and distance moduli (measured independently from redshift information) for each object in $S$ are the only data needed to construct the observable.  
We do not use luminosity distances directly (since their errors are non-Gaussian), nor do we rely on peculiar velocity measurements (which are potentially biased).

By design, $\eta$ is thus a random fluctuating variable 
with zero mean. It attains non-vanishing values if cosmic expansion is not perfectly isotropic and homogeneous. Small departures from zero are expected due to local structure. Significant departures can indicate breakpoints of the Cosmological Principle, specially if these happen at higher redshift bins. Statistically, $\eta$ follows the same distribution as the distance moduli. In the remainder of this paper, we assume the distance moduli to be Gaussian distributed. This is a practical approximation that works well for high signal-to-noise photometric data and for modelling the scatter associated with calibrating the distance-indicator zero point; see however \cite{Dainotti:2024gca} for a discussion of possible subtle deviations from Gaussianity.

The method for estimating the expansion rate fluctuation field at a galaxy's position, as well as the optimal choice of redshift shell thickness to ensure that the quantity is monopole-free, are discussed in detail in Appendix~\ref{app:fit_mult}. 
Here we emphasize that the amplitude of the expansion–rate fluctuation field 
 $\eta$ is   fundamentally observer-dependent as demonstrated in \cite{paper2,paper3}. 
Following the analysis strategy of \cite{paper0,paper2}, we choose to reconstruct the $\eta$ field as inferred by an ideal observer boosted relative to the terrestrial observer, such that the CMB dipole vanishes in his/ her reference frame (the CMB frame). This is accomplished by systematically using redshifts  expressed in the CMB frame in our analysis.   
We emphasize that Doppler boosting directly measured data (e.g., in the heliocentric frame) into the CMB frame does not require knowledge of the cosmological model, the spacetime metric, or the rest frame of the cosmic fluid. It only necessitates knowledge of the Sun's velocity relative to the CMB, which is well-constrained by CMB experiments \cite{Planck:2018nkj}.
Adopting the notation  convention of  \cite{paper1,paper2}, 
the redshift and the expansion rate fluctuation field expressed in the CMB frame are denoted by an overtilde:
 $\tilde{z}, \;{\tilde \eta}({\tilde z}), \cdots$.

The expansion rate fluctuation field is inherently a discrete random field. To enable angular analysis, we construct a two-dimensional, piecewise-continuous representation by coarse-graining the data using the HEALPix tessellation scheme \cite{Gorski:2004by}. Specifically, we partition the sky into pixels and compute the inverse-variance weighted average of the measured $\tilde{\eta}$ values within each pixel $p$:
\begin{equation}
\tilde{\eta}(p) = \frac{\sum_i \tilde{\eta}(\tilde{z}_i, \bm{n}_i) / \sigma_i^2}{\sum_i \sigma_i^{-2}} \;,
\end{equation}
where $\sigma_i$ is the uncertainty in $\tilde{\eta}_i$, related to the distance modulus uncertainty via $\sigma_i = \sigma_\mu(\tilde{z}_i) / 5$.

\begin{figure}
\begin{center}
\includegraphics[scale=0.18]{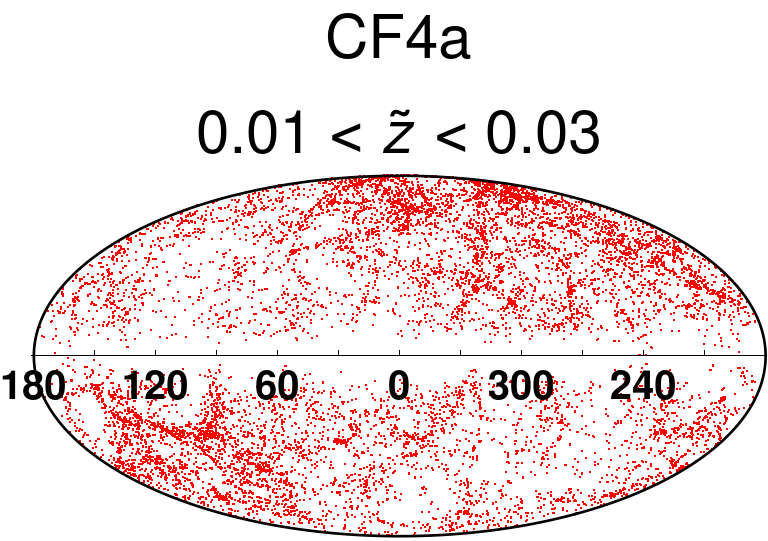}
\includegraphics[scale=0.18]{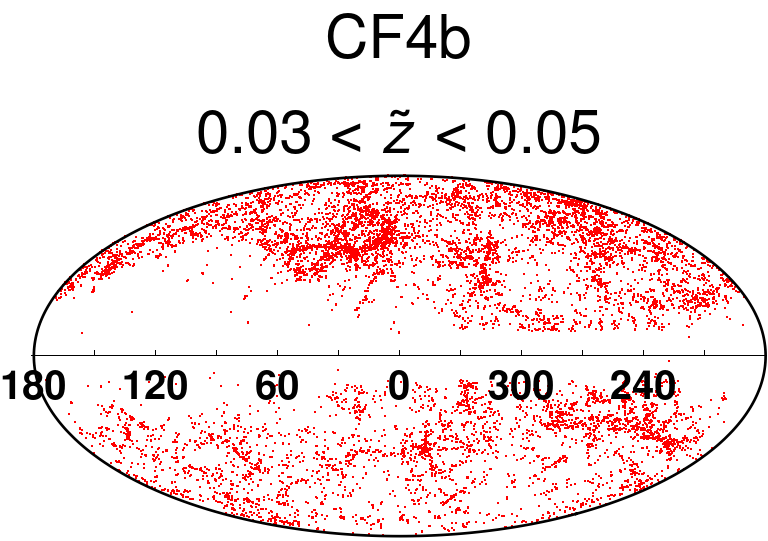}
\includegraphics[scale=0.18]{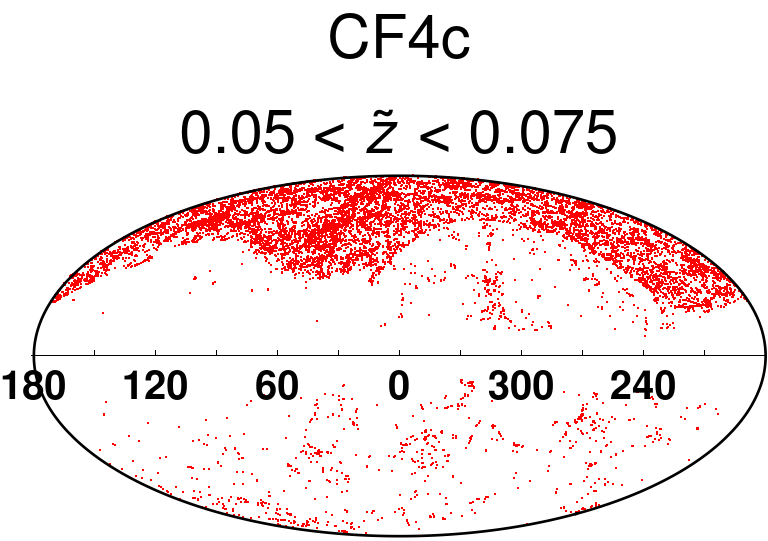}
\includegraphics[scale=0.18]{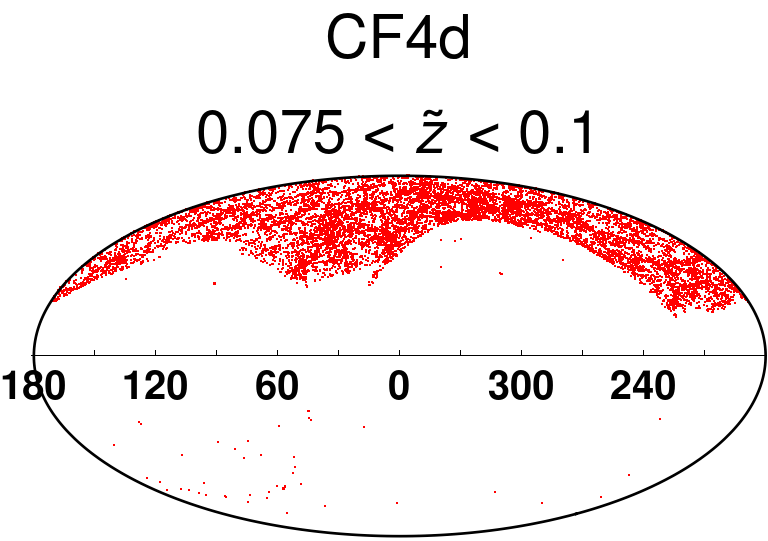}
\\
\includegraphics[scale=0.173]{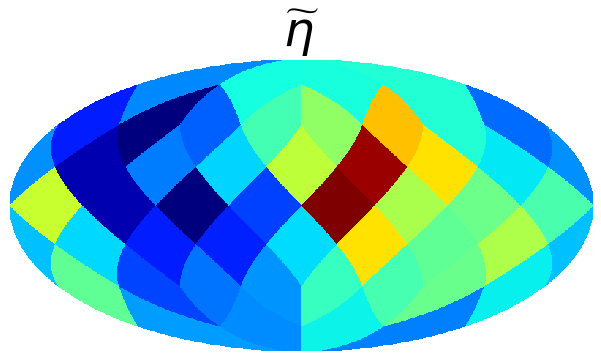}
\includegraphics[scale=0.173]{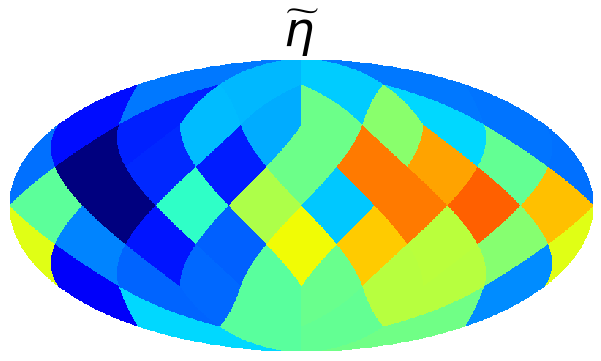}
\includegraphics[scale=0.173]{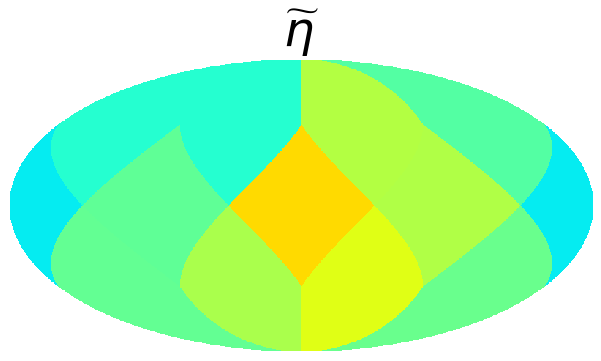}
\includegraphics[scale=0.173]{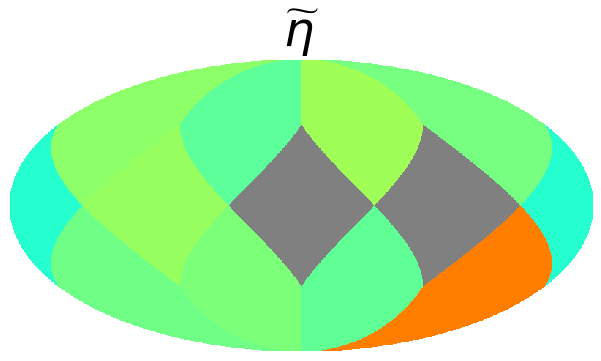}
\\
\includegraphics[scale=0.38]{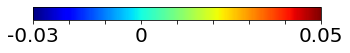}
\\
\includegraphics[scale=0.23]{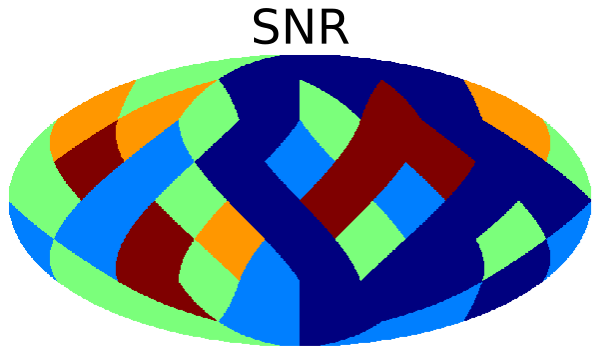}
\includegraphics[scale=0.23]{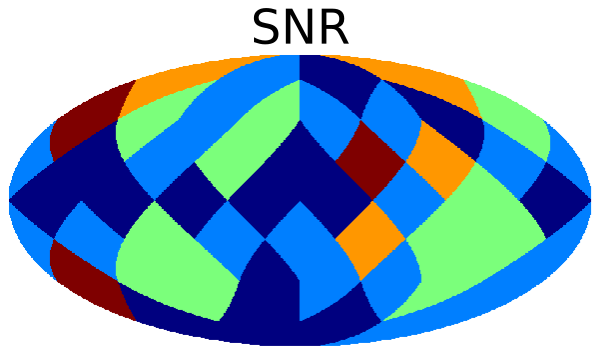}
\includegraphics[scale=0.17]{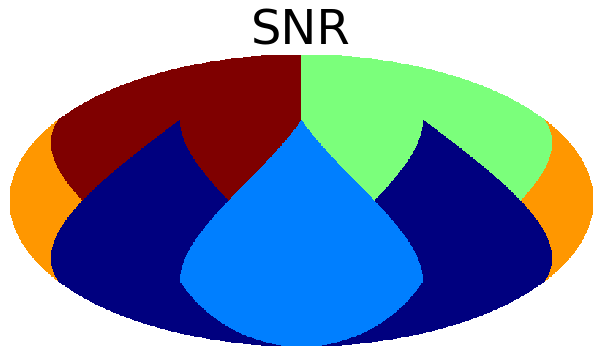}
\includegraphics[scale=0.17]{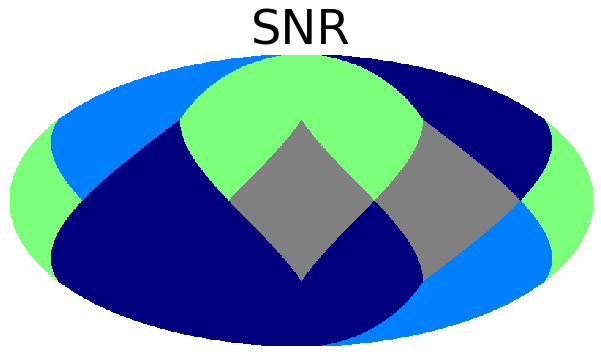}
\\
\includegraphics[scale=0.5]{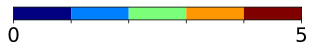}
\\
\includegraphics[scale=0.17]{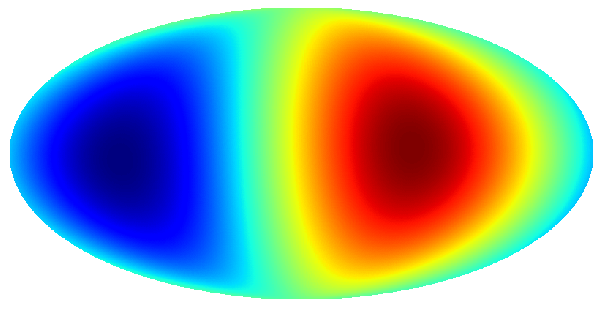}
\includegraphics[scale=0.17]{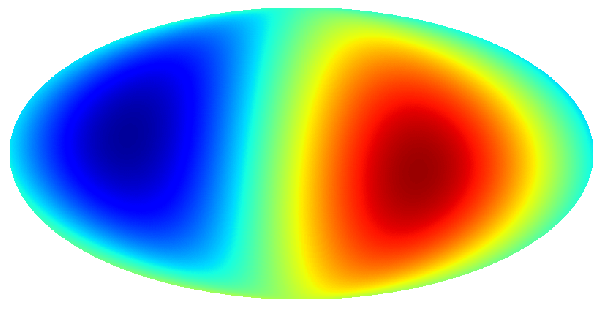}
\includegraphics[scale=0.17]{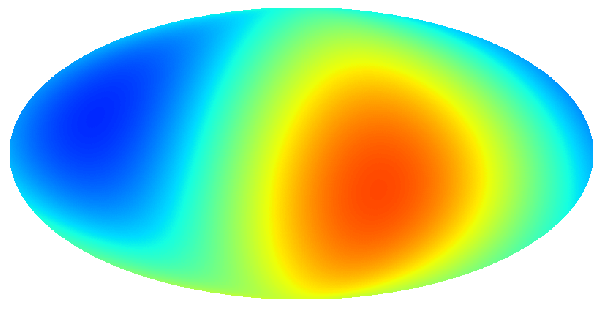}
\includegraphics[scale=0.17]{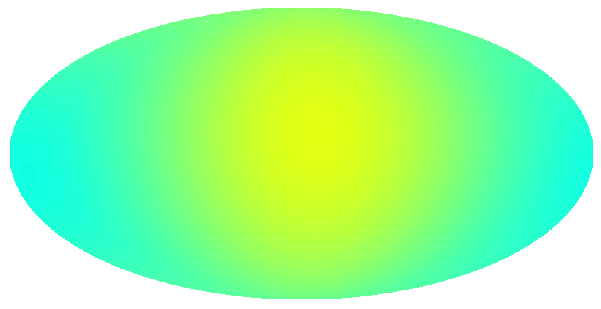}
\\
\includegraphics[scale=0.17]{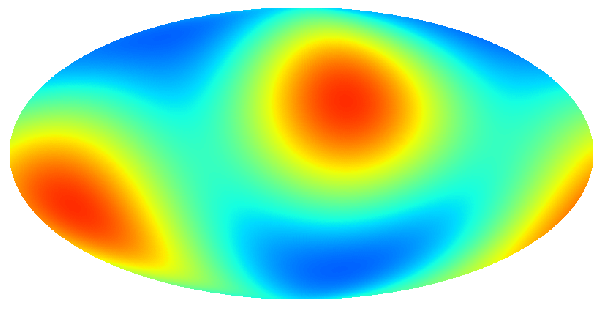}
\includegraphics[scale=0.17]{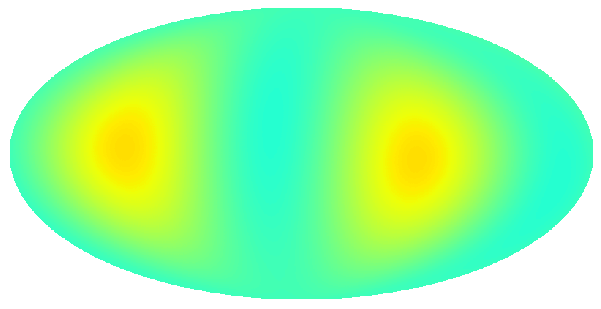}
\includegraphics[scale=0.17]{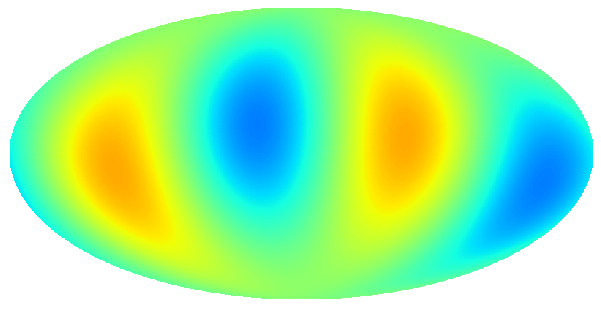}
\includegraphics[scale=0.17]{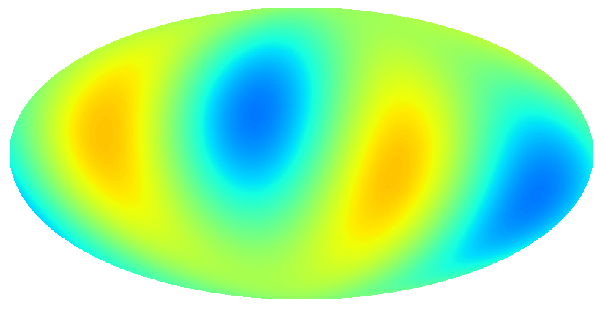}
\\
\includegraphics[scale=0.17]{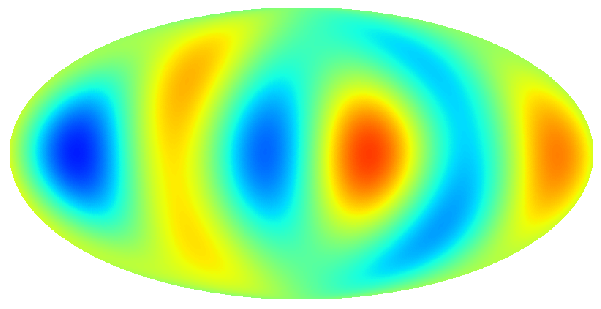}
\includegraphics[scale=0.17]{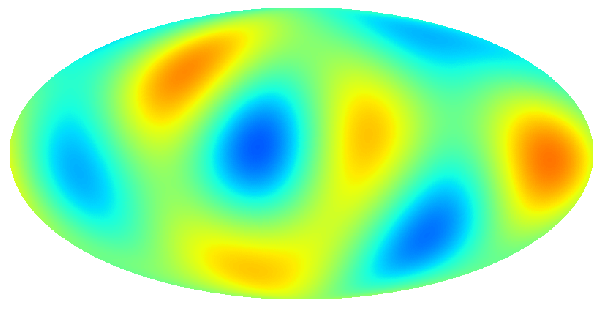}
\includegraphics[scale=0.17]{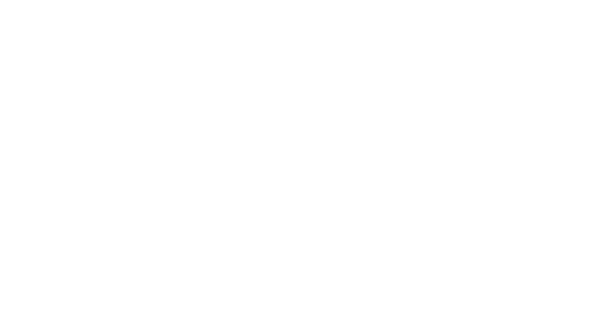}
\includegraphics[scale=0.17]{figures/fig_99emp.png}
\\
\includegraphics[scale=0.17]{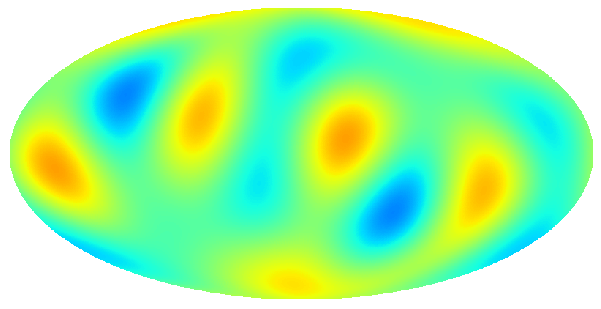}
\includegraphics[scale=0.17]{figures/fig_99emp.png}
\includegraphics[scale=0.17]{figures/fig_99emp.png}
\includegraphics[scale=0.17]{figures/fig_99emp.png}
\\
\includegraphics[scale=0.5]{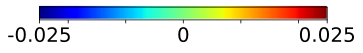}
\caption{
{\it Top row of panels:} Sky distribution of CF4 galaxies in Galactic coordinates. Different subsamples, corresponding to distinct redshift ranges, are shown from left to right: CF4a ($0.01 < \tilde{z} < 0.03$, 11978 galaxies), CF4b ($0.03 < \tilde{z} < 0.05$, 12660 galaxies ), CF4c ($0.05 < \tilde{z} < 0.075$, 13678 galaxies), and CF4d ($0.075 < \tilde{z} < 0.1$, 14486 galaxies).
{\it Second row of panels:} HEALPix-pixelized maps of the expansion rate fluctuation field \(\tilde{\eta}\).
{\it Third row of panels:} Signal-to-noise ratio maps of the fluctuation field (gray pixels indicate empty cells).
{\it Bottom four rows of panels:} Multipolar decomposition of \(\tilde{\eta}\). From top to bottom, the rows show the dipole, quadrupole, octupole, and hexadecapole components.
}
\label{eta_l_map2}
\centering
\end{center}
\end{figure}

Figure \ref{eta_l_map2} presents the 2-dimensional pixelized HEALPix maps of $\tilde\eta$ in Galactic coordinates for the samples CF4a, CF4b, CF4c, and CF4d respectively. The number of HEALPix pixels is set to 48 for the CF4a and CF4b samples, and to 12 for CF4c and CF4d, which exhibit sparser and more anisotropic sky coverage. The quality of the expansion rate fluctuation field reconstruction is quantified by the signal-to-noise ratio (SNR) in each pixel $p$ (third row of panels in Figure~\ref{eta_l_map2}). The SNR is computed as the ratio of the weighted average to its uncertainty, with the latter given by $\sigma(p) = \big( \sum_i \sigma_i^{-2} \big)^{-1/2}$. As expected, the noise decreases with increasing galaxy counts per pixel. The reconstruction deteriorates at higher redshifts due to sparser sampling, and near the Zone of Avoidance, where the Milky Way obscures large portions of the sky.

The two-dimensional expansion rate fluctuation field traced by each sample exhibits significant anisotropy, with fluctuations displaying coherent large-scale angular correlations rather than random noise. In the lowest redshift bin (CF4a), and for the given resolution of the reconstruction (see second row in Figure  \ref{eta_l_map2}), the expansion rate fluctuation peaks at $\tilde\eta \sim 0.05 \pm 0.01$ in a direction near the Galactic plane, indicating that, at fixed distances, redshifts are systematically higher than average in that region.  If interpreted within the framework of the standard cosmological model, this would imply a local variation in the Hubble constant, relative to a fiducial value of $H_0 = 70$km/s/Mpc, at the level of approximately $10\%$, i.e. $7$ km/s/Mpc.
Notably, a similar anisotropic pattern is qualitatively preserved across increasing redshift shells. However, in the highest redshift bin (CF4d), the relevant region of maximal expansion indicated by the lower-redshift shells is not sampled due to the limited sky coverage of the CF4d sample.

\section{Multipolar structure of the anisotropic expansion rate in the local Universe}

 The expansion rate fluctuation field \(\eta\), being a scalar field,  can be naturally and straightforwardly expressed in a spherical harmonic (SH) orthonormal basis:
\begin{equation}
\tilde{\eta}(z, \boldsymbol{n}) = \sum_{\ell=1}^{\infty} \sum_{m=-\ell}^{\ell} \tilde{\eta}_{\ell m}(z)\, Y_{\ell m}(\boldsymbol{n}),
\label{etamodo}
\end{equation}
where the real spherical harmonics are 
$Y_{\ell m}=[(2\ell+1)(\ell -|m|)!]^{1/2} [4\pi (\ell +|m|)!]^{-1/2} P^{|m|}_{\ell }(\cos{\theta})\, S_m(\phi)
$, with $S_m(\phi) = 1$ for $m=0$, $\sqrt{2}\cos(m\phi)$ for $m>0$, and $\sqrt{2}\sin(|m|\phi)$ for $m<0$ \cite{Gorski:1994ye}. The expansion coefficients $\tilde{\eta}_{\ell m}$ are therefore also real.
This decomposition ensures that contributions from distinct angular scales remain uncorrelated, thereby facilitating statistical likelihood analyses by preserving the independence of each scale's contribution.

In order to avoid excessive smoothing from the tessellation scheme, we do not use HEALPix for reconstructing the SH multipoles. Instead, we fit the SH coefficients directly to the galaxy data, as detailed in Appendix \ref{app:fit_mult}. This estimation is unbiased only if there is no mode coupling, i.e. provided that the fit is done up to a maximum multipole ($\ell_{\rm max}$), defined as the highest multipole with statistically significant SNR.  This was verified in Appendix \ref{app_ZOA_effect} by performing Monte Carlo simulations.

Several factors might contribute to the appearance of artificial multipolar signals in an otherwise perfectly uniform FLRW metric. These include random errors in the galaxy distance moduli, the
anisotropic sky distribution of the galaxies—particularly due to the ZoA and uneven sampling between the northern and southern Galactic hemispheres—and the finite thickness of the redshift shells
used to ensure that \(\tilde{\eta}\) is monopole-free. 

Unlike in \cite{paper0}, where we used  computationally intensive Monte Carlo
simulations, we now adopt a faster yet equivalent approach to assess the biasing impact of these various observational and reconstruction constraints.
Since $\tilde{\eta}$ is  a Gaussian random field  with zero mean, its spherical harmonic coefficients are jointly Gaussian random variables. We thus  consider the $\chi^2$ statistics for each multipole $\ell$,
\begin{equation}
\chi^2_\ell=\Big( \tilde{\boldsymbol{\eta}}_\ell-\boldsymbol{\eta}^{\rm M}_\ell \Big)^T\,\boldsymbol{\mathcal{C}}_\ell^{-1}\,\Big(\tilde{\boldsymbol{\eta}}_\ell-\boldsymbol{\eta}^{\rm M}_\ell\Big)\,,
\label{chi2_pval1}
\end{equation}
where $\boldsymbol{\mathcal{C}}_\ell$  is the covariance matrix of the SH coefficients of the multipole $\ell$ (estimated as detailed in Appendix \ref{app:fit_mult}), $\tilde{\boldsymbol{\eta}}_\ell$ denotes the vector of observed SH coefficients, 
\begin{equation}
\tilde{\boldsymbol{\eta}}_\ell=\Big( \tilde\eta_{\ell,-\ell}\,,~\tilde\eta_{\ell,-\ell+1}\,,\dots\,,~\tilde\eta_{\ell,0}\,,\dots\,,\tilde\eta_{\ell,\ell-1}\,,~\tilde\eta_{\ell,\ell}\Big)\,,
\label{vecel}
\end{equation} 
while $\boldsymbol{\eta}^{\rm M}_\ell$  denotes the corresponding vector of predicted SH coefficients in the model under consideration. Note that, while the SH coefficients depend on the coordinates, the $\chi^2_\ell$ is invariant.

For each sample analyzed, we replace the observed distances with the theoretical values expected in a homogeneous $\Lambda$CDM universe, defined by the \textit{Planck} 2018 cosmological parameters~\cite{Planck:2018nkj}, and reconstruct the corresponding expansion rate fluctuation field. We then estimate the best-fitting SH coefficients describing the signal, i.e. the vector $\tilde{\boldsymbol{\eta}}_\ell$. 
The latter is in principle not identically zero because of  systematics induced by the data  and the estimation pipeline.  We then apply the \(\chi^2\) statistic to test against the null hypothesis \(\boldsymbol{\eta}^{\rm M}_\ell = \boldsymbol{0}\). A deviation is  considered statistically significant when the probability (\textit{p}-value) of obtaining a larger \(\chi^2_\ell\) than that computed in eq.~\eqref{chi2_pval1}, assuming a $\chi^2$ distribution with \(2\ell + 1\) degrees of freedom, is less than 0.05.

At all redshifts accessible within the depth of the CF4 catalog, and for any choice of shell thickness \(\delta z \lesssim 0.025\), we find consistently low $\chi^2$  values for all multipoles \(\ell\), corresponding to \(p\)-values exceeding 95\%.  This indicates that, within measurement uncertainties, there is no statistically significant bias introduced by the shell widths adopted in our analysis (CF4a, CF4b, CF4c, CF4d), which were chosen to balance the need for maximizing the  statistical signal with the ability to trace its evolution across different cosmic epochs.

The $\ell \leq 4$  multipoles of the $\tilde{\eta}$ field, traced by the CF4 sample at varying depths,  are shown in the lower panels of Figure \ref{eta_l_map2}. The anisotropic signal is dominated by the dipole component, which accounts for approximately $50\%$  of the total expansion rate fluctuation. Its amplitude  ($\sim 0.025$) and direction $(l,b)\sim(290^\circ,0^\circ)$ remain broadly consistent across the first three redshift bins. In the highest redshift shell, the amplitude of the dipole signal is weaker and noisier due to reduced data coverage, yet its orientation and polarity remain in agreement with those at lower redshifts.

The quadrupole component of the $\tilde{\eta}$ anisotropy field is also statistically significant, with a peak amplitude of approximately $0.02$, comparable to that of the dipole. As in the case of the dipole, the quadrupole's spatial pattern remains stable across all redshift shells. Notably, the direction of the quadrupole maximum consistently fluctuates around that of the dipole, pointing toward $(l,b) \approx (330^\circ, 27^\circ)$ for CF4a, $(304^\circ, 2^\circ)$ for CF4b, $(296^\circ, 8^\circ)$ for CF4c, and $(303^\circ, -10^\circ)$ for CF4d.
The octupole and hexadecapole components, instead, can be reliably reconstructed only at low redshift (see next section), where they contribute non-negligibly to the $\tilde{\eta}$ anisotropy field, with peak amplitudes comparable to that of the quadrupole.

These results confirm and extend to higher redshifts ($0.05 < z < 0.1$) the preliminary finding reported in \cite{paper0}, based on the shallower and sparser CF3 sample: the maximum of the quadrupole component remains consistently aligned with that of the dipole. Moreover, where measurable, the octupole and hexadecapole components exhibit local maxima in the same general direction, which is broadly consistent with the preferred axis identified in \cite{paper0}, located at  $(l, b) \approx (285^\circ \pm 5^\circ,\ 11^\circ \pm 4^\circ)$.

While qualitative agreement is suggestive, it is insufficient to establish the robustness of the results. In the following section, we present a series of quantitative tests designed to assess their statistical significance.

\subsection{Robustness of the results}
\label{roburesu}
We begin by testing the robustness of the observed multipole components of \(\tilde{\eta}\) by assessing their compatibility with being just noise in an otherwise perfectly uniform \(\Lambda\)CDM universe. This hypothesis is tested using the \(\chi^2_\ell\) statistic defined in eq.~\eqref{chi2_pval1}, where \(\boldsymbol{\eta}_\ell\) denotes the vector of best-fitting SH coefficients obtained from the CF4 catalog, and \(\boldsymbol{\eta}^{\rm M}_\ell\) represents the expected result when the same analysis pipeline is applied to a uniform  \(\Lambda\)CDM universe. To construct $\boldsymbol{\eta}^{\rm M}_\ell$, we replace the observed distances in the CF4 samples with the theoretical values predicted by a homogeneous $\Lambda$CDM model at the same redshifts, reconstruct the corresponding $\tilde{\eta}$ field, and estimate its best-fitting SH coefficients.

A given multipole is considered statistically significant (i.e. inconsistent with arising from random noise) in a statistically homogeneous \(\Lambda\)CDM model if the probability of incorrectly rejecting the null hypothesis (i.e., the \textit{p}-value) falls below the conventional 5\% threshold.
In Table \ref{tab_pval1}, we present the $p$-value for each multipole $\ell$, in addition to the joint $p$-value of all the multipoles in each shell up to $\ell_{\rm max}$.

\begin{table}
\centering
\begin{tabular}{|c|cccccc|}
\hline
\multirow{3}{*}{Sample}                                              & \multicolumn{6}{c|}{$p$-value (\%)}            \\ \cline{2-7} 
& \multicolumn{4}{c|}{Uniform $\Lambda$CDM}                                                                                     & \multicolumn{2}{c|}{\begin{tabular}[c]{@{}c@{}}$\Lambda$CDM + bulk\\ ($400$ km/s)\end{tabular}} \\ \cline{2-7} 
& \multicolumn{1}{c|}{Dipole} & \multicolumn{1}{c|}{Quadrupole} & \multicolumn{1}{c|}{Octupole} & \multicolumn{1}{c|}{Joint} & \multicolumn{1}{c|}{Quadrupole}                            & Octupole                           \\ \hline\hline
CF4a                                                                 & \multicolumn{1}{c|}{0}      & \multicolumn{1}{c|}{0}          & \multicolumn{1}{c|}{0.06}     & \multicolumn{1}{c|}{0}        & \multicolumn{1}{c|}{0}                                     & 6.89                               \\ \hline
CF4b                                                                 & \multicolumn{1}{c|}{0}      & \multicolumn{1}{c|}{7.7}        & \multicolumn{1}{c|}{0.3}      & \multicolumn{1}{c|}{0}        & \multicolumn{1}{c|}{6.62}                                  & 0.42                               \\ \hline
CF4c                                                                 & \multicolumn{1}{c|}{0.02}   & \multicolumn{1}{c|}{0.02}       & \multicolumn{1}{c|}{-}        & \multicolumn{1}{c|}{0}        & \multicolumn{1}{c|}{0.02}                                  & -                                  \\ \hline
CF4d                                                                 & \multicolumn{1}{c|}{63.46}  & \multicolumn{1}{c|}{1.44}       & \multicolumn{1}{c|}{-}        & \multicolumn{1}{c|}{2.55}     & \multicolumn{1}{c|}{1.43}                                  & -                                  \\ \hline\hline
\begin{tabular}[c]{@{}c@{}}CF4TF\\ {[}0.01,0.05{]}\end{tabular}      & \multicolumn{1}{c|}{0}      & \multicolumn{1}{c|}{0}          & \multicolumn{1}{c|}{-}        & \multicolumn{1}{c|}{0}        & \multicolumn{1}{c|}{0}                                     & -                                  \\ \hline
\begin{tabular}[c]{@{}c@{}}CF4FP\\ {[}0.01,0.05{]}\end{tabular}      & \multicolumn{1}{c|}{0}      & \multicolumn{1}{c|}{38.13}      & \multicolumn{1}{c|}{0.02}     & \multicolumn{1}{c|}{0}        & \multicolumn{1}{c|}{24.36}                                 & 0.36                               \\ \hline\hline
\begin{tabular}[c]{@{}c@{}}CF4FP\\ {[}0.05,0.1{]}\end{tabular}       & \multicolumn{1}{c|}{2.2}    & \multicolumn{1}{c|}{0}          & \multicolumn{1}{c|}{-}        & \multicolumn{1}{c|}{0}        & \multicolumn{1}{c|}{0}                                     & -                                  \\ \hline
\begin{tabular}[c]{@{}c@{}}CF4TF+CF4SN\\ {[}0.05,0.1{]}\end{tabular} & \multicolumn{1}{c|}{6.21}   & \multicolumn{1}{c|}{54.24}      & \multicolumn{1}{c|}{-}        & \multicolumn{1}{c|}{34.71}    & \multicolumn{1}{c|}{59.48}                                 & -                                  \\ \hline
\end{tabular}
\caption{$p$-values (in $\%$) for each multipole across different samples.  Values are quoted to four decimal places; 0 indicates $p < 0.01\%$. The CF4SN and Pantheon+ samples have a signal only  in the dipole, with a $p$-value $<0.01\%$. The hexadecapole is only available for CF4a with $p$-value $<0.01\%$.}
\label{tab_pval1}
\end{table}

A strong dipolar fluctuation is detected across all redshift shells, independent of sample depth. In the two lowest redshift bins (CF4a and CF4b), the amplitudes are consistent at approximately $2.5 \times 10^{-2}$, and are highly unlikely to result from random distance measurement errors in a uniform $\Lambda$CDM universe $(p < 0.01\%)$. In CF4c, both the amplitude and signal-to-noise ratio decrease, though the dipole remains statistically significant $(p = 0.02\%$). In CF4d, the dipole is weaker and no longer statistically significant; however, its orientation remains consistent with that observed at lower redshifts.

The \(\chi^2\) test also indicates that the presence of a genuine quadrupole cannot be ruled out in any of the samples.
The probability of a spurious signal is $7.70\%$ for CF4b and $1.44\%$ for CF4d, and substantially lower for CF4a and CF4c. While the $p$-value for CF4b indicates only marginal significance, the consistent orientation and polarity of the corresponding multipole across the three other independent shells provide additional support for the robustness of the observed feature.

The octupole signal is statistically significant in both CF4a and CF4b, with $p$-values of $0.06\%$ and $0.3\%$, respectively, indicating a low probability of a false detection. In the higher redshift shells, the signal is noise-dominated and therefore omitted from Figure~\ref{eta_l_map2}.
Similary, the hexadecapole signal is robustly reconstructed  only for CF4a (with a $p$-value  $<0.01\%$), while it is negligible in the other samples.

Another  eventuality to test is whether the higher-order multipoles ($\ell \geq 2$) represent genuine cosmic signals or are instead mode-mixing artifacts, i.e. results from the leakage of a strong $\ell = 1$ signal into higher multipole components.
A physically relevant and plausible scenario in which the expansion rate field exhibits a strong, time-evolving dipole arises from a bulk motion of matter relative to the CMB, embedded within an otherwise uniform $\Lambda$CDM universe. We therefore perturb the uniform expansion rate with a large bulk flow of 400~km/s, oriented along the direction of the dipole estimated for each CF4 subsample, so as to generate a dipole amplitude which is roughly comparable to what we measured in each sample and whose amplitude scales  as $1/\tilde{z}$
(see eq. \eqref{vl_th}). We then process this model with our data analysis pipeline to  calculate $\eta^{\mathrm{M}}_{\ell}$ and test, using eq.  (\ref{chi2_pval1}),  whether this model  can plausibly  generate the observed SH coefficients $\tilde{\eta}_{\ell}$.  The $p-$value that  the observed  higher-order multipoles (\(\ell \geq 2\)) are spurious artifacts generated by the dipole leakage are  shown in Table~\ref{tab_pval1}. 
 
The multipoles for all shells are generally inconsistent with  resulting from mode-mixing.  A marginal exception is the quadrupole in CF4b, which—as previously noted—exhibits an intrinsically weaker signal compared to the other subsamples (CF4a, CF4c, and CF4d) and cannot also be excluded, at more than 92.3\% probability, as a possible result of dipolar  contamination.

Another important concern is whether the signal could be an artifact resulting from potential inhomogeneities in the CF4 catalog, which compiles distances obtained using a variety of different methods.
To address this, we examine whether the multipolar pattern observed in the full sample persists when the expansion rate is reconstructed separately using spiral galaxies, early-type galaxies, and SNIa—three samples with largely uncorrelated sky distributions and independently derived distance moduli.

\begin{figure}
\begin{center}
\includegraphics[scale=0.16]{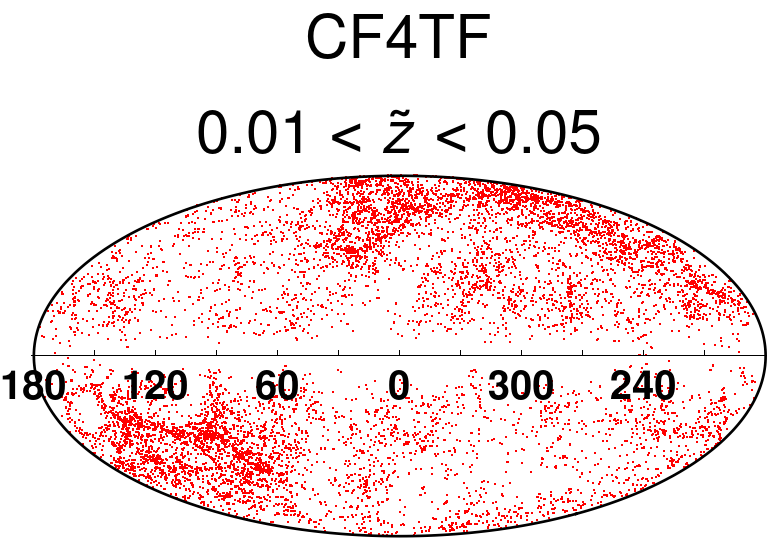}
\includegraphics[scale=0.16]{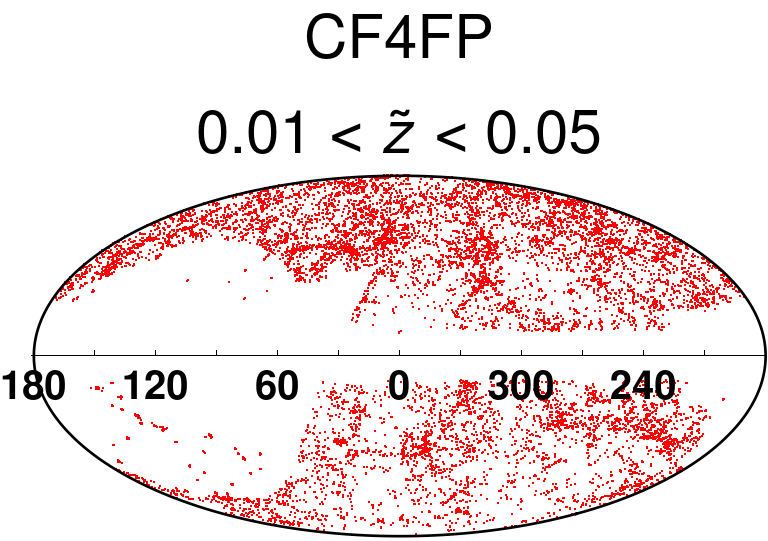}
\includegraphics[scale=0.16]{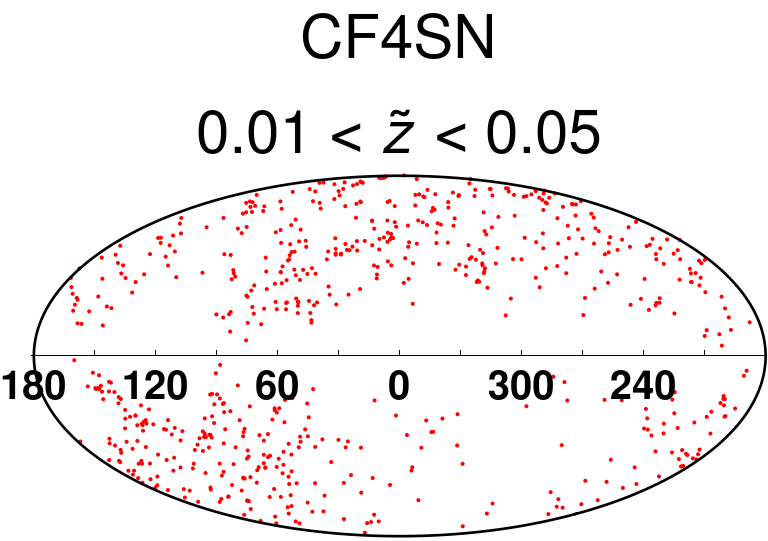}
\includegraphics[scale=0.16]{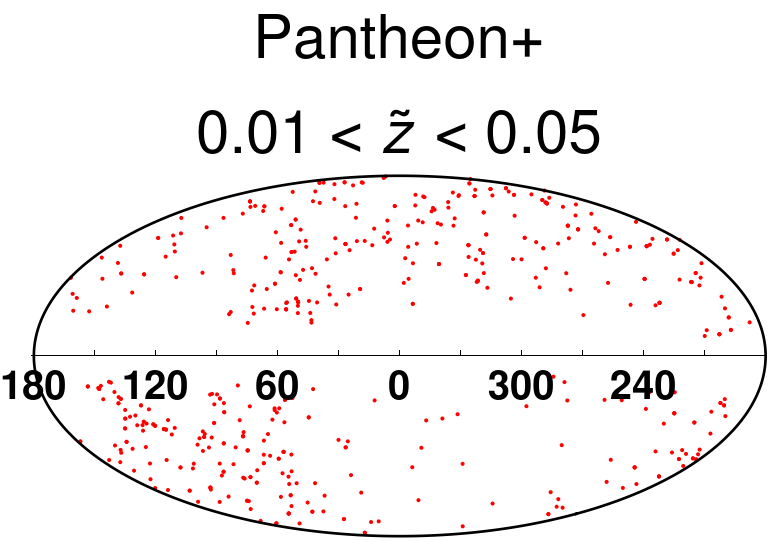}
\\
\includegraphics[scale=0.155]{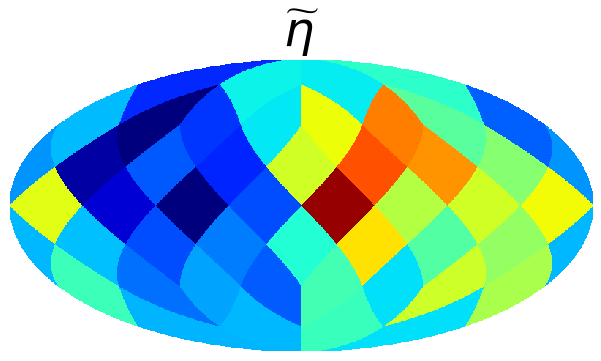}
\includegraphics[scale=0.155]{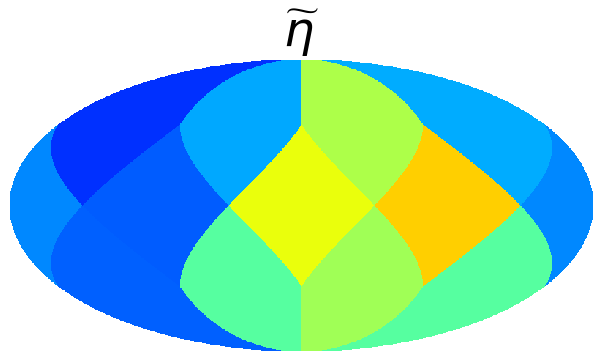}
\includegraphics[scale=0.155]{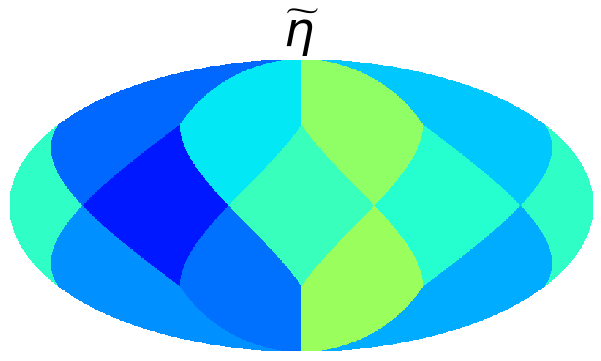}
\includegraphics[scale=0.155]{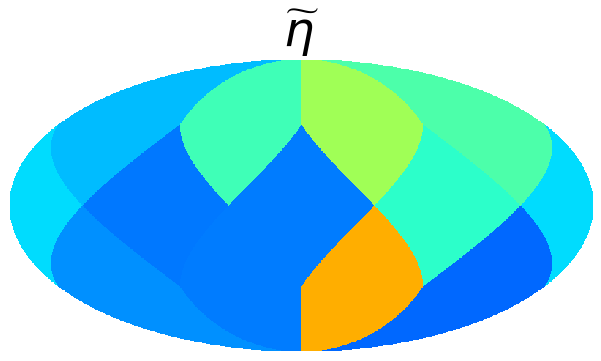}
\\
\includegraphics[scale=0.38]{figures/fig_99c1.png}
\\
\includegraphics[scale=0.155]{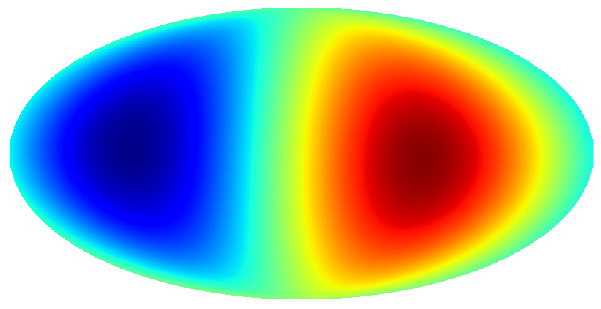}
\includegraphics[scale=0.155]{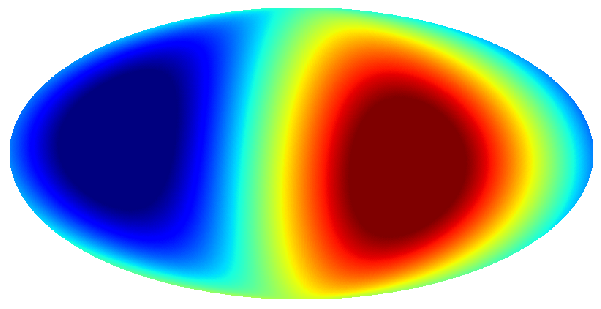}
\includegraphics[scale=0.155]{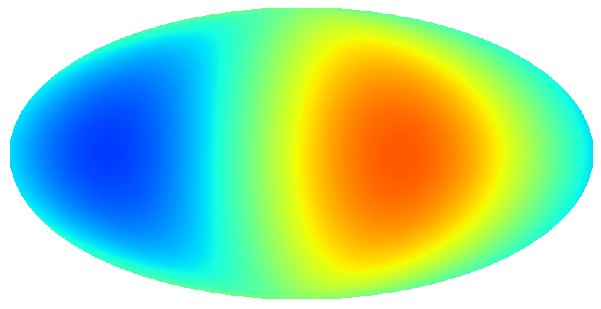}
\includegraphics[scale=0.155]{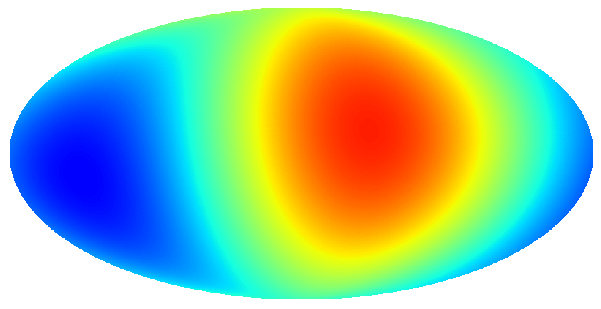}
\\
\includegraphics[scale=0.155]{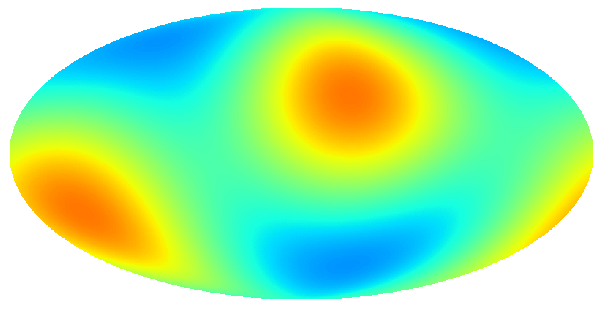}
\includegraphics[scale=0.155]{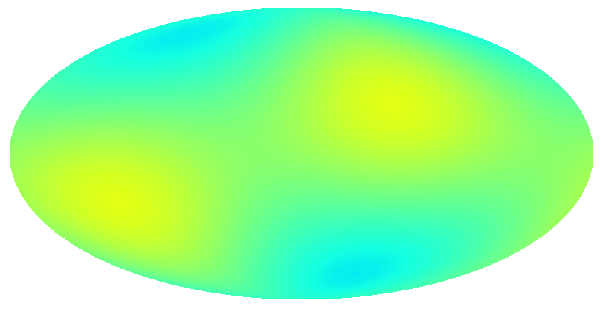}
\includegraphics[scale=0.155]{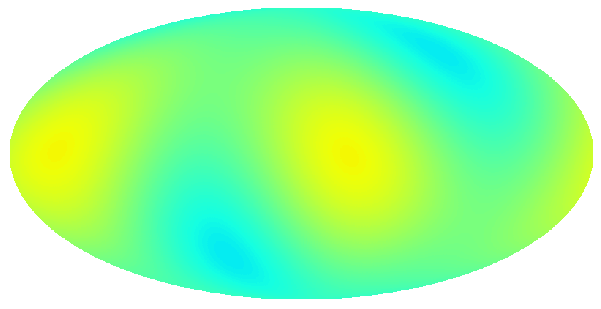}
\includegraphics[scale=0.155]
{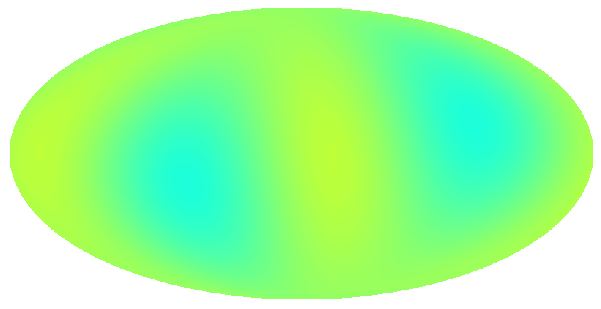}
\\
\includegraphics[scale=0.53]{figures/fig_99c2.png}
\caption{
{\it Top row of panels:} Sky distribution of CF4TF, CF4FP, CF4SN, and Pantheon+ samples in the range $0.01<\tilde z<0.05$.
{\it Second row of panels:} HEALPix-pixelized maps of the expansion rate fluctuation field \(\tilde{\eta}\).
{\it Bottom two rows of panels:} Multipolar decomposition of \(\tilde{\eta}\) into dipole ({\em top}) and quadrupole ({\em bottom}).}
\label{eta_maps_samp}
\centering
\end{center}
\end{figure}

Due to the sparse density of objects in these individual samples, only two single redshift shells, \(0.01 < \tilde{z} < 0.05\) and \(0.05 < \tilde{z} < 0.1\), could be analyzed. Results are shown in Figure~\ref{eta_maps_samp} for   \(\tilde{z} < 0.05\).
All the samples exhibit strong dipoles in the expansion rate (with $p$-values under 5\% of being of random origin), consistently oriented in the same general direction. 
The quadrupole is also detected in both the late- and early-type galaxy samples, with similar orientations; in both cases, the quadrupole maxima align closely and coincide with that observed in the full CF4a and CF4b samples. In the CF4FP and CF4SN samples, the quadrupole amplitude is weaker 
and only marginally significant (with $p$-values of approximately $38\%$ and 
$16\%$, respectively). In the former case, this reduced statistical significance of the signal to noise ratio arises  from the absence of galaxies in the large portion of the sky  toward which the quadrupole maximum points;  In the latter, it is due to the limited number of objects in the sample  (673 SNe).

Interestingly, the Pantheon+ sample, the sparsest of all those shown in Figure \ref{eta_maps_samp}, also exhibits the same dipolar and quadrupolar patterns identified in the galaxy samples. However, its quadrupole intensity is very weak, becoming apparent only when guided by the structures already revealed in the denser datasets.  All the Pantheon+ objects are also included in the CF4SN sample -- indeed they constitute a large fraction of this sample, around $63\%$ at $0.01<\tilde{z}<0.05$. This raises the question of whether the difference in intensity of the quadrupole observed in the CF4SN and Pantheon+ is statistical in nature, resulting solely from the different sample sizes. 

We therefore selected the subset  of common entries between the two catalogs--240 SNIa with \( 0.01 < \tilde{z} < 0.03 \) and 148 with \( 0.03 < \tilde{z} < 0.05 \) and assign to them the distance moduli quoted in the CF4SN sample  ( the column ``pantheon+'' of the catalog); we refer to this data set as CF4SN\(^{+}\). 
We further define a new sample, the CF4SN$^{-}$ sample, in which each SN in common between the CF4 and the Pantheon+ catalogs is assigned a distance modulus obtained by averaging all independent observations\footnote{These are listed in the columns "csp1", "ganesh", "amanullah", "prieto", "hicken", "folatelli", "walker", "stahl",
"twins", and "avelino" of the "All CF4 SNIa Samples" in the Extragalactic Distance Database.} of that object, excluding the Pantheon+ determination.

\begin{figure}
\begin{center}
\includegraphics[scale=0.17]{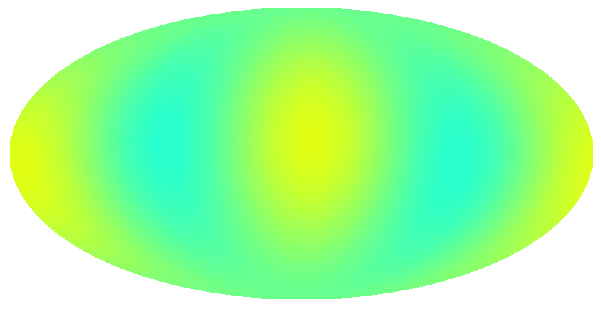}
\includegraphics[scale=0.17]{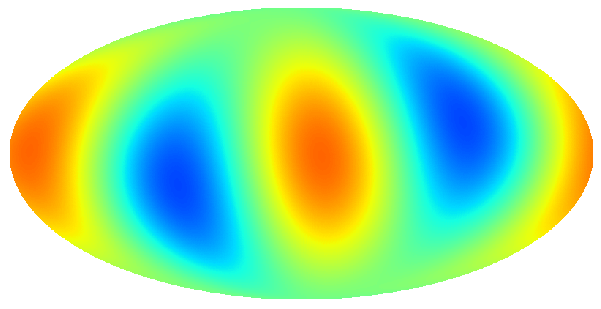}
\includegraphics[scale=0.17]{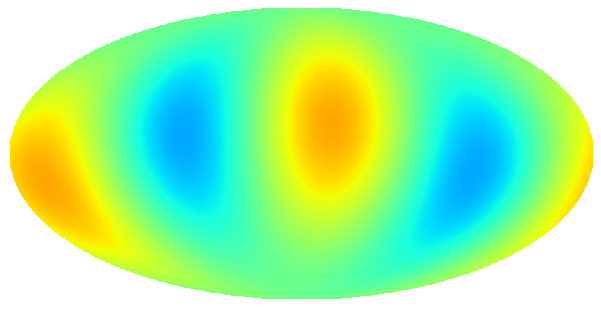}
\\
\includegraphics[scale=0.17]{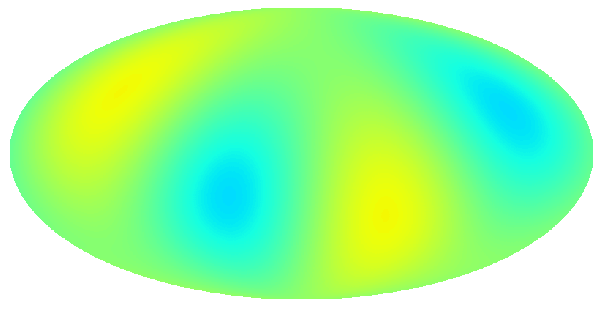}
\includegraphics[scale=0.17]{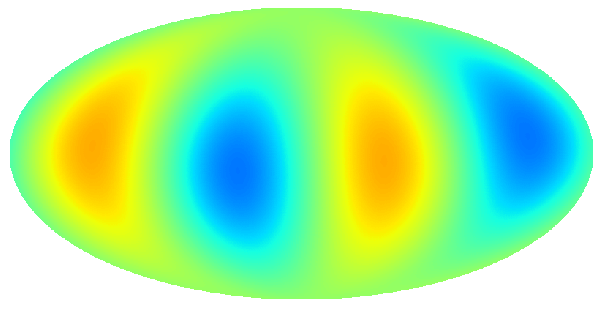}
\includegraphics[scale=0.17]{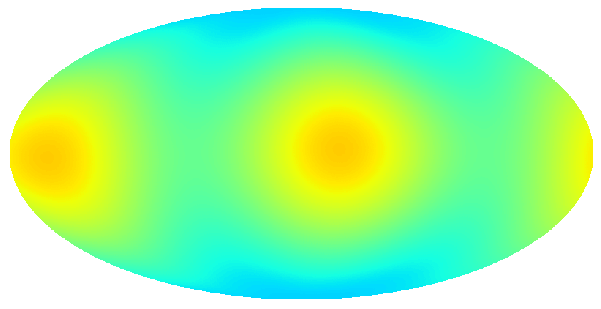}
\\
\includegraphics[scale=0.17]{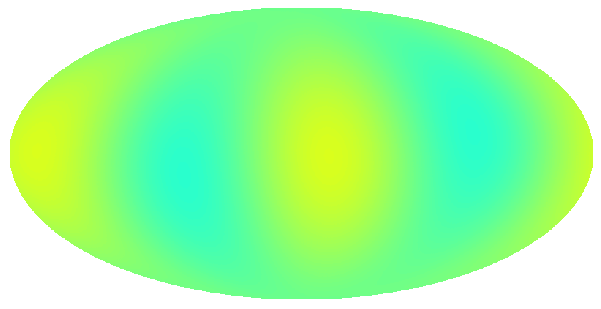}
\includegraphics[scale=0.17]{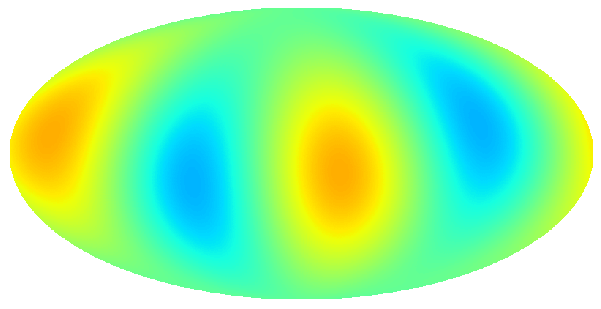}
\includegraphics[scale=0.17]{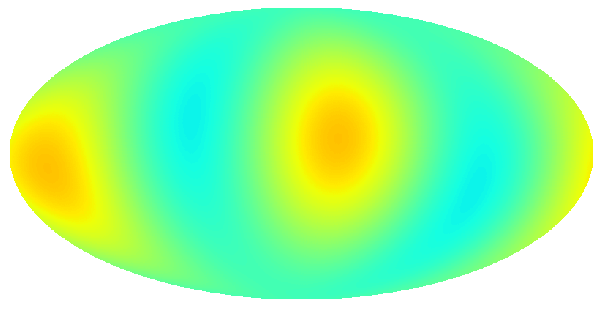}
\\
\includegraphics[scale=0.5]{figures/fig_99c2.png}
\caption{
Maps of the quadrupole reconstructed in the redshift ranges $0.01 < \tilde{z} < 0.03$ ({\em first row}), $0.03 < \tilde{z} < 0.05$ ({\em second row}), and $0.01 < \tilde{z} < 0.05$ ({\em third row}), using the same SNe but different determinations of their distance moduli: Pantheon+ ({\em first column}), CF4SN$^+$ ({\em second column}), and CF4SN$^-$ ({\em third column}).
}
\label{quad_sim_indip}
\centering
\end{center}
\end{figure}

The quadrupole component reconstructed from the distance moduli reported in the Pantheon+,  CF4SN\(^{+}\) and CF4SN\(^{-}\) samples is displayed in Figure~\ref{quad_sim_indip}. Owing to the limited sampling of these datasets, the statistical significance of the detection remains marginal.  Nevertheless, it is noteworthy that  all samples exhibit a quadrupolar pattern in the expansion 
rate fluctuation field that is mutually consistent and broadly aligned with those inferred from independent tracers, such as early- and late-type galaxies (cf. Figure~\ref{eta_maps_samp}). In contrast, the quadrupole derived from the CF4SN\(^{+}\)  and CF4SN\(^{-}\) samples appear systematically stronger than that obtained from Pantheon+.
As we have explicitly verified, this does not arise from the absence of a covariance matrix for the distance-modulus uncertainties in CF4 (unlike Pantheon+, which provides a full covariance matrix that is explicitly included in our analysis). Rather, it originates from the homogenization procedure applied to the various samples in the CF4 compilation, where the distance modulus is obtained as an MCMC average over multiple independent observations of the same SN.
This is confirmed in Figure \ref{quad_sim_indip}, where the quadrupolar amplitude of the expansion rate—calculated using, for the same objects in the Pantheon+ catalog, the average distance moduli reported by independent teams—is larger.
This analysis suggests that the various SN samples consistently trace the same quadrupolar structure detected in the CF4 galaxy samples. However, due to the sparseness of the SN data and the resulting low signal-to-noise ratio, the intensity of the quadrupole moment remains a subject for further investigation.

\begin{figure}
\begin{center}
\includegraphics[scale=0.21]{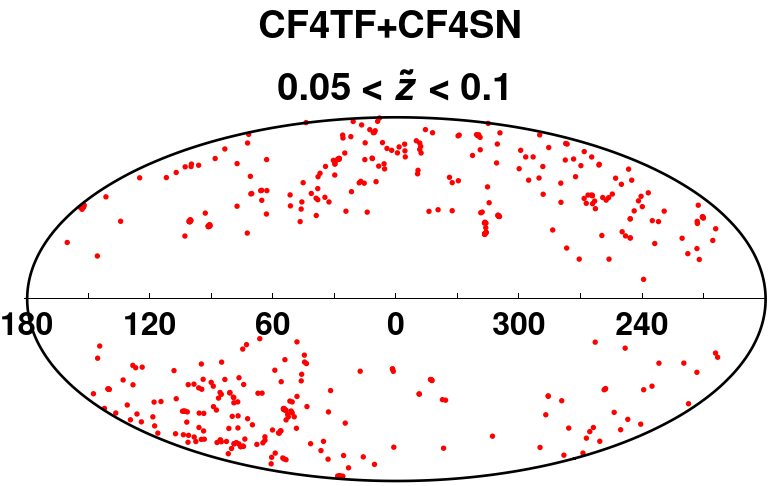}
\includegraphics[scale=0.21]{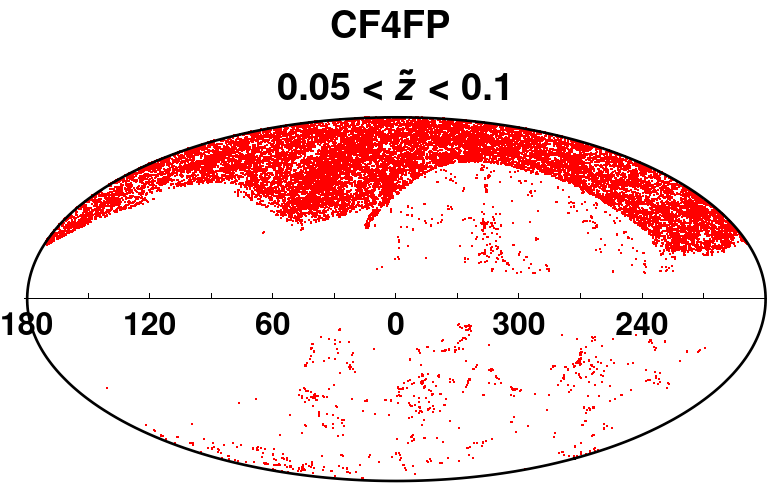}
\\
\includegraphics[scale=0.2]{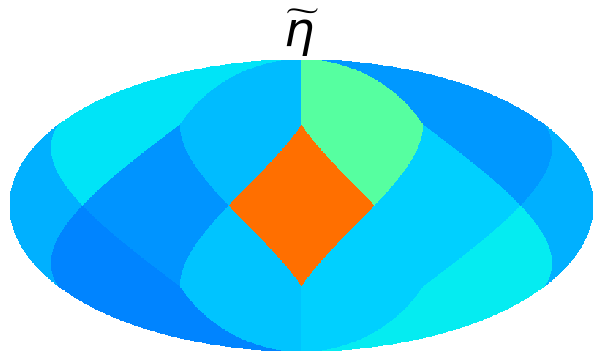}
\includegraphics[scale=0.2]{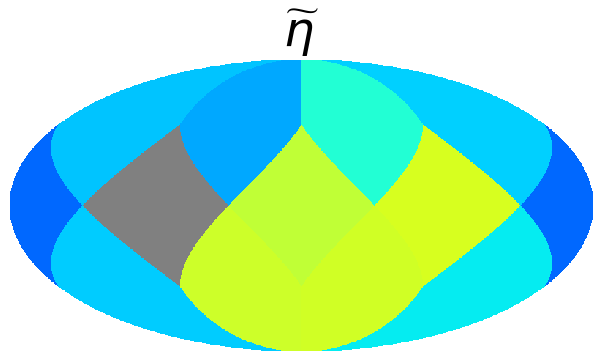}
\\
\includegraphics[scale=0.38]{figures/fig_99c1.png}
\\
\includegraphics[scale=0.2]{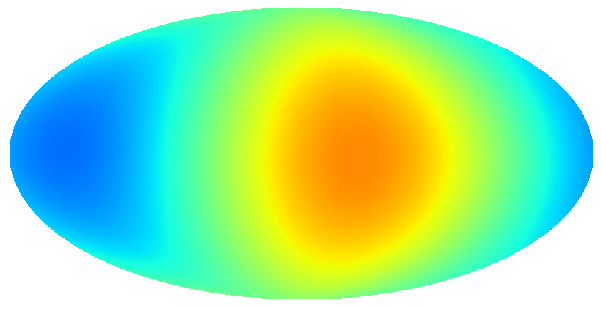}
\includegraphics[scale=0.2]{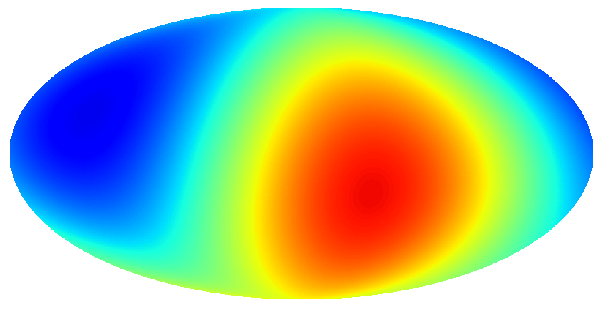}
\\
\includegraphics[scale=0.2]{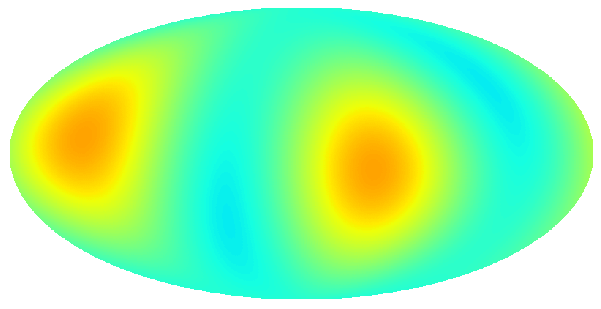}
\includegraphics[scale=0.2]{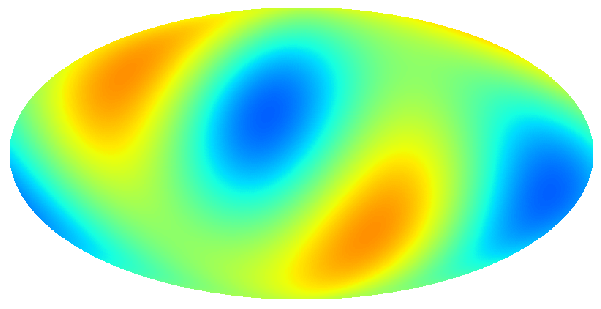}
\\
\includegraphics[scale=0.5]{figures/fig_99c2.png}
\caption{
{\it Top row of panels:} Sky distribution of CF4TF+CF4SN (combined)  and CF4FP  samples in the range $0.05<\tilde z<0.1$.
{\it Second row of panels:} HEALPix-pixelized maps of the expansion rate fluctuation field \(\tilde{\eta}\).
{\it Bottom two rows of panels:} Multipolar decomposition of \(\tilde{\eta}\) into the dipole ({\em top}) and quadrupole ({\em bottom}).}
\label{eta_mult_510}
\centering
\end{center}
\end{figure}

In the deepest redshift range investigated (\(z > 0.05\)), the richest individual sample is CF4FP, while combining CF4TF and CF4SN is necessary in order to maximize the SNR. The CF4FP sample exhibits a significant dipole and quadrupole, as shown in Table~\ref{tab_pval1} and Figure \ref{eta_mult_510}. Consistently, the combined CF4TF+CF4SN sample displays multipoles with similar amplitudes and orientations of those detected at low redshift (see Figure \ref{eta_mult_510}), although only the dipole amplitude reaches a statistically significant level, with a low probability ($\approx6\%$)  of being a noise artifact. 
Interestingly, both the dipole and quadrupole of the full CF4 sample are more significant than those of CF4FP alone, suggesting that the CF4TF+CF4SN subset reinforces the observed signal. The $p$-values improve from \(2.2\%\) to \(0.6\%\) for the dipole, and from \(0.002\%\) to \(0.00001\%\) for the quadrupole.

We have performed several complementary tests to identify and quantify potential biases that could mimic the observed signals. Notably, we investigated the effect of the ZoA on the observed directions of the multipoles, since they all appear to point intriguingly toward the region obscured by dust in the Galactic disk.  Methods and results are detailed in Appendix \ref{app_ZOA_effect}. We find no evidence that the anisotropic distribution of galaxies on the sky  has any substantial impact on the orientation of the inferred multipoles. A strong conclusion is that  anisotropies in the data distribution -- or in the measurement uncertainties -- increase the variance of the SH coefficients but do not bias the results. 

We also  examined the effect of distance-dependent selection biases, which arise due to the flux (or apparent magnitude) limits imposed in the construction of the samples. At a given redshift, this introduces an upper cutoff on the absolute magnitude, thereby biasing the average measured distance. We describe the test strategy and results in Appendix~\ref{app_malm}. We find that this effect impacts only the monopole term, \(\mathcal{M}\), to which the expansion rate fluctuation observable \(\eta\) is explicitly designed to be insensitive. As a result, the multipoles $\ell\ge 1$  of \(\eta\) remain unaffected (see Figure~\ref{Malm_sim1} and Figure \ref{Malm_sim2}).

Finally, the results remain virtually unchanged also when we replace the redshift of individual objects with the average redshift of the galaxy groups to which they may belong (see Appendix \ref{app_group}).

\subsection{Axial symmetry of the expansion rate fluctuation field}\label{sec_cf4_fit_etal}

The structure of the multipolar components of the local expansion rate fluctuation field confirms — and places on a firmer statistical foundation — the preliminary results previously obtained from the analysis of the CF3 catalog~\cite{paper0}. Of particular interest now is whether the extended CF4 sample also confirms that the three lowest multipoles — the dipole, quadrupole, and octupole — not only reach their maxima in nearly the same direction, but also exhibit an overall axisymmetric pattern. In that case,
the expansion rate field can be orthogonally decomposed using only the Legendre expansion:
\begin{equation}
\tilde{\eta}(z, \cos\theta) = \sum_{\ell=1}^{\ell_{\text{max}}} \tilde{\eta}_{\ell}(z) P_{\ell}(\cos \theta),
\label{Lcoeff}
\end{equation}
with coefficients given by
\begin{equation}
\tilde\eta_\ell(z) = \frac{2\ell + 1}{2} \int_{-1}^{1} \tilde{\eta}(z, \cos\theta)\, P_\ell(\cos\theta)\;\mathrm{d}(\cos\theta),
\end{equation}
and where \(\theta\) is the angle between the line of sight and the assumed axis of symmetry, and $P_\ell$ are the Legendre polynomials.

Rather than relying on a qualitative assessment, as in our preliminary analysis of the CF3 sample, we now address this question quantitatively. The technique we adopt is based on the fact that, if the field \(\tilde{\eta}\) is axially symmetric, the spherical harmonic coefficients \(\tilde{\eta}_{\ell m}\) satisfy the relation:

\begin{figure}
\includegraphics[scale=0.35]{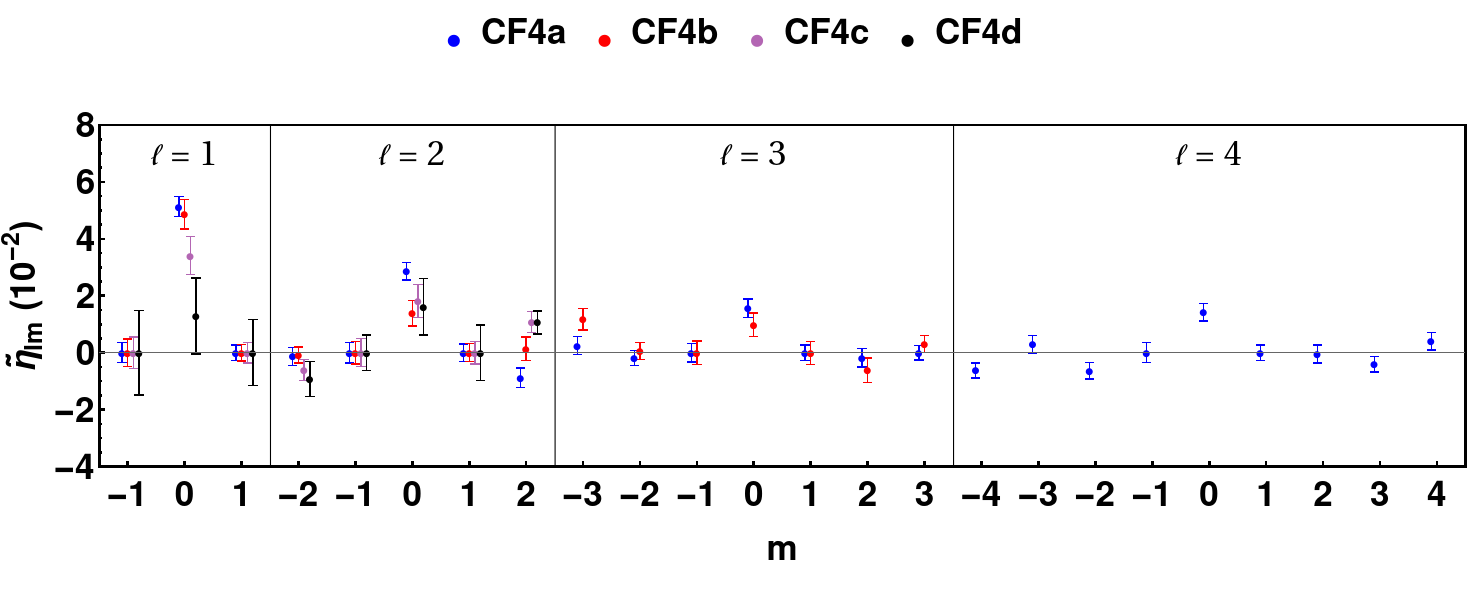}
\caption{The coefficients $\tilde\eta_{\ell m}$ for CF4 at different shells in redshift (CF4a, CF4b, CF4c, and CF4d). For each multipole, the z-axis is rotated to be the direction of its maximum (which is the closest to the dipole direction, see Table \ref{tab_max_dir}).}
\label{alm_CF4}
\centering
\end{figure}

\begin{table}
\centering
\begin{tabular}{|c|cc|cc|cc|cc|}
\hline
\multirow{2}{*}{Sample} & \multicolumn{2}{c|}{Dipole}    & \multicolumn{2}{c|}{Quadrupole} & \multicolumn{2}{c|}{Octupole}  & \multicolumn{2}{c|}{Hexadecapole} \\ \cline{2-9} 
& \multicolumn{1}{c|}{$l$} & $b$ & \multicolumn{1}{c|}{$l$}  & $b$ & \multicolumn{1}{c|}{$l$} & $b$ & \multicolumn{1}{c|}{$l$}   & $b$  \\ \hline \hline
CF4a                    & \multicolumn{1}{c|}{292} & 3   & \multicolumn{1}{c|}{331}  & 27  & \multicolumn{1}{c|}{297} & 1   & \multicolumn{1}{c|}{333}   & 7    \\ \hline
CF4b                    & \multicolumn{1}{c|}{287} & -9  & \multicolumn{1}{c|}{289}  & -3  & \multicolumn{1}{c|}{291} & 14  & \multicolumn{1}{c|}{-}     &   -   \\ \hline
CF4c                    & \multicolumn{1}{c|}{311} & -18 & \multicolumn{1}{c|}{296}  & 8   & \multicolumn{1}{c|}{-}   &   -  & \multicolumn{1}{c|}{-}     &   -   \\ \hline
CF4d                    & \multicolumn{1}{c|}{350} & 10  & \multicolumn{1}{c|}{303}  & -10 & \multicolumn{1}{c|}{-}   &  -   & \multicolumn{1}{c|}{-}     &   -   \\ \hline
\end{tabular}
\caption{The direction in Galactic coordinates (given in degrees) of the maximum of each multipole. For $\ell>1$ we show the direction of the maximum which is the closest to the dipole direction.}
\label{tab_max_dir}
\end{table}

\begin{figure*}
\begin{center}
\includegraphics[scale=0.23]{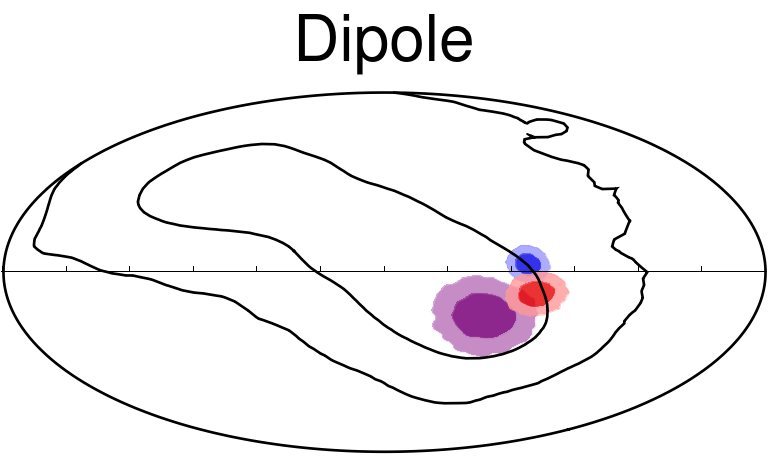}
\includegraphics[scale=0.23]{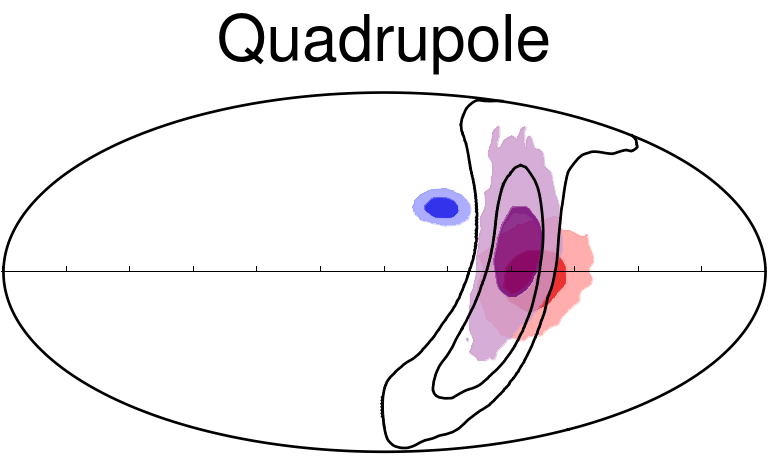}
\\
\includegraphics[scale=0.23]{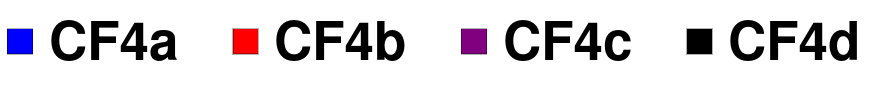}
\caption{The direction in Galactic coordinates of the maximum of the dipole and quadrupole of the CF4 samples at different redshifts  ($68\%$ and $95\%$ confidence levels).}
\label{Fig_lbdq}
		\centering
	\end{center}

\end{figure*}

\begin{equation}
    \tilde\eta_{\ell m}=\frac{4\pi}{2\ell+1}\tilde \eta_\ell Y_{\ell m}(\boldsymbol{n}_a)
\label{etalml}
\end{equation}
where $\boldsymbol{n}_a$ is a unit vector in the direction of the axis of symmetry and $\tilde\eta_\ell$ is the unique number specifying the multipolar decomposition (the Legendre coefficient) at a given order $\ell$. If the polar axis of the spherical coordinate system is aligned with the axis of symmetry $\boldsymbol{n}_a$, then all the $\tilde\eta_{\ell m}$ vanish except for $m=0$ and $\tilde\eta_{\ell 0} = [4\pi/(2\ell+1)]^{1/2}\tilde\eta_\ell$.

Figure \ref{alm_CF4} shows the coefficients $\tilde\eta_{\ell m}$ for CF4a, CF4b, CF4c and CF4d. For each multipole, we choose coordinates such that the z-axis points in the direction of the multipole's maximum, which is close to the dipole maximum. 

It is evident from Figure~\ref{alm_CF4} that the quadrupole and octupole components are consistent with the hypothesis of axial symmetry. The coefficients \(\tilde{\eta}_{20}\) and \(\tilde{\eta}_{30}\) are significantly different from zero (and positive) in all redshift shells, indicating that the axis of symmetry aligns with the direction of maximum amplitude for each respective multipole. In contrast, the remaining coefficients with \(m \neq 0\) consistently fluctuate around zero across different redshift shells. This behavior suggests that any deviations from axial symmetry are random and effectively average out over the full volume considered.

The directions of the maxima of the multipoles for the various CF4 subsamples at different depths are summarized in Table~\ref{tab_max_dir}. The dipole and quadrupole directions, along with their associated uncertainties, as estimated by Monte Carlo simulations, are shown in Figure~\ref{Fig_lbdq}. 
This supports the assumption of overall axial symmetry as a reasonable zeroth-order model for describing the structure of the local Universe.

We therefore  consider a single “global axis of symmetry” common to all multipoles and redshift shells. 
The orientation of this axis is treated as a free parameter, which we determine through a joint likelihood analysis of the samples. We use multiple CF4 samples, each corresponding to a different redshift shell, with its own set of estimated SH coefficients (all the multipoles in Figure \ref{eta_l_map2}). Assuming that all multipoles are axially symmetric about a common direction, we fit the Legendre coefficients using eq.~\eqref{etalml}, taking \(\boldsymbol{n}_a\) to be the same for all samples,  while allowing the amplitudes \(\tilde{\eta}_\ell\) to vary independently. The results are presented in the second to fifth rows of Table~\ref{tab_etal_cf4}.
The inferred direction of $\boldsymbol{n}_a$ is $(l,b) = (299^\circ, 5^\circ)$.

We can test the validity of the assumption of  ``global'' axial symmetry for the expansion rate fluctuation field, by comparing how well the $\tilde{\eta}$ signal is reconstructed using two different approaches: the full SH decomposition via the coefficients \(\tilde\eta_{\ell m}\), and the reduced decomposition using only the Legendre expansion $\tilde{\eta}_{\ell}.$
This comparison is carried out using  \(\chi^2\) statistics, i.e. by determining the sets of coefficients \(\tilde{\eta}_{\ell m}\) and \(\tilde{\eta}_{\ell}\) that best fit the data, and evaluating the goodness of the fits. 

If we assume that the CF4 sample exhibits no significant angular fluctuations—that is, the expansion rate field is fully described by a single constant (the monopole only, corresponding to one degree of freedom in the statistical test)—we obtain a minimum chi-square value of \(\chi^2_{\text{min}} = 47574.8\) ($\chi^2_{\rm red}=0.901$).

By contrast, the best-fitting axially symmetric multipolar model—which approximates the data in each of the four redshift shells using a fixed global axis of symmetry—has a total of \(5 + 4 + 3 + 3 = 15\) free parameters. These  correspond to the number of significant multipoles retained per shell under axial symmetry: 5 for the lowest shell, then 4, 3, and 3 for the subsequent shells. This model yields a significantly improved fit with \(\chi^2_{\text{min}} = 46578.7\) ($\chi^2_{\rm red}=0.882$).

For comparison, the best-fitting full spherical harmonic model, in which each multipole component can independently vary in amplitude and orientation across the four redshift shells, yields a slightly lower \(\chi^2_{\text{min}} = 46376.4\) ($\chi^2_{\rm red}=0.878$), but at the cost of a much larger number of parameters: \(25 + 16 + 9 + 9 = 59\). This highlights the efficiency of the axially symmetric model in capturing the structure of the data with far fewer degrees of freedom.

\begin{figure}
\centering
\includegraphics[scale=0.37]{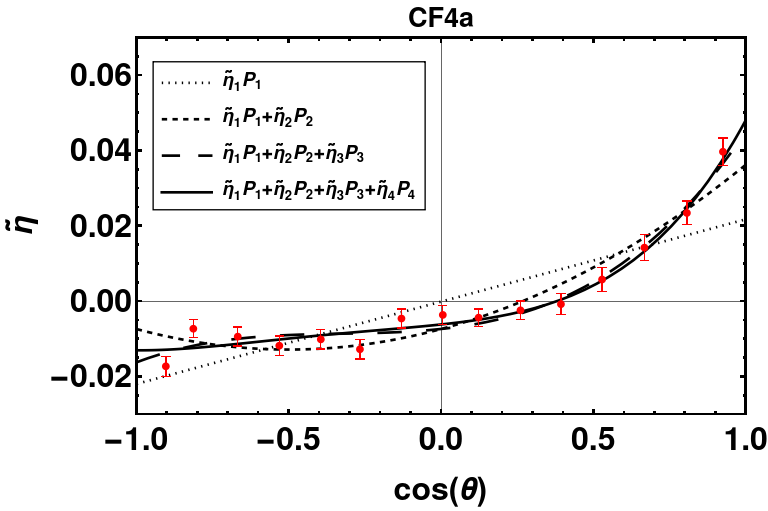}
\includegraphics[scale=0.37]{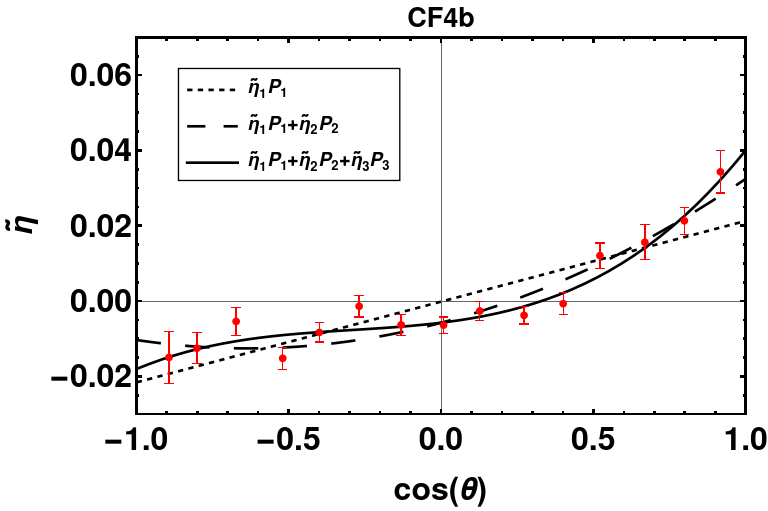}
\\
\includegraphics[scale=0.37]{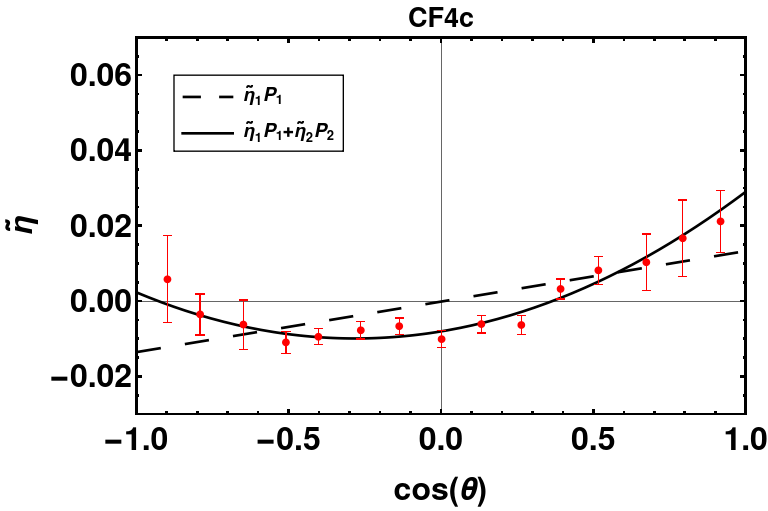}
\includegraphics[scale=0.37]{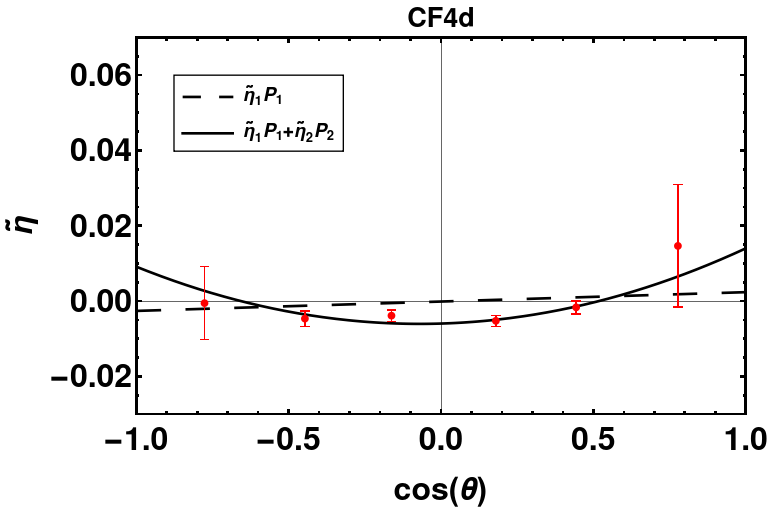}
\caption{The expansion rate fluctuation field $\tilde\eta$ for CF4a, CF4b, CF4c, and CF4d samples, in open spherical sectors, where  $\theta$ is the angular separation from the  axisymmetry direction $(l=299^\circ, b=5^\circ)$.}
\label{fig:eta_cos_CF4}
\centering
\end{figure}

Figure~\ref{fig:eta_cos_CF4} graphically displays the expansion rate fluctuation field \(\tilde{\eta}\) estimated from the CF4 sample, as a function of \(\cos\theta\) (recall that $\theta$ is the angular separation from the  axisymmetry direction), along with the predictions from the axisymmetric model using progressively higher-order multipolar approximations. It is evident that the residual fluctuations of \(\tilde{\eta}\) around the axisymmetric model decrease as higher multipoles are included. The dipole model alone provides a poor fit to the observed fluctuations, while the inclusion of higher-order terms—particularly the quadrupole—yields a substantial improvement across all four redshift shells analyzed.

The role of the quadrupole is especially noteworthy: its contribution is critical in capturing the structure of the expansion rate field and is even visually discernible in the deepest redshift bins accessible with the CF4 dataset, \(0.05 < \tilde{z} < 0.1\) (see the second row of Figure~\ref{fig:eta_cos_CF4}).

Table \ref{tab_etal_cf4} presents the best fitting Legendre coefficients of $\tilde\eta$ for different redshift shells. The dipole remains nearly constant up to $\tilde{z} = 0.05$, but its amplitude is roughly halved in the third shell and it is statistically compatible with vanishing in the deepest redshift shell. In contrast, the amplitude of the quadrupole stays constant  in the deepest shell analyzed. The octupole also shows no weakening over the limited depth that current data allows us to explore (up to $\tilde{z} = 0.05$.)

\begin{table}[h]
\centering
\begin{tabular}{|c|c|c|c|c|}
\hline
\begin{tabular}[c]{@{}c@{}}Sample\end{tabular} & $\tilde\eta_1$ $(10^{-2})$& $\tilde\eta_2$ $(10^{-2})$& $\tilde\eta_3$ $(10^{-2})$& $\tilde\eta_4$ $(10^{-2})$\\ \hline\hline
CF4a                                            & $2.19\pm0.13$    & $1.45\pm0.17$    & $0.88\pm0.21$    & $0.32\pm0.25$    \\ \hline
CF4b                                            & $2.14\pm0.21$    & $1.12\pm0.24$    & $0.77\pm0.29$    & -    \\ \hline
CF4c                                             & $1.34\pm0.22$    & $1.58\pm0.34$      & -    & -   \\ \hline
CF4d                                             & $0.25\pm0.25$    & $1.17\pm0.52$      & -    & -   \\ \hline
\end{tabular}
\caption{The Legendre coefficients of the expansion rate fluctuation field $\tilde\eta_{\ell}$ for CF4 samples.}
\label{tab_etal_cf4}
\end{table}

The axially symmetric hexadecapole ($\tilde{\eta}_4$) in CF4a shows a weaker signal compared to the full hexadecapole, because the chosen axis of symmetry is not perfectly aligned with the maximum of the hexadecapole. Unlike other multipoles, the hexadecapole is more sensitive to the orientation of the axis due to its smaller angular scale. Given these considerations, the hexadecapole contribution is small enough that, in the next analysis step, it will be neglected.

It is interesting to compare the results in Table \ref{tab_etal_cf4} with those obtained in \cite{paper0}, where we analyzed the CF3 sample using a different reconstruction pipeline. The CF3 sample is over three times sparser and only extends to about half the depth, covering redshifts up to $\tilde{z} < 0.05$. 
The dipole estimates for the two samples are highly consistent in the range 
$0.01 < \tilde{z} < 0.03$, showing only negligible differences. 
In contrast, for $0.03 < \tilde{z} < 0.05$, the CF4 dipole exhibits a significantly stronger signal,  differing from the CF3 result by $2.8\sigma$.  The quadrupole estimates, however, remain consistent, with differences of about $0.3\sigma$ and $1\sigma$ across the two intervals.  Similarly, the octupole estimates agree within $1.6\sigma$ and $1.2\sigma$, respectively.

\section{Interpreting expansion rate anisotropies in the framework of standard cosmology}

We now address the physical interpretation of the multipole moments of the  expansion rate fluctuation field $\tilde{\eta}$, as measured in the local universe.  These multipoles are analyzed within the framework of the standard cosmological  model, where they are understood as arising from local gravitational perturbations  of an underlying homogeneous and isotropic FLRW background.

\subsection{Bulk motion of matter}

The dipolar anisotropy observed in the expansion rate field $\tilde\eta$ can be attributed, within the standard model, to coherent peculiar motions of galaxies ($\boldsymbol{v}$) superimposed on an otherwise uniform Hubble flow.  In linearly perturbed FLRW models, the expansion rate fluctuation field in the CMB frame (the frame we use in this work) is related to the radial component of the peculiar velocity (at the time of emission) of galaxies, at low redshifts  by (see eq. (9.1) in \cite{paper2})
\begin{equation}
    \tilde\eta=\log\left(\frac{\tilde z}{\tilde d_L}\right)-\tilde{\mathcal{M}}\approx\log H_0-\tilde{\mathcal{M}}+\frac{\boldsymbol{v}\cdot\boldsymbol{n}}{\tilde z \ln10} \,.
\label{eta_vnH}
\end{equation}
In eq. \eqref{eta_vnH}, the first two terms have only a monopole component, and  for $\ell>0$,
\begin{equation}
    \tilde\eta_{\ell m}=\frac{(\boldsymbol{v}\cdot\boldsymbol{n})_{\ell m}}{\tilde z \ln 10}.
\label{etalmvlm2}
\end{equation}
Therefore, the bulk motion of a spherical volume $V(R)$ of radius $R$ can  be related to the dipole of the expansion rate fluctuation field  as \cite{paper2}
\begin{equation}
v_b =\frac{\ln10}{V(R)}\,{\int_0^{\tilde z(R)}\tilde\eta_1\; \tilde z \; \mathrm{d}V}\,,
\label{vl_th}
\end{equation}
where $\tilde z(R)\approx H_0 R$. 
Note that recovering the bulk flow with our fully model-independent method does not require assuming any value of $H_0$. Furthermore, since the expansion rate fluctuation field $\tilde{\eta}$ method is independent of the monopole of the expansion rate field, it is inherently robust against distance-dependent selection effects and against biases arising from the anisotropic sky distribution of the objects (see Appendices \ref{app_ZOA_effect} and \ref{app_malm}).

The bulk motion within a volume of radius $\tilde z < 0.05$ for CF4  is $ v_b = 520\pm 91\,\mathrm{km/s}$, directed along the axis of symmetry $(l, b) = (299^\circ, 5^\circ)$. This amplitude is about $2.3\sigma$ higher than our previous estimate in the same region using the CF3 sample (\( v_b = 307 \pm 23 \,\mathrm{km/s} \)) \cite{paper0}, primarily because the dipole in the range $0.03 < \tilde z < 0.05$ is significantly larger in CF4 than in CF3 (in the shallower volume $0.01<z<0.03$ the bulk flows estimated in CF4 or CF3 are in very good agreement).  Note that the larger error bars arise from the strongly anisotropic angular distribution of the CF4 sample at higher redshifts (see Appendix \ref{app_ZOA_effect}). This anisotropy prevents a reliable extraction of the bulk amplitude with meaningful signal-to-noise beyond $z \sim 0.05$ (also because, in our reconstruction technique, the uncertainties increase approximately linearly with redshift (see eq.~(\ref{vl_th})).

Our results confirm that the CF4 sample exhibits a large bulk flow, in agreement with previous findings reported in the literature. Specifically, we find good  agreement with the analysis of \cite{Whitford:2023oww}, who examined the same  CF4 sample using an alternative bulk--flow reconstruction method but adopting  a similar limiting radius to ours ($173\,h^{-1}\,\mathrm{Mpc}$). They obtained a 
bulk velocity of $v_b = 428 \pm 108 \,\mathrm{km/s}$, corresponding to a  $0.7\sigma$ level of agreement with our estimate, in the direction  $(l, b) = (297^\circ, 5^\circ)$. Despite an angular uncertainty of order  $15^\circ$, this direction is very well aligned with the axis of symmetry of 
the $\tilde{\eta}$ field.  
Similarly, our results are consistent (at the $1.4\sigma$ level in amplitude)  with those of \cite{Watkins:2023rll}, who, within a radius of  $150\,h^{-1}\,\mathrm{Mpc}$, reported a bulk velocity of 
$v_b = 387 \pm 28 \,\mathrm{km/s}$ in the direction 
$(l, b) = (297^\circ \pm 4^\circ, -6^\circ \pm 3^\circ)$. 
Moreover, our measurement is consistent with the bulk velocity reported by \cite{Duangchan:2025uzj}, who found a bulk amplitude of $v_b \approx 460 \pm 35$ km/s at a distance of $150h^{-1}$ Mpc, differing from our estimate by only $0.6\sigma$.

Part of the residual discrepancy is systematic in origin and arises from differences in the adopted definitions of bulk motion across various studies. In the present work, we adopt the definition given in eq. (2) of \cite{Nusser:2015oma}, which is directly interpretable in terms of the dipole component of the expansion rate field. By contrast, \cite{Whitford:2023oww, Watkins:2023rll} employ the definition given in eq. (1). The two formulations are equivalent in the limiting case of a velocity field that is spatially uniform within the considered volume. 
Other minor fluctuations may result from the use of subsamples of the CF4 data in previous studies (whereas we analyze the full dataset), or from the specific value of $H_0$ adopted (while our analysis is independent of this choice).
Nevertheless, the deviations introduced by these differing methods, conventions and samples are statistically marginal, thereby reinforcing the robustness of the expansion rate fluctuation field and its multipole decomposition.

\begin{figure}
\centering
\includegraphics[scale=0.42]{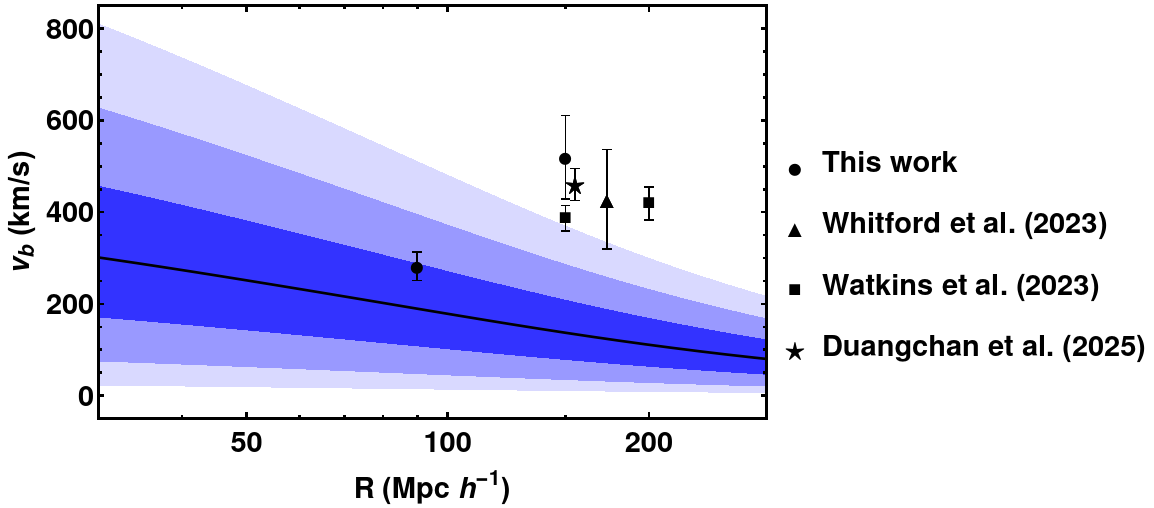}
\caption{Comparison of bulk velocity with the prediction from $\Lambda$CDM. The shaded regions indicate the $68\%$ (dark blue) and $95\%$ (light blue) confidence levels. Black solid circles show our estimated bulk velocity from CF4, while other markers show estimates from the literature.}
\label{CF4_vbulk}
\centering
\end{figure}

In Figure \ref{CF4_vbulk}, we present a comparison between the measured amplitude of the bulk velocity and the predictions of the standard $\Lambda$CDM model, adopting the cosmological parameters from the Planck best-fit solution. Consistent with earlier studies, we find that the amplitude inferred at the boundary of our sample, corresponding to scales of $R = 150$–$200\,h^{-1}\,\mathrm{Mpc}$, exhibits a level of tension with $\Lambda$CDM expectations at approximately the $\gtrsim 3\sigma$ significance level. By contrast, on smaller scales, $R < 100\,h^{-1}\,\mathrm{Mpc}$, the observed coherent flows are fully compatible with theoretical predictions. Importantly, the inclusion of our methodology within the broader set of existing bulk-flow estimators reinforces the robustness of this finding: the apparent excess power on large scales cannot be ascribed to statistical systematics or methodological artifacts, but rather appears to be a genuine feature present in the data.

Regarding the possible origin of this large bulk motion, we refer to \cite{Hoffman:2017ako}, who investigated the structure of the local Universe (\( z \lesssim 0.05 \)) using the CF2 sample. Their analysis indicates that the principal component of the quadrupole of the peculiar velocity field is predominantly influenced by the Shapley supercluster,  located at \((l, b) = (312^\circ, 31^\circ)\) and \( z \approx 0.048 \), while the dipole, reflecting  the bulk velocity,  is primarily influenced by the so-called  ``cosmic repeller'', a large under-dense region \cite{Hoffman:2023pac, Bohringer:2019tyj} located  at \((l, b) = (93^\circ, -18^\circ)\) and \( z \approx 0.053 \), with the corresponding motion directed toward the opposite point, \((273^\circ, 18^\circ)\). Interestingly, the axis of symmetry found in this work lies between the Shapley supercluster and the direction opposite to the cosmic repeller.

One crucial point to highlight is that, if the Shapley supercluster is the dominant structure in the region, the multipoles in the deepest redshift shell ($0.05<\tilde z<0.1$) should reverse their polarity and their amplitude should decay \cite{paper2,Sarma:2025yfw}. However, Figures \ref{eta_l_map2} and \ref{alm_CF4} show that the structure of the dipole and quadrupole components (both in amplitude and orientation) remain consistent in the deepest redshift shells, which survey the region beyond the Shapley supercluster. This provides strong evidence for the existence of a coherent, large-scale structure beyond $z \approx 0.05$, within which the local volume is embedded.

A more in-depth analysis is required to stress-test the robustness of the standard-model scenario—for example, a comprehensive comparison of the quadrupole and higher-order moments with Standard Model predictions. Such an analysis falls outside the scope of this paper and is addressed in a separate work \cite{paper_letter}.
However, given the hints of a possible tension that emerge when our findings are interpreted within the Standard Model framework, we now adopt a fully model-independent approach to analyze the observed fluctuations in the expansion rate field and to extract their physical meaning: the covariant cosmographic (CC) framework.

\section{Interpreting the anisotropies in the framework of covariant cosmography} \label{sec_cosmp}

The covariant cosmographic (CC) framework consists of a set of line-of-sight–dependent functions, defined at the observer's location, which are directly related to the matter four-velocity and the local geometry of spacetime \cite{paper1,paper2,paper3} (see also
\cite{Kristian_Sachs_1966,ellis_2009,MacCallum_Ellis_1970,Clarkson_theses_2000,Clarkson:2010uz,heinesen_2021,
Heinesen:2024gdi,Macpherson:2025qec,Adamek:2024hme}). They appear in the expansion of the luminosity distance in terms of redshift and provide a fully model-independent description of the observed expansion rate signal in any sky direction—i.e., without assuming a specific cosmological line element a priori.

Using the definition in eq. \eqref{defeta1}, $\eta$ (in the matter frame) is related to the CC  functions ($\mathbb{H}_o$,  $\mathbb{Q}_o$, $\mathbb{J}_o$, $\mathbb{R}_o$ and $\Sbb_o$) by
\cite{paper2,paper3}
\begin{align}
  \eta (z,\boldsymbol{n}) &=-\mathcal{M}(z) + \log \mathbb{H}_o(\boldsymbol{n}) -\frac{1-\mathbb{Q}_o(\boldsymbol{n})}{2 \ln 10}\, z
  +\frac{7-\mathbb{Q}_o(\boldsymbol{n}) \big[10+9\mathbb{Q}_o(\boldsymbol{n})\big] +4\big[\mathbb{J}_o(\boldsymbol{n})-\mathbb{R}_o(\boldsymbol{n})\big]}{24\ln 10}\, z^2 
   \notag \\ \label{eta_exp_1}
  &~ +\frac{1}{24 \ln10}\bigg\{-5 \Jbb_o (\boldsymbol{n}) \big[2 \Qbb_o(\boldsymbol{n})+1\big]+2 \big[\Jbb_o(\boldsymbol{n})-\Rbb_o(\boldsymbol{n})\big] \big[\Qbb_o(\boldsymbol{n})-1\big]
  \\ \notag
  &\qquad\qquad\quad +\Qbb_o(\boldsymbol{n}) \Big[9+2 \Qbb_o(\boldsymbol{n}) \big[5 \Qbb_o(\boldsymbol{n})+8\big]
  +6 \Rbb_o(\boldsymbol{n})\Big]+2 \Rbb_o(\boldsymbol{n})-5-\Sbb_o(\boldsymbol{n})\bigg\}\,z^3+
   \mathcal{O}(z^4). 
\end{align}
Here $\mathbb{H}_o$,  $\mathbb{Q}_o$, $\mathbb{J}_o$, $\mathbb{R}_o$ and $\Sbb_o$ are the generalized covariant Hubble, deceleration, jerk, curvature, and snap  at the event of observation $o$ ($z=0$). Moreover, 
in \cite{paper2,paper3}, we showed how the multipoles of $\eta$ are related to the multipoles of the CC functions 
for the case of axial symmetry.

Interestingly, these are not functional degrees of freedom, but rather a finite set — specifically, a limited number of covariant multipoles for each CC function.
In the most general case, this set comprises 86 degrees of freedom when the luminosity distance is expanded up to order \( O(z^4) \). However, as shown in \cite{paper3}, by assuming axial symmetry (such that each multipole has only one associated degree of freedom) and retaining only the dominant multipoles, this number reduces to 12.

An important aspect in theoretical treatments of the reliability of the CC formalism is the practical distinction between its formal and operational definitions (see e.g., \cite{Koksbang:2024nih}). While the higher-order CC parameters are formally defined as successive derivatives of the luminosity distance with respect to redshift (in the matter frame) evaluated at the observer’s position, their actual estimation relies {\em not} on derivatives at the observer, but on fitting. Specifically, the luminosity distance is treated as an explicit function of redshift over a finite interval, and the CC parameters are obtained as the best-fit coefficients that minimize the discrepancy between the model and observed distance-redshift data within that interval (see also \cite{paper3}).
While these parameters are defined locally, they are estimated non-locally to reduce sensitivity to local noise in the surroundings of the observer and to be more sensitive to the large-scale structure of the gravitational field.  In this sense, for example, although the Hubble constant $H_0$ is formally defined as the first-order derivative of redshift with respect to distance at the observer's position in the standard model, it is empirically determined as the slope of the distance–redshift relation over a finite redshift interval. This technique offers three main advantages. Theoretically, it avoids reliance on model-dependent estimates of matter density fields or on the  ill-defined problem of averaging over an a priori unknown background spacetime. Observationally, it yields more stable estimates across a broader redshift range, with reliability that can be verified a posteriori through goodness-of-fit tests or consistency checks. Moreover this approach also provides an estimate of the minimum radius around the observer within which data must be excluded from the analysis, if one is to recover the large scale geometry of space.  Physically, this means that one can quantify directly from data the characteristic scale at which the CC formalism breaks down due to the failure of the fluid limit approximation on which it relies.

The first step consists in estimating the Legendre coefficients of the expansion rate fluctuation field $\tilde\eta_\ell$ by means of the maximum likelihood method, described in Appendix \ref{app_fit_eta_l}. 
The highest multipole included in the Legendre expansion is the octupole ($\ell=3$).
Indeed, as discussed in Section \ref{sec_eta}, the higher multipoles are weak and their amplitudes are statistically insignificant.

While we estimate all the multipoles including the monopole \( \tilde{\mathcal{M}} \), we do not include the monopole in the fitting process -- so that the results are independent of the absolute calibration of distances. This approach helps to avoid biases arising from zero-point calibration systematics across different samples. The inclusion of the monopole will be explored in future studies. This conservative choice leads to degeneracies between certain multipoles of the cosmographic parameters. In particular, the monopole of \( \mathbb{H} \) cannot be constrained, and the quadrupole of \( \mathbb{H} \) cannot be directly measured. Instead, only the ratio between the quadrupole and the monopole of \( \mathbb{H} \) can be determined.
Furthermore, at third order in the expansion of $\tilde{\eta}_\ell$, instead of directly measuring the dipole and octupole of the snap, only the combinations $\mathbb{S}_1 + 4\mathbb{Q}_1(2\mathbb{J}_0 - \mathbb{R}_0)$ and $\mathbb{S}_3 + 4\mathbb{Q}_3(2\mathbb{J}_0 - \mathbb{R}_0)$ can be estimated (see Eqs. (B.1)--(B.4) in \cite{paper3}).
Therefore, the number of parameters that need to be fitted is $8$, including the velocity of the matter observer with respect to CMB frame $v_o$ (see eq. (4.10) in \cite{paper2}):
\begin{equation}
\boldsymbol{X}=\Big\{ \Hbb_2/\Hbb_0,\;\Qbb_0,\;\Qbb_1,\;\Qbb_3,\;\Jbb_2,\;\mathbb{S}_1 + 4\mathbb{Q}_1(2\mathbb{J}_0 - \mathbb{R}_0),\;\mathbb{S}_3 + 4\mathbb{Q}_3(2\mathbb{J}_0 - \mathbb{R}_0),\;v_o\Big\},
\end{equation}
where the subscript indicates the multipole moment $\ell$. We first estimate the Legendre coefficients $\tilde{\eta}_\ell$ in each redshift bin, as described in Section \ref{sec_cf4_fit_etal}, and then use these to infer the cosmographic parameters by minimizing the $\chi^2$ statistic,
\begin{equation}                \chi^2(\boldsymbol{X})=\sum_{i=1}^{N_{b}}\boldsymbol{\Delta\tilde\eta}^T(i)\,\boldsymbol{\psi}(i) \,\boldsymbol{\Delta\tilde\eta}(i) \,,
\label{chi2_cp}
\end{equation}
where $\boldsymbol{\Delta\tilde\eta}(i)$ is a vector with an index $\ell$, with elements $\Delta\tilde\eta_\ell(i) = \tilde\eta_\ell(i)-\tilde\eta_{\ell}^{\rm (model)}(\tilde z_i,\boldsymbol{X})$, where $\tilde\eta_\ell(i)$ is the measured multipole of $\tilde\eta$ for the $i$-th shell, and $\tilde\eta_{\ell}^{\rm (model)}(\tilde z_i,\boldsymbol{X})$ is the theoretical model (see Eqs. (B.1)--(B.4) in \cite{paper3} with eq. (4.10) in \cite{paper2}). The matrix $\boldsymbol{\psi}(i)$ is the inverse of the covariance matrix between the multipoles in the $i$-th shell, and is calculated as shown in Appendix \ref{app_fit_eta_l}.

The number of spherical shells $N_b$, used to reconstruct the expansion rate field, depends on the richness of the sample. This choice is guided by the need to balance two competing requirements: each shell must contain a sufficiently large number of objects to reduce statistical noise, while also being thin enough to avoid biasing the recovered CC parameters. 

\begin{figure}
\begin{center}
\includegraphics[scale=0.4]{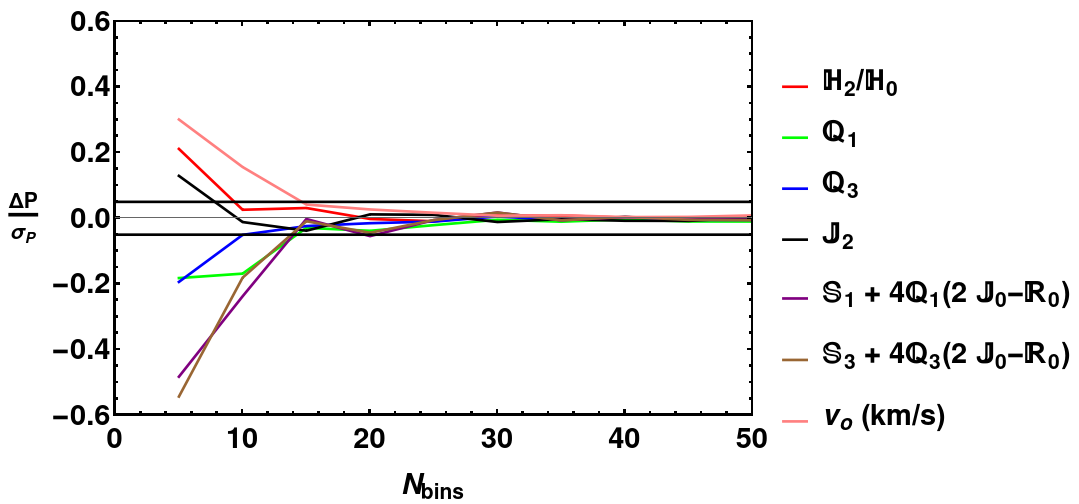}
\caption{The relative error in recovering the input CC  parameters ($P$) as a function of $N_b$, the number of shells in which the expansion rate signal is reconstructed. Results  are based on Monte Carlo simulations of the full CF4 sample in the redshift range $0.01<\tilde z<0.1$. The black lines represents $\pm0.05 \sigma$.}
\label{ccpar_bins}
\centering
\end{center}
\end{figure}

Figure~\ref{ccpar_bins} illustrates how the difference between the input and recovered values of the cosmographic parameters of the CF4 sample varies with $N_b$. 
To facilitate graphical visualization of the data--model comparison, we use the minimum number of $N_b$ in the analysis that avoids introducing bias. We find that this optimal number of shells is $N_b = 15$ for the CF4, CF4TF, and CF4FP samples; $N_b = 10$ for CF4SN; and $N_b = 9$ for Pantheon+ (see Section~\ref{sec:data2}). As the plot shows, any different sampling strategy relying on a larger number of bins will not modify the estimated parameters.
The bin sizes are not equal across all shells. At lower redshifts, the distribution of objects is more isotropic, resulting in reduced uncertainties in the multipole measurements. This improved precision permits the use of finer redshift bins; accordingly, we adopt a binning scheme in which $\delta \tilde z$ scales proportionally with $\tilde z$.

Among the CC parameters to be fitted, the monopole $\Qbb_0$ is the only one for which we impose a prior. Although $\Qbb_0$ enters the expressions for $\tilde\eta_1$ and $\tilde\eta_2$, these multipoles provide only weak constraining power. In contrast, the monopole $\mathcal{M}$—excluded from the analysis of the expansion rate fluctuation field—is the component most sensitive to $\Qbb_0$. For this reason, we impose a Gaussian prior on $\Qbb_0$, centered at $\Qbb_0 = -0.5$ with a standard deviation of $\sigma_{\Qbb_0} = 0.5$.
The physical motivation for this choice is that both simulations and analytical arguments (see \cite{paper2}) indicate that the monopole of the CC deceleration parameter in perturbed FLRW models is largely insensitive to the specific structure of density fluctuations. Instead, it is primarily determined by the background cosmological value, which is around $-0.5$ in a flat $\Lambda$CDM model.

\begin{figure}
\begin{center}
\includegraphics[scale=0.4]{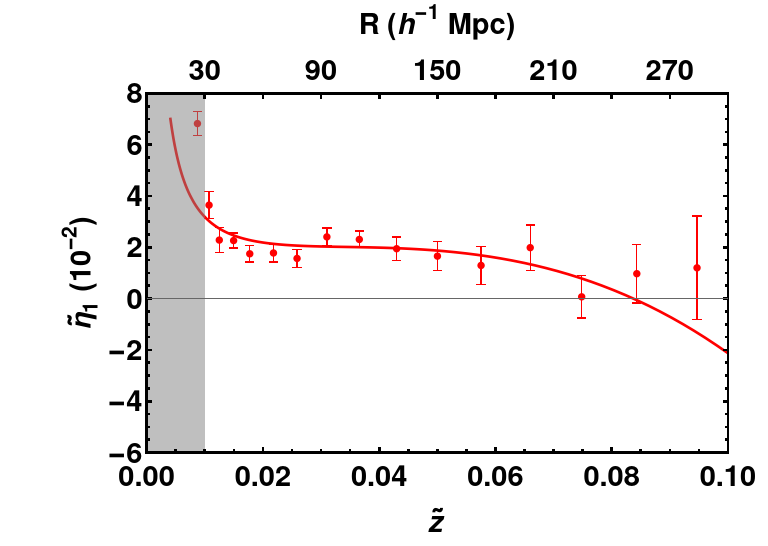}
\includegraphics[scale=0.4]{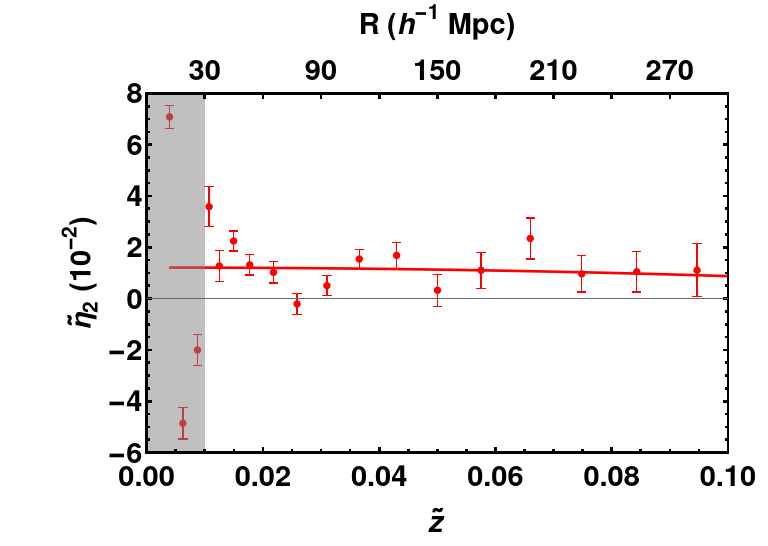}
\includegraphics[scale=0.4]{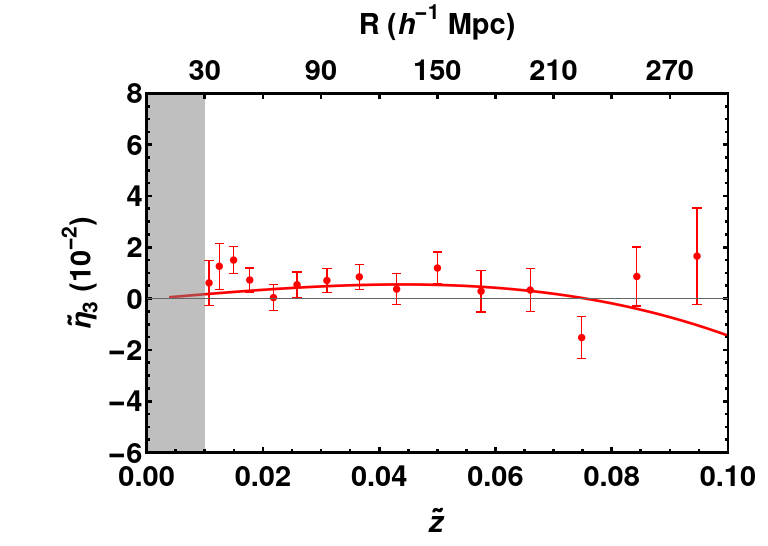}
\caption{The observed axisymmetric multipoles of the expansion rate fluctuation field in CMB frame $\tilde \eta_\ell$ up to $\ell=3$ for CF4 in redshift range $0.01<\tilde z<0.1$, with the corresponding best-fit covariant cosmographic predictions.  The best-fit cosmographic parameters are shown in Table \ref{tab_cop_tot}.}
\label{eta_l_z}
\centering
\end{center}
\end{figure}

\subsection{Results}\label{results_ch7}
We begin by examining the CF4 sample, which facilitates the tomographic mapping of the local Universe through numerous redshift shells. In Figure \ref{eta_l_z}, we present the estimated multipoles for CF4 along with the best-fit model. The corresponding estimated CC  parameters are listed in the second column of Table \ref{tab_cop_tot} for CF4.

The dipole of the expansion rate fluctuation field $\tilde{\eta}_1$ appears to be constant up to $\tilde{z} \sim 0.06$ (see Figure \ref{eta_l_z}). Its amplitude is modulated by  the observer’s velocity $v_o$ with respect to the matter frame, and the intrinsic metric anisotropy captured by the CC parameter $\Qbb_1$. Its redshift constancy indicates that this dipole cannot be eliminated simply by boosting the observer. Indeed, the dipole contribution sourced  in an isotropic universe by the observer's motion has an amplitude that scales inversely with redshift, as shown by eq. (4.10) in \cite{paper2}.
This indicates the existence of an intrinsic dipole anisotropy arising from the expansion dynamics, whose characteristics are predominantly determined by the covariant deceleration parameter.

\begin{table}
\centering
\scalebox{1}{
\begin{tabular}{|c|c|c|c|c|}
\hline
Sample                           & CF4             & CF4TF           & CF4FP           & Pantheon+     \\ \hline\hline
$\Hbb_2/\Hbb_0$ ($10^{-2}$)      & $2.8\pm0.4$     & $2.3\pm0.7$     & $2.6\pm0.9$     & $0.0\pm1.3$   \\ \hline\hline
$\Qbb_1$                         & $2.0\pm0.4$     & $1.3\pm0.7$     & $1.5\pm0.7$     & $0.6\pm1.0$   \\ \hline
$\Qbb_3$                         & $0.9\pm0.3$     & $1.2\pm0.7$     & $0.5\pm0.5$         & $0.3\pm0.9$   \\ \hline\hline
$\Jbb_2$                         & $5\pm10$        & $3\pm47$        & $-8\pm17$       & $-6\pm42$     \\ \hline\hline
$\Sbb_1+4\Qbb_1(2\Jbb_0-\Rbb_0)$ & $3745\pm1012$   & $1959\pm2984$   & $1177\pm1683$   &  $2373\pm2893$ \\ \hline
$\Sbb_3+4\Qbb_3(2\Jbb_0-\Rbb_0)$ & $1901\pm947$    & $3604\pm5062$   & $-508\pm1507$   & $224\pm2721$  \\ \hline\hline
$v_o$ (km/s)                     & $188\pm22$      & $204\pm35$      & $328\pm63$      & $267\pm80$    \\ \hline\hline
$\chi^2_{\nu}$ ($\nu$)               & $0.891$ (52793) & $1.017$ (10017) & $0.836$ (42016) & $0.955$ (619) \\ \hline
\end{tabular}}
\caption{The covariant cosmographic parameters
and $v_o$ for different samples using the range $0.01<\tilde z<0.1$. The last row shows the reduced $\chi^2$ of the cosmographic model with the number of objects in each sample.
}
\label{tab_cop_tot}
\end{table}

While the observed dipole does not provide convincing evidence of a polarity inversion,  the best-fitting expansion rate model does suggest a decreasing amplitude beyond a redshift of $\tilde{z}=0.06,$  and possibly even a sign change at greater distances, although the large error bars make it challenging to assess this conclusion with a high level of confidence.  The high redshift decrease in  the best-fitting dipole model shown in Figure \ref{eta_l_z} (upper panel) is induced by the fact that the third term in the expansion of $\tilde{\eta}_1$ is negative (see eq. (B.1) in \cite{paper3}). 
This  implies that the dipole of the snap parameter is positive.

The quadrupole $\tilde\eta_2$ also appears to be constant up to $\tilde z \sim 0.06$, indicating that the quadrupole of $\Hbb$, where  $\Hbb_2/\Hbb_0=(2.8\pm0.4) \times 10^{-2}$, is the dominant contribution compared to the quadrupole of the jerk parameter whose contribution is expected to become more important with increasing redshift (see eq. (B.2) in \cite{paper3}).  Despite the larger error bars—and in contrast to the dipole—the best-fit curve also indicates a remarkably constant quadrupole amplitude across the entire redshift range investigated, with no sign of weakening at greater depths.

The best-fit model for the octupole $\tilde\eta_3$ is non-monotonic: it increases up to  $\tilde z \sim 0.05$ (bottom panel of Figure~\ref{eta_l_z}), then decreases thereafter. The initial rise suggests a positive value of the octupole of the covariant deceleration parameter, while the subsequent decline is a smoking gun of  the positive sign of the octupole of the snap (see eq. (B.3) in \cite{paper3}).

The  change in behavior of the dipole and octupole around $\tilde{z} = 0.1$ may indicate the presence of a structure, a phenomenon typically associated with a spherical structure at around $\tilde z=0.1$
(see Section 6 in \cite{paper2}). However, the quadrupole does not exhibit a similar  abrupt change in behavior, making it challenging to draw definitive conclusions. More precise measurements at redshifts greater than $0.06$ would be necessary to strengthen this assessment.

\begin{figure}
\begin{center}
\includegraphics[scale=0.45]{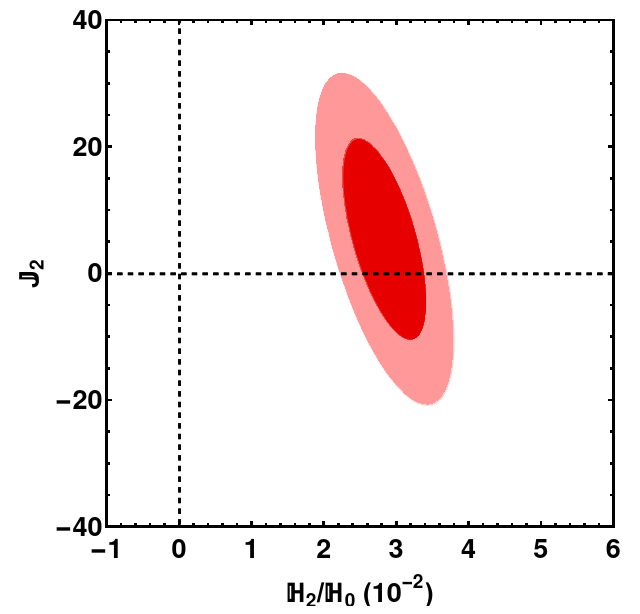}
\includegraphics[scale=0.45]{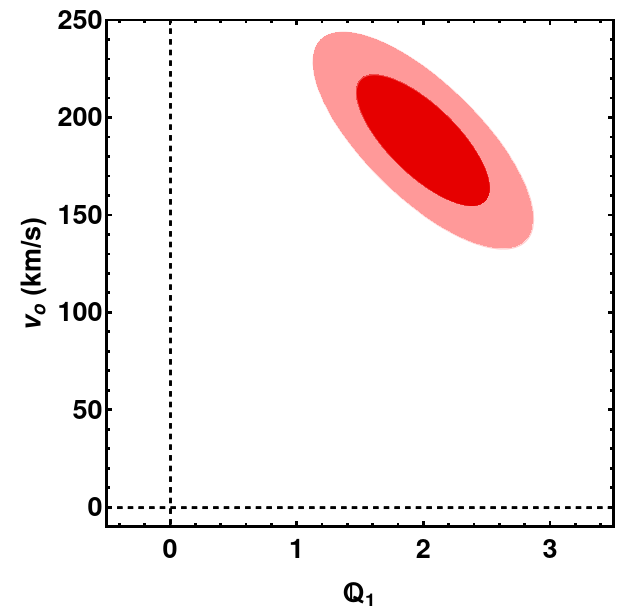}
\includegraphics[scale=0.45]{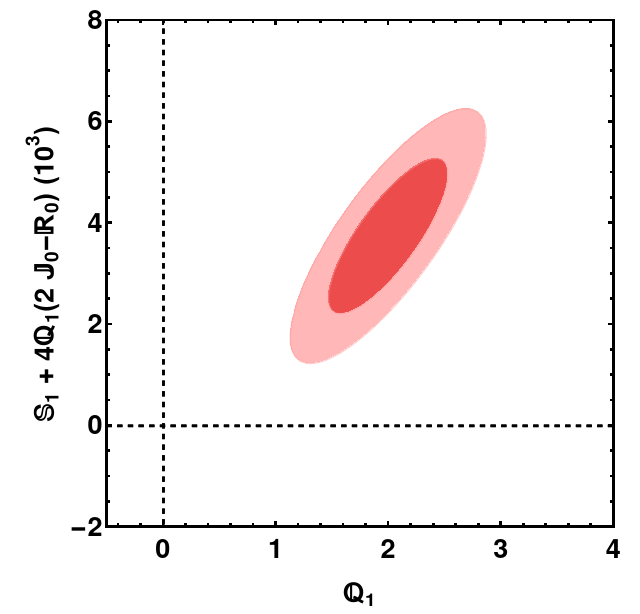}
\includegraphics[scale=0.45]{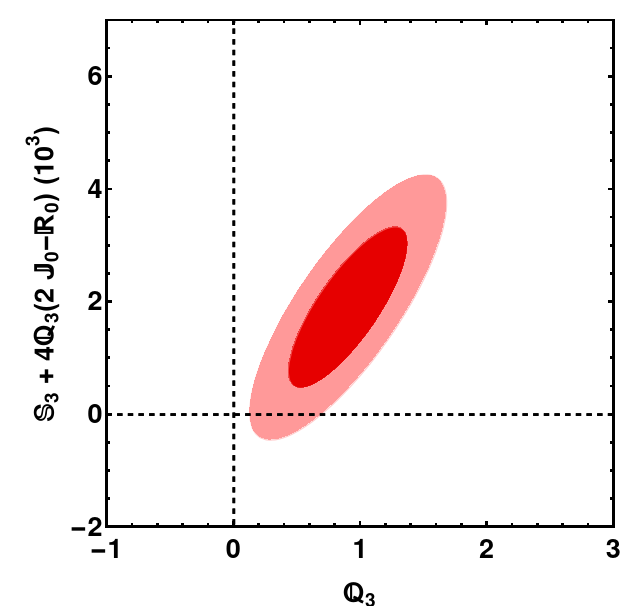}
\caption{Marginalized posterior distribution of the cosmographic parameters for the CF4 sample,
showing the $68\%$ and $95\%$ confidence levels.}
\label{ellps}
\centering
\end{center}
\end{figure}

Table~\ref{tab_cop_tot} summarizes the best-fit CC parameters, in the interval $0.01<z<0.1$,  along with their $1\sigma$ uncertainties, obtained from the quantitative analysis of the CF4 sample, while Figure \ref{ellps}  shows the marginalised posterior distribution of the CC parameters.
We also report results for specific CF4 subsamples CF4TF and  CF4FP in Table~\ref{tab_cop_tot}. Note that we restrict the analysis to galaxy samples. As previously discussed, while the Pantheon+ sample does exhibit indications of the same anisotropic pattern in the expansion rate observed in the galaxy data, the signal is weak and blurred by noise, preventing the determination of the parameters with meaningful precision (see also Figure \ref{quad_sim_indip} and the associated discussion in Section \ref{roburesu}).

We find that the quadrupole-to-monopole ratio  $\mathbb{H}_2/\mathbb{H}_0$ is consistently nonzero across all three samples, with CF4 yielding the highest amplitude at $(2.8 \pm 0.4) \times 10^{-2}$, which is consistently recovered when the analysis is performed over the independent subsamples  CF4FP and CF4TF.  These results suggest a robust detection that persists across different distance indicators (TF and FP), indicating that the anisotropy is not driven by a specific tracer population. The covariant deceleration dipole $\mathbb{Q}_1$ is also significant in all samples, especially in the full CF4 dataset, where its deviation from zero is established at the $\sim5\sigma$ level. This points to a strong directional dependence in the CC deceleration function, and to its   major role in shaping the observed axially-symmetric anisotropy in the expansion rate field. 
The octupolar component $\mathbb{Q}_3$ is weaker, but still nonzero in CF4 and CF4TF, while CF4FP shows a more modest value. This suggests that higher-order anisotropies exist but are less prominent and more sensitive to the choice of distance estimator.

In contrast, the jerk quadrupole $\mathbb{J}_2$ is poorly constrained in all three samples, with uncertainties exceeding the central values. As a result, no definitive conclusions can be drawn about third-order anisotropies in the expansion from these measurements alone. Also the structural combinations $\mathbb{S}_1 + 4\mathbb{Q}_1(2\mathbb{J}_0 - \mathbb{R}_0)$ and $\mathbb{S}_3 + 4\mathbb{Q}_3(2\mathbb{J}_0 - \mathbb{R}_0)$, which encode nonlinear couplings between dipole and octupole terms, are moderately significant in CF4 but suffer from large uncertainties in the CF4TF and CF4FP subsamples. While nonlinear structure is likely present in the full dataset, it is difficult to isolate with higher precision in the individual subsets.

Finally, the inferred {velocity  $v_o$} (with respect to the CMB frame) of the matter fluid element representing the observer  is nonzero across all datasets, with CF4 giving $188 \pm 22$ km/s, CF4TF yielding $204 \pm 35$ km/s, and CF4FP reporting a higher, although noiser, value of $328 \pm 63$ km/s. Figure \ref{ellps}  shows that this parameter is highly degenerate, exhibiting a strong anticorrelation with the dipolar components of the deceleration parameter.

\subsection{Robustness of the results}

The reduced $\chi^2$ values for all samples (see Table~\ref{tab_cop_tot})—arising from the joint fit to the observed redshift evolution of $\tilde{\eta}_1$, $\tilde{\eta}_2$, and $\tilde{\eta}_3$ (see Figure \ref{eta_l_z})
—remain close to unity, indicating that the covariant cosmographic model offers a satisfactory description of the data. Nevertheless, the quality of the reconstruction of the covariant cosmographic parameters—and, ultimately, the consistency between the observed expansion rate and that predicted within the framework of covariant cosmography—will be further assessed using several complementary strategies beyond the standard goodness-of-fit test.

\begin{figure}
\begin{center}
\includegraphics[scale=0.17]{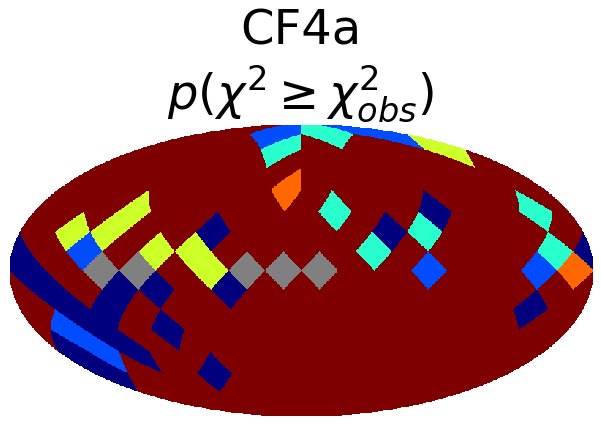}
\includegraphics[scale=0.17]{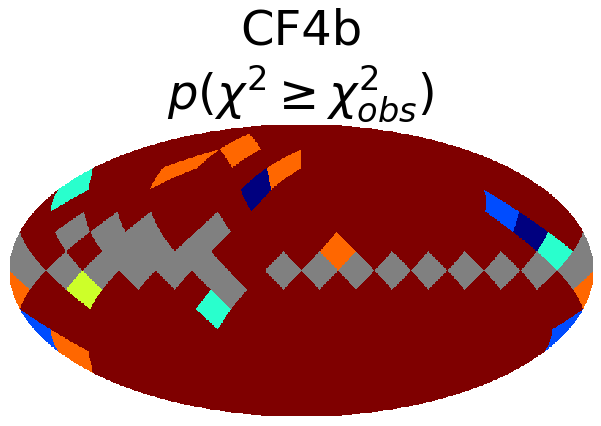}
\includegraphics[scale=0.17]{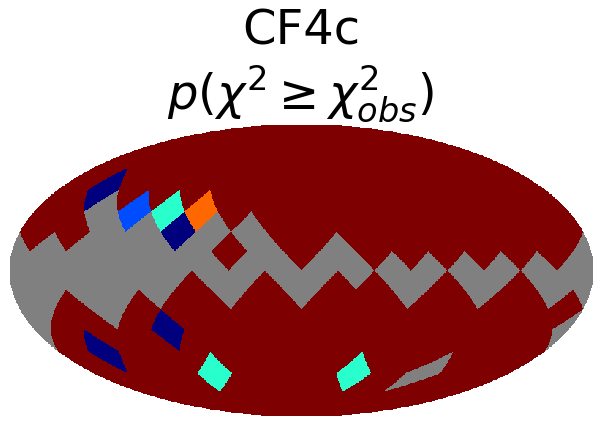}
\includegraphics[scale=0.17]{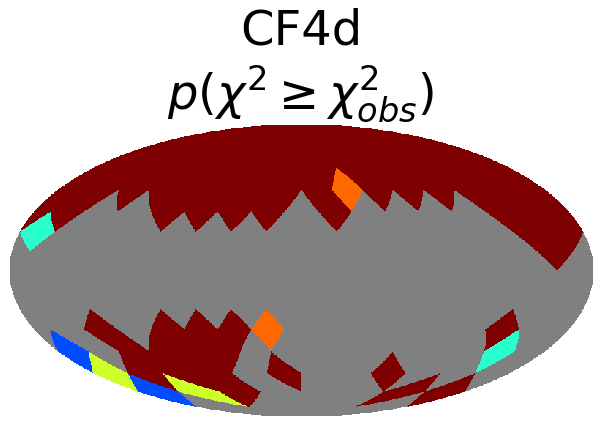}
\\
\includegraphics[scale=0.4]{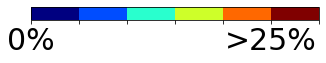}
\caption{
The probability, computed in each individual HEALPix pixel, of incorrectly rejecting the null hypothesis that the covariant cosmographic reconstruction of the expansion rate (see eq.~\eqref{eta_exp_1}) accurately reproduces the value measured along that line of sight.
Four different redshift ranges are shown. We calculate the CC functions using the parameters reported in Table \ref{tab_cop_tot}. 
In the color bar, dark red pixels indicate probabilities above 25$\%$. Grey pixels indicate that no galaxies are present, and no estimate of the expansion rate fluctuation field $\tilde{\eta}$ could be made.
The number of pixels is 192 ($N_{\rm side}=4$).}
\label{chi2min_cf4_2d}
\centering
\end{center}
\end{figure}
\begin{figure}
\begin{center}
\includegraphics[scale=0.4]{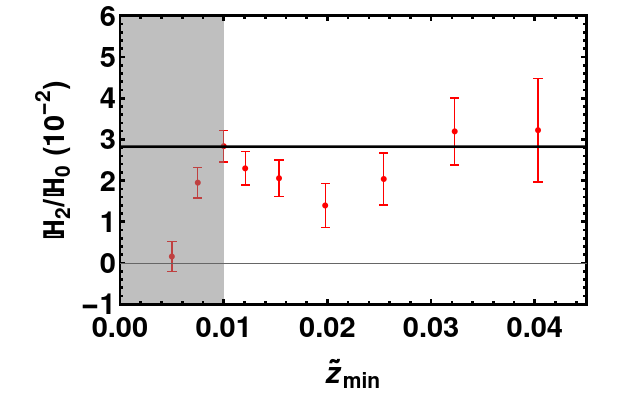}
\includegraphics[scale=0.4]{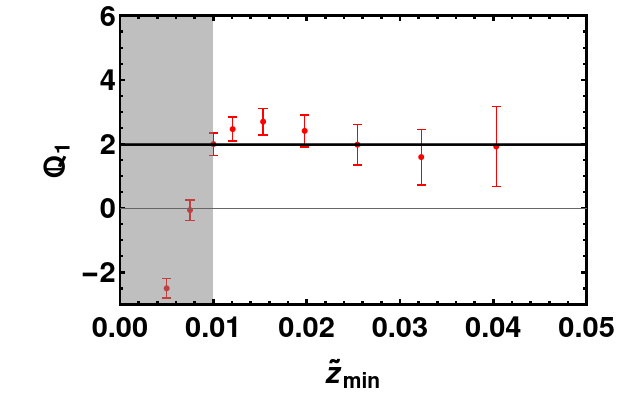}
\includegraphics[scale=0.4]{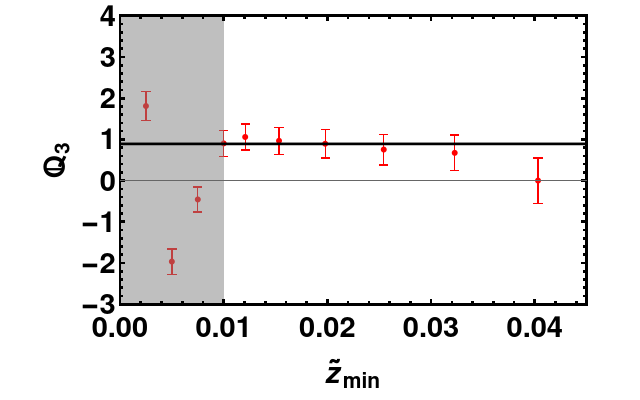}
\includegraphics[scale=0.4]{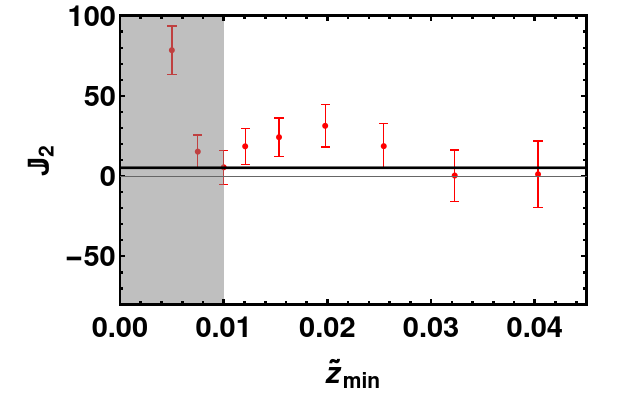}
\includegraphics[scale=0.4]{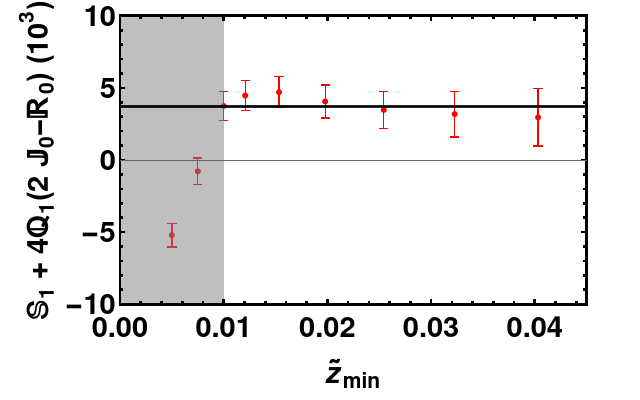}
\includegraphics[scale=0.4]{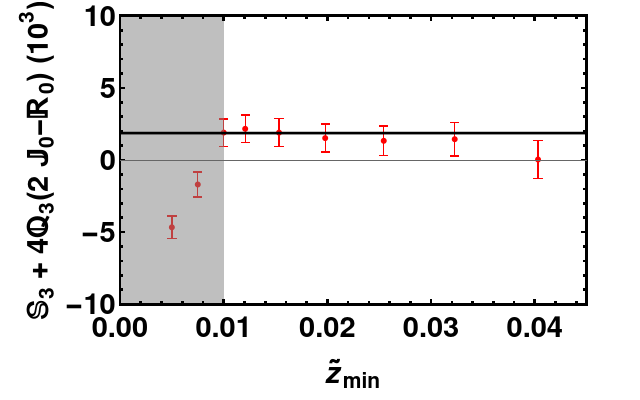}
\includegraphics[scale=0.4]{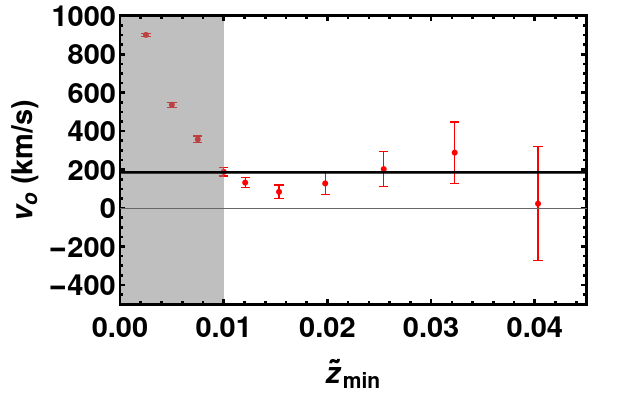}
\caption{The best-fit covariant cosmographic parameters and the observer's velocity $v_o$ as a function of the lower redshift cut ($\tilde z_{\min}$) for the CF4 sample. The black lines presents the best values chosen for Table \ref{tab_cop_tot} and correspond to those estimated  using $\tilde z_{\min}=0.01$. The shaded area  indicates the range  excluded to estimate the parameters quoted in Table \ref{tab_cop_tot}. Note that the multipoles of the CC parameters are constants unlike those of the expansion rate fluctuation field  shown in Figure \ref{eta_l_z}.}
\label{param_zmin}
\centering
\end{center}
\end{figure}

A first test of the reliability of the recovered CC parameters focuses on how accurately they describe angular fluctuations across the sky. To this end, we tessellate the sky into 192 HEALPix pixels and compare the observed expansion rate fluctuation field in each direction with the field reconstructed from the best-fitting CC parameters using eq. 
\eqref{eta_exp_1}. The level of agreement is quantified along each line of sight by computing the probability of obtaining a minimum $\chi^2$ value larger than the one observed. The results are shown in Figure~\ref{chi2min_cf4_2d} across four different redshift shells. Only a small number of isolated pixels fail to pass the conventional $5\%$ threshold, highlighting the overall accuracy of covariant cosmography in reproducing the anisotropies in the distance–redshift relation within each redshift shell and over all the sky.

\begin{figure}
\begin{center}
\includegraphics[scale=0.45]{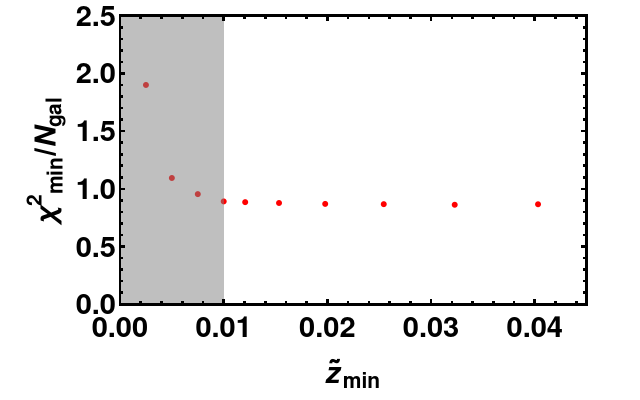}
\caption{The minimum $\chi^2$ shown in eq. \eqref{chi2_cp} over the number of galaxies as a function of $\tilde z_{\rm min}$ for CF4.}
\label{chi2min_cf4}
\centering
\end{center}
\end{figure}

Since the CC parameters are the coefficients of a redshift expansion around the position of the matter observer, we also investigate their stability as a function of the sample depth. This allows us to verify that, despite their local definition, they are not solely sensitive to anisotropies in the immediate vicinity of the observer, but are also capable of capturing large-scale features across different directions of the sky.

Figure \ref{param_zmin} displays the best-fit covariant cosmographic parameters and the velocity $v_o$ of the matter observer over the  redshift interval $\tilde{z}_{\rm min}\le \tilde z \le 0.1$ as a function of $\tilde z_{\rm min}$. Interestingly,  the parameters remain generally stable for any choice of the lower redshift cutoff used in the likelihood analysis which is greater than $\tilde z_{\rm min}=0.01$.  Their values systematically deviate when smaller redshift cuts are applied. This suggests that  covariant cosmography consistently captures the cosmic expansion phenomenology, failing only when trying to  explain the kinematics  
when including data below $\tilde z=0.01$ (the grey regions in Figure \ref{param_zmin}).  

The lack of predictability of the formalism when very local data ($\tilde z<0.01$) are included in the fit  can be appreciated also from a different angle. In Figure \ref{chi2min_cf4}, we show the minimum $\chi^2$ values corresponding to each best-fitting covariant cosmographic model from the analysis of the CF4 dataset.
It is evident that while the goodness of fit remains stable for larger values of $\tilde{z}_{\rm min}$, it progressively worsens as the lower redshift cut is applied for $\tilde{z} < 0.01$. The mismatch between theory and observations when including very local data in the immediate surroundings of the observer,  can also be seen in the grey region in Figure \ref{eta_l_z}.

We conclude with a comparison to previous results in the literature. Only a few studies exist, namely those of \cite{Dhawan:2022lze,Cowell:2022ehf}.
However, a direct comparison with our findings is difficult for several reasons. From a methodological perspective, those authors analyze the CC parameters by introducing external, non-physically justified assumptions, such as multiplying them by distance-dependent suppression terms in order to make them redshift dependent, rather than treating them as pure scalar constants estimated at the time of observation (at $z=0$). In addition, they fix the direction of the parameters (the dipole of  $\Qbb$ and the quadrupole of $\Hbb$), instead of allowing the data itself to constrain them. 
Moreover, these analyses do not estimate the CC parameters in the matter comoving frame, where the CC formalism is fundamentally defined (see also the discussion in Section~4.2 of \cite{paper2}). As a result, they also fail to provide an estimate of the geodesic observer’s motion, $v_o$. Finally, they apply corrections for peculiar velocities that effectively erase much of the signal expected from metric distortions.

On the observational side, those authors focus on describing anisotropies over a very large redshift baseline (up to $z \sim 2$). As a consequence,  the smoothing scale implicit in the fluid approximation---an essential ingredient of the CC formalism (see also the discussion in the next section)---is significantly larger than the one adopted in this analysis. This makes their results not directly representative of the specific anisotropy configuration in the local universe. 

\subsection{Modelling cosmic matter in the CC framework}

Including very local data ($\tilde z < 0.01$) compromises the ability of the covariant cosmographic model to describe expansion rate fluctuations across the entire volume considered ($\tilde z < 0.1$). The fundamental physical reason for this lies in the breakdown of the smooth continuum approximation on which the formalism is based.
The CC framework assumes that the cosmic matter field can be modeled as a smooth pressure-free fluid (`dust') on sufficiently large scales.
On smaller scales, however, the data are dominated by 
the `thermal' motions (i.e. velocity dispersion)  of galaxies, which reflect local gravitational fluctuations rather than the large-scale cosmic 
gravity. This not only prevents the calculation of CC parameters using local derivatives but also reveals another limitation: the formalism does not provide any {\it a priori}  way to determine, in advance, the scale below which the CC framework breaks down. In other words, there is no way to determine in advance the smallest scale at which the anisotropies in the expansion rate no longer accurately represent those of the cosmic volume under study. This issue is similar to a limitation of the standard cosmological model: the assumption of the Cosmological Principle does not specify the scale below which the matter density ceases to be statistically representative of the universe as a whole.

Remarkably, this breakdown scale can be identified \textit{a posteriori} using the data.  Figure \ref{eta_l_z} shows that this transition consistently appears at $z\sim 0.01$  across all multipole decompositions of the expansion-rate fluctuation field: the dipole, quadrupole, and octupole all begin to diverge coherently from theoretical predictions below this redshift.
As a consequence, the anisotropic expansion-rate structure of the surveyed volume can be effectively described by just seven CC parameters, as soon as the noisy local data are excluded from the analysis.

We have argued that the CC parameters cannot be measured directly from the observer’s position; instead, they must be inferred by fitting the redshift evolution of the expansion rate field over a broad interval. Ensuring that these estimates are independent of the chosen redshift boundaries is therefore essential for achieving genuine convergence to a cosmologically meaningful value. Figures \ref{param_zmin} and \ref{chi2min_cf4} demonstrate that the best-fit multipoles of the CC parameters, when estimated over redshift intervals with progressively higher lower bounds $z_{\rm min}$, converge toward stable values up to the limit of the data at  $z=0.1$.

These results serve as a caution against defining the matter comoving observer in the covariant cosmography framework by averaging data within spherical volumes centered on the Local Group, with radii of approximately $30\,h^{-1}\,\mathrm{Mpc}$ (where, for example,  the CF4 sample typically includes nearly 3000 objects.) A more physically motivated estimate of the spatial scale over which one must average to define the velocity of the matter element representing the CC observer can be inferred through a model-dependent argument based on standard cosmological information.
We know that the velocity of the matter observer is $v_o = 188 \pm 22$~km/s in the direction $(l, b) = (299^\circ, 5^\circ)$. Using Cosmicflows-3 sample, \cite{Tully:2019ngb} determined the contributions to the Local Group (LG)  velocity from mass enclosed within a sphere of radius $29h^{-1}$ Mpc, and from the surrounding shell extending from $29h^{-1}$ Mpc to $75h^{-1}$ Mpc (see Table~\ref{Table_v0}).
Given the observed CMB dipole, we can deduce that the component of the LG's motion generated by mass beyond $29h^{-1}$ Mpc is approximately $v_{29}=320 \pm 70$~km/s in the direction $(l, b) = (292^\circ, 4^\circ)$, while that generated by structures beyond $75h^{-1}$ Mpc is about  $v_{75} = 100 \pm 70$~km/s in the direction $(l, b) = (302^\circ, -38^\circ)$. The fairly good alignment between these directions and the measured $v_o$, and the fact that
$v_{75}<v_o<v_{29}$ suggests that the effective averaging length corresponding to the size of the fluid element representing the CC observer lies somewhere in the range $29h^{-1}$ Mpc $< R~(\mathrm{Mpc})$ $75h^{-1}$ Mpc. This holds under the simplifying assumption that the contribution of shells external to the volume $V$ provides, within statistical noise, a fair representation of the bulk velocity of $V,$ rather than merely the velocity of the Local Group.

\begin{table}[]
\centering
\begin{tabular}{|c|c|c|c|}
\hline
\begin{tabular}[c]{@{}c@{}}Local group velocity\\ components\end{tabular} & \begin{tabular}[c]{@{}c@{}}$v$\\ (km/s)\end{tabular} & $l$ & $b$ \\ \hline\hline
Near ($<29$ $h^{-1}$Mpc)                                                  & 388                                                  & 256 & 49  \\ \hline
Far ($29-75$ $h^{-1}$Mpc)                                                & 255                                                  & 289 & 19  \\ \hline\hline
Total velocity                                                            & 631                                                  & 276 & 30  \\ \hline
Total -- Near                                                    & 320                                                  & 292 & 4   \\ \hline
Total -- Near -- Far                                              & 100                                                  & 302 & $-38$ \\ \hline
\end{tabular}
\caption{
Portions of the Local Group  velocity generated by nearby mass concentrations (first row) and by more distant structures (second row). The third row shows the total LG velocity. The fourth and fifth rows display the residual LG velocity after subtracting the contributions from mass within $29$ $h^{-1}$Mpc and up to  $75$ $h^{-1}$Mpc, respectively. All values are taken from Table~2 of \cite{Tully:2019ngb}. The typical uncertainties are approximately $70$~km/s in velocity magnitude and $\sim 10^\circ$ in direction.}
\label{Table_v0}
\end{table}

\section{Conclusion}\label{conc_ch4}

We have introduced the expansion rate fluctuation field $\eta$ \cite{paper0} as a scalar Gaussian observable that, through its natural multipolar decomposition, provides an unbiased framework for identifying and classifying departures from isotropy in the redshift–distance relation, while enabling a transparent interpretation of the signal.

In \cite{paper0}, we reconstructed the expansion rate fluctuation field and its multipolar decomposition using the Cosmicflows-3 and Pantheon samples out to $150\,h^{-1}$ Mpc. Building on that foundation, we now push the analysis further with the updated CF4 and Pantheon+ datasets, extending the reach to twice that scale, $300\,h^{-1}$ Mpc. 

On the methodological side, we subjected the expansion rate observable to a battery of robustness tests, with Monte Carlo simulations, to ensure its unbiasedness against angular anisotropies in the galaxy distribution—including the Zone of Avoidance deficiency, distance-dependent selection biases, and the choice of redshift-bin size in tomographic analyses of the universe—and further optimized its implementation, relative to our earlier work \cite{paper0}, to maximize the signal-to-noise ratio.

On the data-analysis side, the depth and richness of the CF4 sample enable us to probe the fluctuation field with higher fidelity, allowing us to estimate multipole amplitudes up to the octupole ($\ell_{\rm max}=3$). Interestingly, when restricted to volumes comparable to those in \cite{paper0}, the new data reaffirm our earlier findings. Yet, they also reveal subtle refinements: the CF4 sample traces a somewhat stronger dipole, while the quadrupole and octupole remain consistent across both datasets. This continuity, alongside the incremental differences, underscores both the robustness of the previous results and the added resolution provided by the new observations.

Overall the observed multipoles display axial symmetry around the same axis  found by analysing the CF3 sample, approximately $(l=299^\circ, b=5^\circ)$. 
Notably  we show that an axisymmetric model for the expansion rate fluctuation field  effectively reduces the $\chi^2$ with a significantly smaller number of parameters compared to the full harmonic expansion, while also greatly simplifying the numerical analysis and physical interpretation. This axially symmetric configuration is no longer tentative evidence -- it is a feature of the local Universe now established on a firmer and more robust basis.

Strikingly, a stable alignment of the maxima of the measured multipoles for $\ell \leq 3$ continues to be observed in the new volume accessed by CF4 data, specifically in the distance range $150 h^{-1}$ Mpc $< r < 300 h^{-1}$ Mpc. However, the power only in the dipole and quadrupole signals is statistically significant in these regions (the octupole has a signal-to-noise ratio of just 2). Remarkably, the multipoles remain aligned in the same direction and exhibit structures consistent with those observed in shells at shallower redshifts.

We showed how to reconstruct the bulk motion of galaxies expected in the $\Lambda$CDM scenario from the information encoded in the dipole component of the expansion rate fluctuation field. Since this is a model-independent observable, our reconstruction of the bulk velocity is not only independent of the assumed value of the Hubble constant $H_0$, but also robust against characteristic biases that affect its estimation, such as the Malmquist bias.

We find that even when peculiar velocities are averaged over large spherical volumes centered on the observer, the bulk velocity remains larger than expected from $\Lambda$CDM, reaching $\gtrsim 3\sigma$ at a depth of $150\,h^{-1}\,\text{Mpc}$. This excess is directly related to the unexpectedly large dipole observed in the expansion rate fluctuation field $\eta$.
These findings confirm the results of other studies using CF4, which also report tensions with the standard model (e.g. \cite{Whitford:2023oww, Watkins:2023rll, Hoffman:2023pac, Duangchan:2025uzj}).

These discrepancies motivate our analysis of the observed fluctuations in the expansion rate using covariant cosmography, a model-independent, non-perturbative method that does not rely on peculiar velocity measurements.
Our findings indicate that the multipoles of the expansion rate fluctuation field are primarily shaped by a pronounced quadrupole in the covariant Hubble parameter $\mathbb{H}$ (detected at a $S/N$ ratio of $\sim 7$), along with contributions from the dipole $\mathbb{Q}_1$ and octupole $\mathbb{Q}_3$ components of the covariant deceleration parameter, detected at $S/N$ levels of $\sim 5$ and $\sim 3,$ respectively. Moreover, combinations of higher-order CC  parameters (Snap, Jerk, and curvature) also contribute to the observed anisotropies in the expansion rate, with a $S/N$ greater than 2.
We point out that the covariant analysis, in common with the standard approach to measure the Hubble ($H_0$) and deceleration parameters ($q_0$) in the standard model, applies throughout the redshift range investigated since it does not rely on estimates of derivatives at the observer, but on fitting an expansion in redshift to the data.

The positive value for the axially symmetric quadrupole of $\mathbb{H}$ indicates that the cosmic fluid is being stretched along the axis of symmetry. In  \cite{paper2},
we showed that a single spherical ``attractor'' (over-density) can reproduce this feature. However we also demonstrated that the same attractor provides opposite signs for the dipole and the octupole of $\mathbb{Q}$, which contradicts what is observed in this analysis. This implies that a single  Shapley-like  attractor cannot explain, alone,  the anisotropic expansion rate fluctuation field traced by the CF4.

Interestingly, the possible companion existence of a cosmic repeller  with a radius $\sim140$ -- $250\,h^{-1}$ Mpc influencing the local expansion rate, has been  suggested by various studies: \cite{Hoffman:2023pac} based on the CF4 sample and 
 \cite{Bohringer:2019tyj} based on the CLASSIX galaxy cluster survey.  
Similar to the attractor, a single spherical repeller cannot produce the same polarity  for $\Qbb_1$ and $\Qbb_3$. Additionally, it cannot explain the positive sign of the quadrupole of $\mathbb{H}$, since, in the case of a repeller, the minimum of the quadrupole of $\mathbb{H}$ would align with the axis of symmetry.
This highlights the urgency of developing a physically consistent geometric model for the gravitational field in the local Universe that can account for the observed fluctuations in the expansion rate. 

Given the large length scale over which the anisotropy signal remains coherent, and the polarity of the multipoles \roy{[is?]} correlated,  it is essential to move beyond traditional linear perturbative analyses framed in the standard model and investigate how the CC parameters relate to the intrinsic parameters of fully relativistic, non-standard metric models that depart from FLRW. We have already begun this exploration in a companion paper \cite{Sarma:2025yfw}, where we examine axially symmetric metric models constructed by placing the observer off-center in a Lamaître-Tolman-Bondi (LTB) geometry.

Looking ahead, it will be of great interest to apply this analysis strategy to samples  with more data,  improved distance precision,  and greater depth, such as those from the Zwicky Transient Facility (ZTF) survey \cite{Rigault:2024kzb} or DESI \cite{DESI:2025fxa}. This could significantly enhance our understanding of the scale at which the ``end of greatness'' occurs \cite{Kirshner2002}, referring to the transition scale to a uniform Universe.

\acknowledgments We would like to thank Julian Adamek, Ruth Durrer,  Michele Mancarella, Federico Piazza, and Brent Tully for useful discussions. BK, CM and JB are supported by the {\it Agence Nationale de la Recherche} under the grant ANR-24-CE31-6963-01, and the French government under the France 2030 investment plan, as part of the Initiative d’Excellence d'Aix-Marseille Université -  A*MIDEX (AMX-19-IET-012).
RM is supported by the South African Radio Astronomy Observatory and the National Research Foundation (grant no. 75415). JS acknowledges support from the Taiwan National Science and Technology Council, grant No. 112-2811-M-002-132.

\appendix

\section{Maximum Likelihood method for estimating \texorpdfstring{$\eta$}{eta} and its multipoles}\label{app:fit_mult}

In this appendix, we discuss the estimation of the $\eta$ field and its multipolar structure. 
Being a monopole-free quantity, the expansion rate fluctuation field $\eta$ inherently encodes non-local information. As shown in eq. (\ref{defeta1}), it is formally defined on the surface of a sphere with radius $z$. In practice, however, it is computed over a spherical shell of finite thickness $\delta z$. To ensure statistical independence and avoid correlations between shells, the estimates are performed separately within non-overlapping spherical shells.

In idealized scenarios, with no noise and full-sky or complete data coverage, the orthogonality of spherical harmonics ensures that multipoles are mathematically separable. In practice, however  
their individual estimation is unbiased only if the contributions from neglected multipoles with $\ell > \ell_{\rm max}$ are small compared to the corresponding estimation uncertainties.
Therefore, to avoid mode coupling due to truncation, and obtain an unbiased estimate, the monopole term $\mathcal{M}$ in eq. (\ref{defeta1}) 
 must determined together with the higher-order multipoles of $\eta$ within each shell. Once the monopole for each shell is obtained, the $\eta$ field is derived by subtracting it from the total signal.

For discrete points on a sphere, the monopole $\mathcal{M}$ and the multipoles of $\eta$ are found by fitting all measurements up to a maximum multipole $\ell_{\rm max}$, without applying any smoothing. 
For each spherical shell $S$ centered on the observer, with  radius $z$,   thickness $\delta z$ and containing $N_{\rm g}$ objects,  we estimate the intermediate quantity 
\begin{equation}
    \nu_i \equiv \log \frac{z_i}{d_L^i}= \log z_i +5 -\frac{\mu_i}{5}\,,
\label{nudef1}
\end{equation}
where $z_i$  is the redshift of the $i-$th galaxy in the shell, $d_L^i$ is its luminosity distance and $\mu_i$ is its observed distance modulus. 
We assume that this estimate is subject to Gaussian error, which is a valid assumption provided that the distance modulus $\mu$ is also Gaussian distributed.

We then look for the best fitting SH multipolar model that accommodates the data  by minimizing the $\chi^2$ statistic 
\begin{equation}
    \chi^2=\boldsymbol{\Delta\nu}^T\, \boldsymbol{\Psi} \,\boldsymbol{\Delta\nu}\;,
\end{equation}
where $\boldsymbol{\Psi}$ is the precision matrix (the inverse of the covariance matrix of the distance modulus $\mu$ multiplied by $1/25$), and
\begin{equation}
\boldsymbol{\Delta\nu}=\boldsymbol{\nu}^{(\rm obs)}-\boldsymbol{\nu}^{\rm (model)}\;,
\end{equation}
where $\boldsymbol{\nu}^{\rm (model)}=\boldsymbol{\mathcal{Y}}\boldsymbol{a}$. Here $\boldsymbol{a}$ is a vector with size $(\ell_{\rm max}+1)^2$, which contains the SH coefficients $\hat a_{\ell m}$ (the maximum cutoff scale $\ell_{\text{max}}$ is determined iteratively using a trial-and-error approach until no residual signal remains in the higher multipoles). To estimate the SH coefficients 
we use a practical  indexing scheme introduced by \cite{Gorski:1994ye} (see also \cite{Mortlock:2000zw}) based on an index $j(\ell,m)$, such that $j(\ell,m)=\ell^2+\ell+m+1$, so that the inverse relations are $ \ell=\text{integer}\big(\sqrt{j-1}\big)$  and $m=j-\ell^2-\ell-1$. This allows us to define a matrix  $\boldsymbol{\mathcal{Y}}$, of size $ N_{\rm g}\times(\ell_{\rm max}+1)^2$, whose  elements are $\mathcal{Y}_{ij}=Y_{\ell(j) m(j)}(\theta_i,\phi_i)$, where $(\theta_i,\phi_i)$ are the angular coordinates of the $i$-th object in the shell.

We search for the minimum of the $\chi^2$ statistic by looking for the point (in the space of $\hat a_{\ell m}$) where its first derivative is zero. The result can be expressed in matrix form 
\begin{equation}
    \boldsymbol{a}=\boldsymbol{\psi}^{-1} \boldsymbol{s}
\quad \text{where}\quad
\boldsymbol{\psi}=\boldsymbol{\mathcal{Y}}^{T}\,\boldsymbol{\Psi}\,\boldsymbol{\mathcal{Y}}\quad \text{and} \quad
\label{psi_mult2}
\boldsymbol{s}=\boldsymbol{\Psi}\,\boldsymbol{\nu}^{\rm (obs)}\,\boldsymbol{\mathcal{Y}}\,.
\end{equation}
The covariance matrix of $\boldsymbol{a}$ is simply 
$\boldsymbol{\mathcal{C}} = \boldsymbol{\psi}^{-1}$. 
The variance of each SH coefficient is given by the corresponding 
diagonal element of $\boldsymbol{\mathcal{C}}$.
The multipole coefficients of the field \( \nu \) are related to the characteristic expansion rate fluctuation quantities as
\begin{align}
\mathcal{M} = \frac{\hat{a}_{00}}{2\sqrt{\pi}} \quad \text{and} \quad \eta_{\ell m} = \hat{a}_{\ell m} ~~(\ell > 0)\,.
\end{align}
Consequently, the expansion rate fluctuation at the location of an object with redshift \( z_i \) is given by $ \eta(z_i) = \nu_i - \mathcal{M}$.

\subsection{Fitting axially symmetric multipoles}\label{app_fit_eta_l}

If the multipoles exhibit axial symmetry about a single axis, a Legendre expansion of the expansion rate fluctuation fied, eq. \eqref{Lcoeff}) suffices, and the relevant expansion coefficients can be estimated using a method analogous to that used for the full spherical harmonic decomposition. In this case, however, the size of the vector $\boldsymbol{a}$  is $(\ell_{\rm max}+1)$. Also $\boldsymbol{\mathcal{Y}}$ is replaced by the $ N_{\rm g}\times(\ell_{\rm max}+1)$ matrix $\boldsymbol{\mathcal{P}}$, where $\mathcal{P}_{ij}=P_{j-1}(\cos\theta_i)$ and $\theta_i$ is the angular separation of the $i$-th object from the axis of symmetry.   The monopole is the first element in $\boldsymbol{a}$ ($\mathcal{M}=\hat a_0$), and $\hat a_\ell=\eta_\ell$ for $\ell>0$.

\subsection{Potential biases in the estimation}

This estimation of the spherical harmonics presents potential risks. If we limit the fitting to a specific multipole $\ell_{\rm max}$, the result may be biased if the $\eta$ signal contains contributions from multipoles higher than $\ell_{\rm max}$. The amplitude of the bias induced by mode couplings can be analytically investigated  by taking the expectation value of the best fitting coefficients  $\boldsymbol{a}$:
\begin{equation}
    E\big[\hat{a}_j\big]=a_j+b_j\,,
\end{equation}
where
\begin{equation}
    b_j=\sum_{k=L_{\rm max}+1}^{\infty}\sum_{q=0}^{L_{\rm max}}a_{k}\, \mathcal{C}_{q j}\sum_{n=1}^{N_{\rm g}}\sum_{r=1}^{N_{\rm g}}\Psi_{nr}\, \mathcal{Y}_{nq}\, \mathcal{Y}_{rk}\,,
\label{bias_alm}
\end{equation}
and $L_{\rm max}=(\ell_{\rm max}+1)^2$. 
The systematic effect is proportional to the amplitude of the neglected coefficients $a_k$ corresponding to multipoles higher than the maximum multipole included in the fit.
Interestingly, eq.~\eqref{bias_alm} indicates that the bias diminishes with an increasing number of measurements, and as the spatial distribution of both the measurements and their associated errors becomes more isotropic across the sky.
In this case, the sum over the orthonormal basis functions will closely approximate their integral over the sphere, which vanishes due to orthogonality. Furthermore, under the same observational conditions, eq.~\eqref{psi_mult2} implies that the covariance matrix \( \boldsymbol{\mathcal{C}} \) tends to become diagonal, indicating that correlations between different spherical harmonic coefficients vanish.

An additional source of bias can arise from the assumption of a constant monopole within each shell. Spurious multipoles can be introduced when the shell thickness is large, and the distribution of measurements is not uniform both across the sky and in redshift within the shell. To assess this effect, we simulate a uniform \(\Lambda\)CDM universe,  
using the same shell configuration and thickness as in the actual analysis. We then compute the resulting artificial multipoles and compare them to the statistical uncertainties.
The analysis of these mock catalogs suggests that, for visualizing the signal, an optimal 
choice for the shell thicknesses—given the sampling and sky distribution of the CF4 sample—
is $[0.01, 0.03]$, $[0.03, 0.05]$, $[0.05, 0.075]$, and $[0.075, 0.1]$. 
This partitioning is fine enough to keep spurious multipoles negligible relative to the associated 
errors, yet wide enough to yield statistically significant and unbiased signals for the low multipoles.

\section{Selection biases}

Here, we examine, using Monte Carlo simulations,  the impact of survey geometric  and photometric 
selection criteria on the estimation of the expansion rate 
fluctuation field \( \eta \).

\subsection{Anisotropic distribution of galaxies on the sky}\label{app_ZOA_effect}

The first question we address is whether the recovered amplitude of the expansion rate fluctuation field—particularly its lowest multipoles, which are the focus of this paper—is biased by the anisotropic angular distribution of galaxies in the CF4 sample. This effect is most relevant at high redshift, where the catalog suffers from non-uniform sky coverage. The incomplete coverage arises both from the CF4 catalog being a compilation of surveys targeting different regions of the sky, and from the obscuration caused by the Milky Way plane, which prevents the detection of galaxies lying behind it.

To assess potential biases, we use the covariant cosmographic model that best fits the CF4 data (second row of Table \ref{tab_cop_tot}) as the reference input. Based on this model, we generate 1000 Monte Carlo realizations of mock catalogs, where Gaussian noise is added to the model distances. Two cases are investigated to assess the impact of  anisotropies: (i) retaining the actual angular distribution of galaxies in the CF4 sample, and (ii) redistributing the galaxies isotropically over the sky. We then evaluate the accuracy of the recovered multipole amplitudes as a function of redshift.

\begin{figure}
\centering
\includegraphics[scale=0.4]{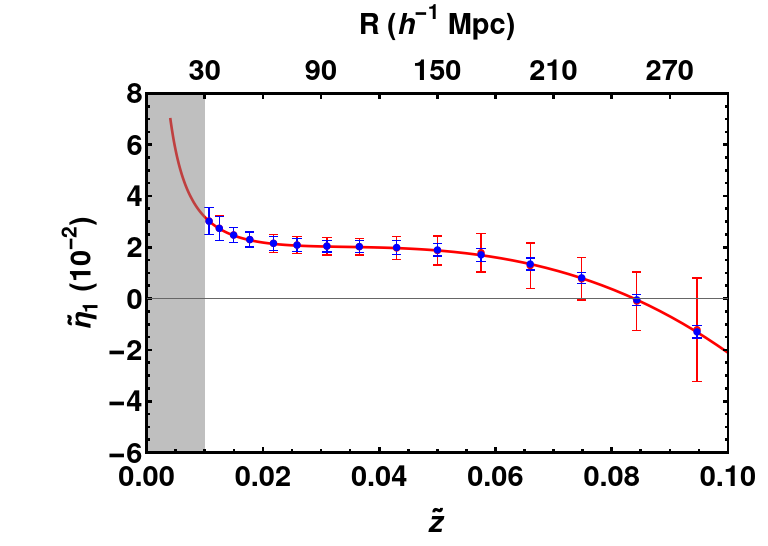}
\includegraphics[scale=0.4]{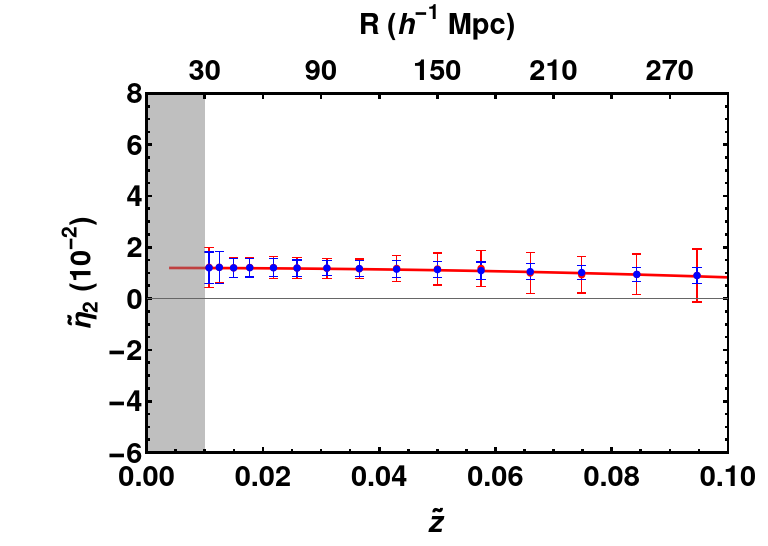}
\includegraphics[scale=0.4]{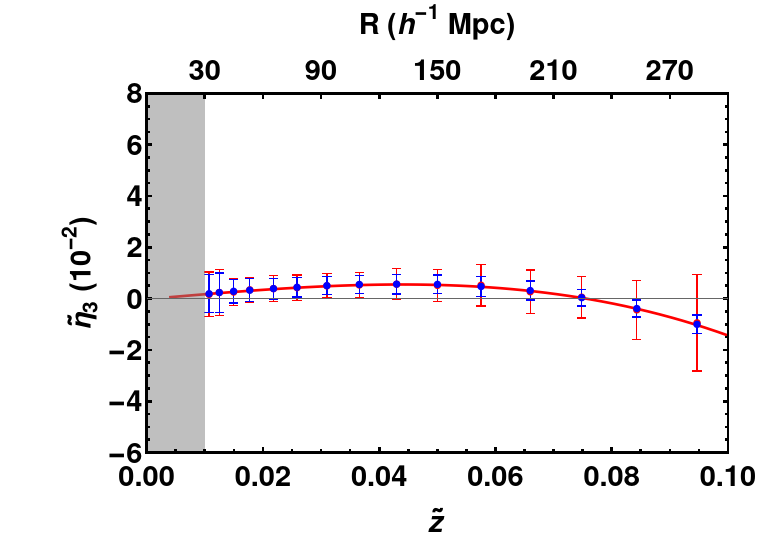}
\caption{
The red curves represent the axially symmetric multipoles of the expansion rate fluctuation field $\tilde{\eta}$ expected 
in the covariant cosmographic model which is the best fit to the CF4 data. The red points show the recovered measurements from Monte-Carlo simulations obtained   by perturbing the distances with Gaussian noise around the CC model, while keeping the angular positions of the objects as reported in the CF4 catalog. Each dot corresponds to the average over 1000 simulations, with error bars indicating the standard deviation of the measurements. The blue points show the recovered multipoles when the angular coordinates of the simulated objects are instead distributed isotropically across the sky.}
\label{fig:eta_cos_CF42}
\centering
\end{figure}

Figure \ref{fig:eta_cos_CF42} shows the input model for each multipole $\ell$ of the expansion rate field $\tilde{\eta}$ (red curves), together with the recovered Legendre coefficients (points) at different redshifts. When the tomographic shells are sufficiently thin, the estimated multipoles remain unbiased across all redshift bins: the averages from the anisotropic mock catalogs accurately reproduce the input values (see also Appendix \ref{app:fit_mult}). However, the anisotropic sky distribution  does strongly affect the size of the uncertainties. This is evident from the systematic increase in the error bars with redshift when comparing multipoles reconstructed from anisotropic (red points) versus isotropic (blue points) mock catalogs.

A second issue requiring investigation arises from the intriguing observational evidence that the symmetry axis of the expansion rate fluctuation field $\tilde{\eta}$ aligns with a region of the sky that is heavily obscured by dust extinction in the Milky Way’s disk. The anisotropic sampling of the catalogs—particularly the extensive ZoA, where galaxy observations are severely limited—raises concerns about potential biases in the reconstruction of the lowest multipole directions, especially the dipole and quadrupole. Could the observed alignment and apex direction (\(l = 299^\circ\), \(b = 5^\circ\)) be artifacts caused by the ZoA?

We consider a perfectly isotropic \(\Lambda\)CDM background model and superimpose an expansion rate field characterized by a dipole—induced by a bulk flow with amplitude \(v_{\text{bulk}} = 400\) km/s—and an axially symmetric  quadrupole in the covariant Hubble parameter, with \(\mathbb{H}_2 / \mathbb{H}_0 = 0.02\). 
We align both the dipole axis and the quadrupolar maximum with different directions on the sky: one pointing toward the ZoA ($l=300^\circ, b=0^\circ$), one toward the north Galactic cap, and one toward the south Galactic cap. Since the quadrupole is symmetric, the last two orientations are equivalent. For each configuration, we generate 500 mock catalogs by assigning to each galaxy at redshift $z$ in the CF4a and CF4c samples, a distance drawn randomly from this model, while incorporating typical distance modulus uncertainties representative of the CF4 dataset.

\begin{figure}
\begin{center}
\includegraphics[scale=0.3]{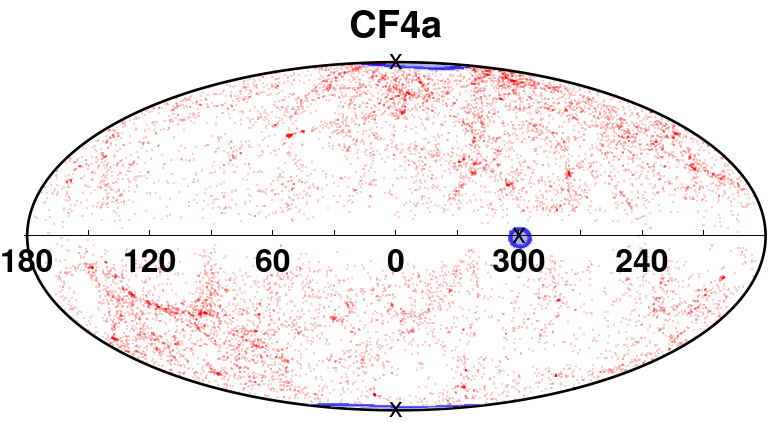}
\includegraphics[scale=0.3]{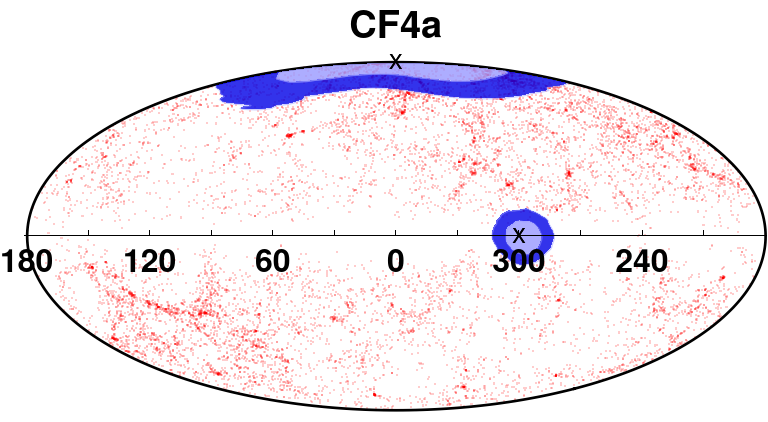}
\\
\includegraphics[scale=0.3]{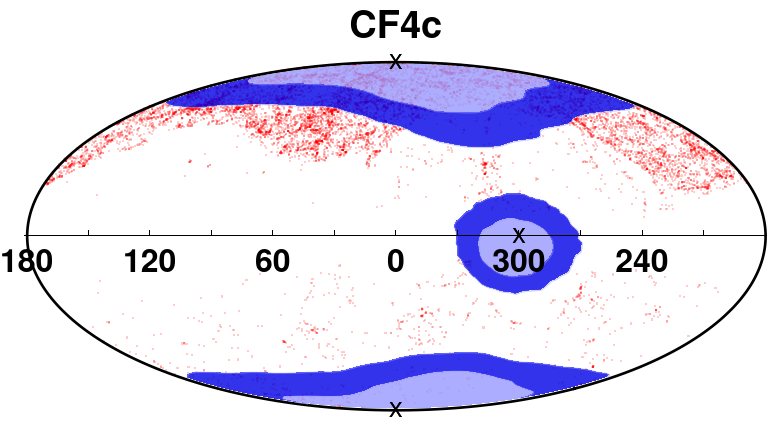}
\includegraphics[scale=0.3]{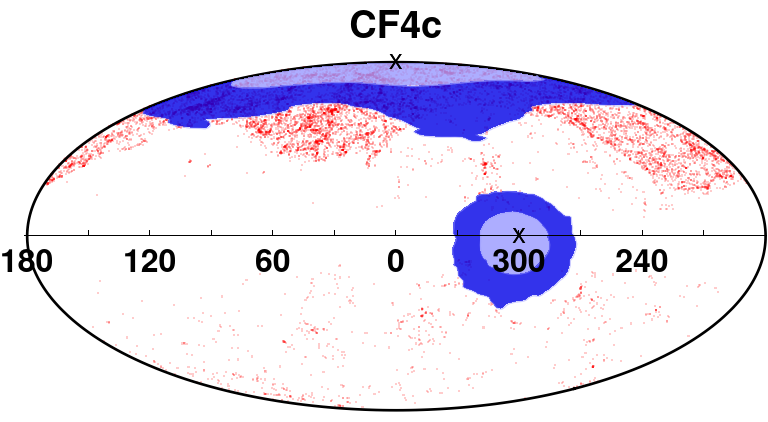}
\caption{The likelihood of recovering the direction of the dipole {\em (left column)} and the maximum of the quadrupole {\em (right column)}. The input bulk flow amplitude is $v_{\rm{bulk}}=400$ km/s and $\mathbb{H}_2/\mathbb{H}_0=0.02$, with a direction pointing towards the cross. (The input quadrupole is axially symmetric). The number of measurements in CF4a is 11978 and 13678 in CF4c.}
\label{lbDQ_sim1}
\centering
\end{center}
\end{figure}

We apply the \(\eta\) reconstruction pipeline to these simulations, and recover the directions of both the dipole and quadrupole components by means of a maximum likelihood analysis, as described in Appendix \ref{app:fit_mult}. Despite the differing anisotropic sky distributions in CF4a and CF4c, and the presence of a prominent ZoA in both, there is no evidence that anisotropic sampling introduces a bias in the recovery of the input dipole and quadrupole directions (see the results in Figure \ref{lbDQ_sim1}). However, anisotropy does degrade the angular resolution, increasing the size of the 68\% and 95\% confidence regions. In particular, although CF4c contains 15\% more galaxies than CF4a, its more anisotropic sampling leads to larger directional uncertainties.

\begin{figure}
\begin{center}
\includegraphics[scale=0.17]{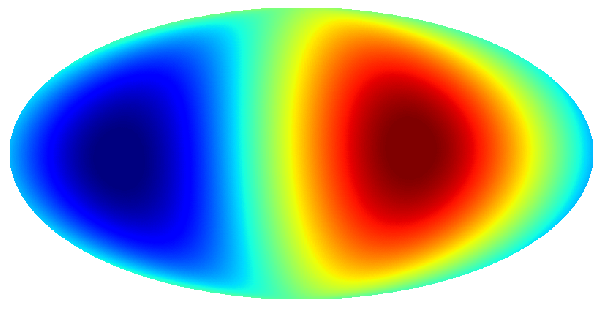}
\includegraphics[scale=0.17]{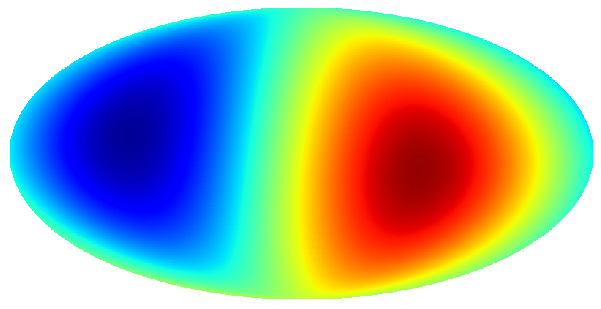}
\includegraphics[scale=0.17]{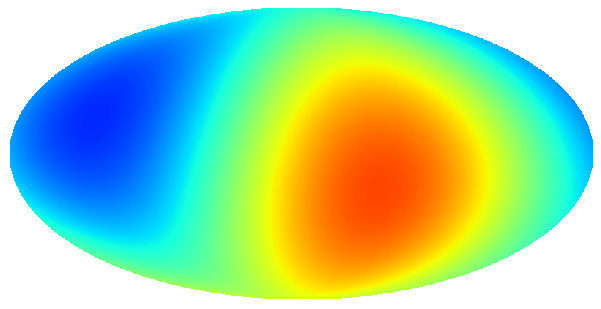}
\includegraphics[scale=0.17]{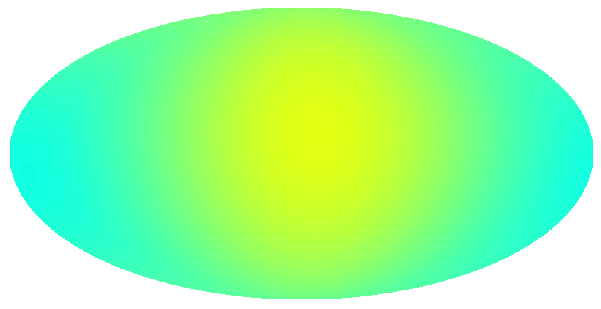}
\\
\includegraphics[scale=0.17]{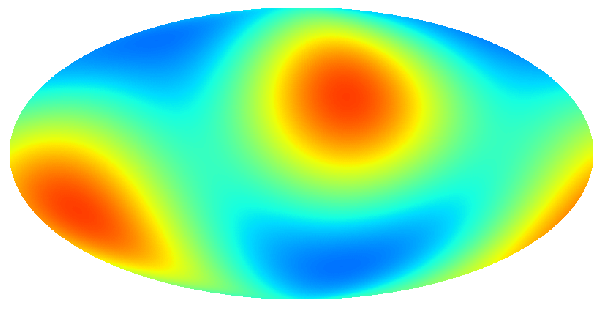}
\includegraphics[scale=0.17]{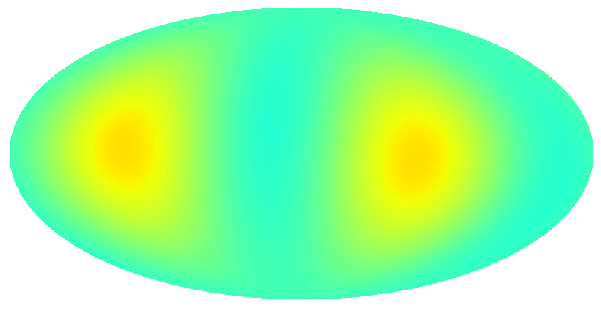}
\includegraphics[scale=0.17]{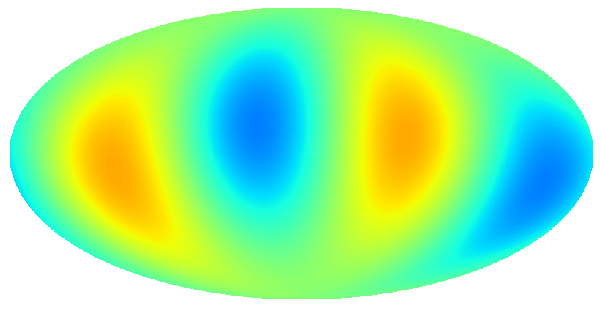}
\includegraphics[scale=0.17]{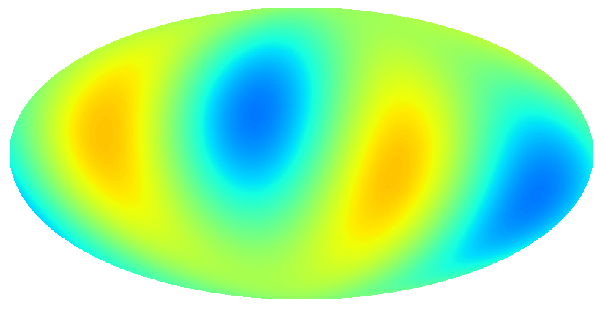}
\\
\includegraphics[scale=0.17]{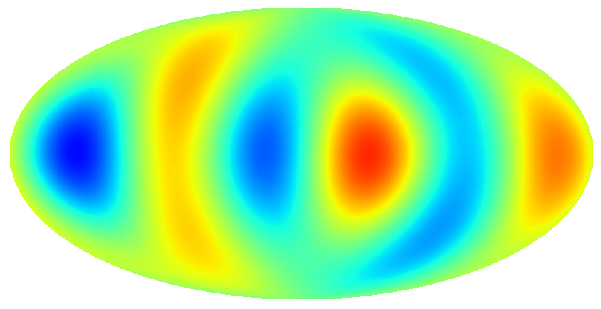}
\includegraphics[scale=0.17]{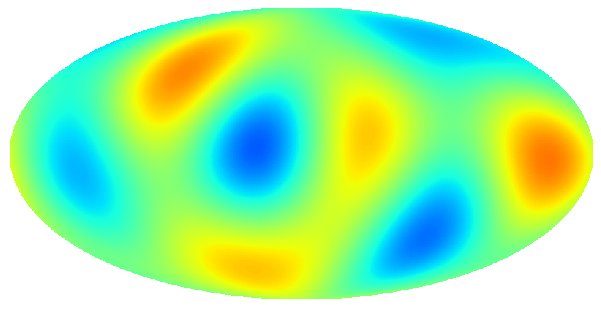}
\includegraphics[scale=0.17]{figures/fig_99emp.png}
\includegraphics[scale=0.17]{figures/fig_99emp.png}
\\
\includegraphics[scale=0.17]{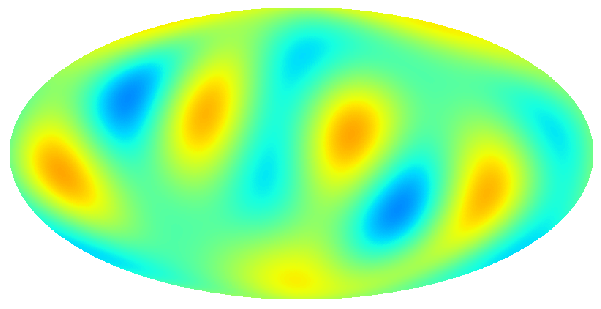}
\includegraphics[scale=0.17]{figures/fig_99emp.png}
\includegraphics[scale=0.17]{figures/fig_99emp.png}
\includegraphics[scale=0.17]{figures/fig_99emp.png}
\\
\includegraphics[scale=0.5]{figures/fig_99c2.png}
\caption{Multipolar decomposition of $\tilde{\eta}$ for CF4a, CF4b, CF4c, and CF4d (from left to right), after removing objects in the ZoA, $|b| < 20^\circ$. From top to bottom, the rows show the dipole, quadrupole, octupole and hexadecapole components.
}
\label{eta_l_map2zoa}
\centering
\end{center}
\end{figure}

To further investigate this issue, we examine whether measurements near the ZoA introduce any bias, possibly arising from gradients in the absorption corrections applied to galaxy magnitudes, which could affect the multipoles of the expansion rate fluctuation field. To test this, we exclude from the reconstruction all galaxies with Galactic latitude $|b|<20^\circ$ in the CF4a, CF4b, CF4c, and CF4d samples, and then remeasure the multipoles. The results, shown in Figure~\ref{eta_l_map2zoa}, are virtually identical in structure and amplitude to those obtained without the cut (Figure~\ref{eta_l_map2}). This demonstrates that the multipoles are not significantly affected by any potential bias in the measured distances of galaxies close to the ZoA.

\subsection{Flux limit}\label{app_malm}

Galaxy surveys used to calibrate distance indicators are generally magnitude-limited, meaning that, at a given distance,  intrinsically brighter galaxies are overrepresented in the sample. This selection effect leads to Malmquist bias, a systematic underestimation of galaxy distances when distance indicators are applied (e.g. \cite{Strauss:1995fz}).

This effect depends solely on distance and not on direction. As a consequence, the monopole-free expansion rate fluctuation field $\eta$ is by construction completely insensitive to such distance-dependent selection effects -- provided that the sample is selected using a uniform flux cut across the entire sky.

What is the impact of an anisotropic magnitude cut, where the flux limit varies across the sky?
Since the original information for the individual subsamples merged into the CF4 compilation is unavailable, we simulate extreme scenarios to estimate the amplitude of any residual selection effects in the multipoles of $\eta$. 
Assuming a perfectly isotropic universe (\( \eta = 0 \)), we assign each CF4 galaxy a distance based on the distance–redshift relation of the standard  \(\Lambda\)CDM model. 

\begin{figure}
\begin{center}
\includegraphics[scale=0.35]{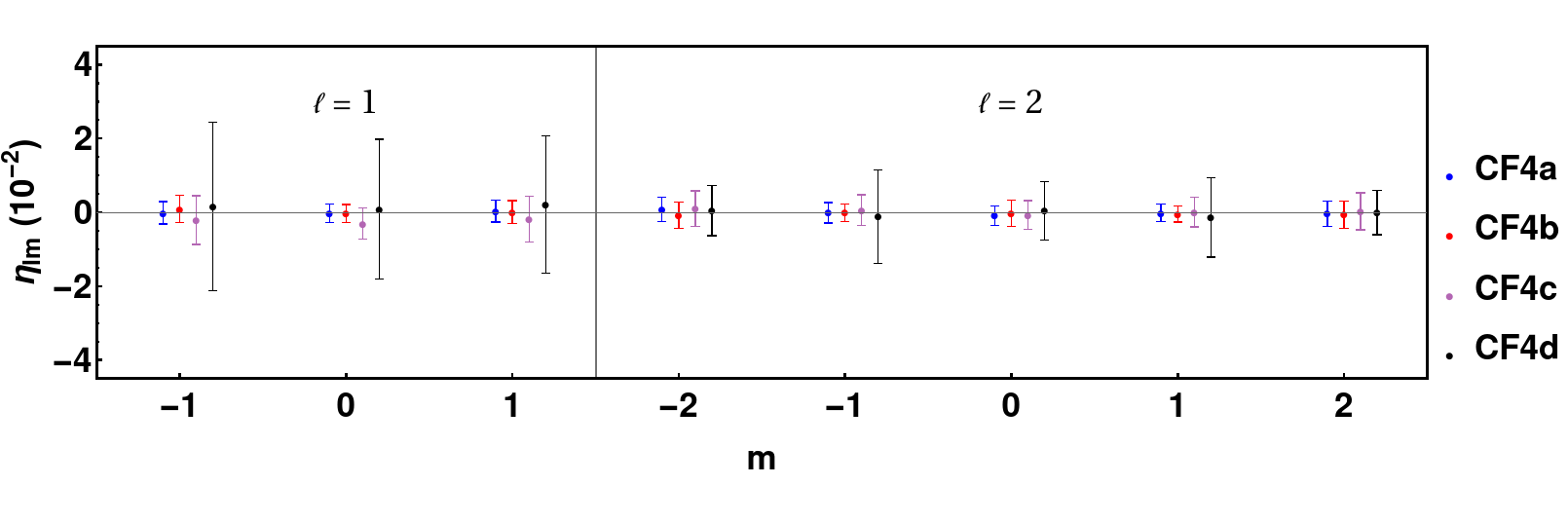}
\includegraphics[scale=0.45]{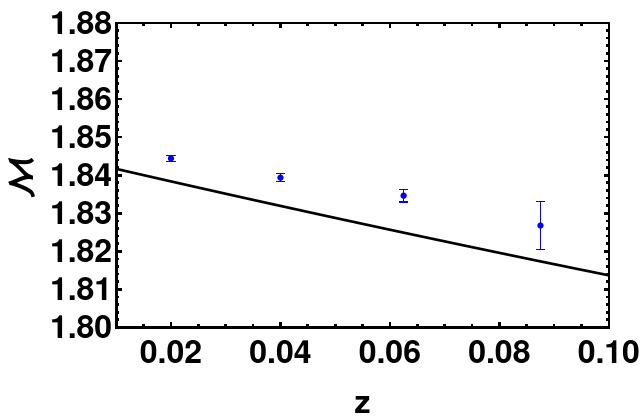}
\caption{The recovered multipoles of $\eta$ at four different redshift depths are shown in the upper panel, and the monopole $\mathcal{M}$ is shown in the lower panel, both obtained from simulations with an isotropic flux limit ($m_{\text{max}} = 22$).}
\label{Malm_sim1}
\centering
\end{center}
\end{figure}

\begin{enumerate}
\item We generate absolute magnitudes $\big(\hat{M}\big)$ for CF4 galaxies according to the Schechter luminosity function
\begin{equation}
    \Phi(M)\propto  10^{-0.4(M-M_*)(\alpha+1)}\exp\left[-10^{-0.4(M-M_*)}\right] \,,     
\end{equation}
where $M_*$ and $\alpha$, are assumed to be $-20.83 +5\log(H_0/100),\, -1.2 $ respectively, as given by \cite{Blanton:2000ns} for $r$ band. 
   
\item For each galaxy, we generate a true absolute magnitude $M_T$ by sampling from a Gaussian distribution with mean $\hat{M}$ and standard deviation $0.4$ (approximately the dispersion of the Tully–Fisher relation). 
    
\item We compute the true apparent magnitude $m_T$ using the relation $m_T = \mu_T + M_T$, where $\mu_T$ is the distance modulus corresponding to the redshift of each galaxy (from CF4), assuming $\Lambda$CDM. No observational error is assumed in $m_T$.
    
\item If $m_T > m_{\text{max}}$,  where $m_{\text{max}}$ is the chosen limiting apparent magnitude of the sample, we re-sample $M_T$ until $m_T \leq m_{\text{max}}$.
    
\item We estimate the distance modulus as $\hat{\mu} = m_T - \hat{M}$, and use this to compute the SH coefficients.
    
\item We repeat this process 500 times.
    
\item For each multipole coefficient, we compute the average and standard deviation across the 500 realizations. These values are used as the points and error bars in Figure \ref{Malm_sim1}.

\end{enumerate}

In Figure \ref{Malm_sim1} we show the multipoles recovered from a simulation where the magnitude-limit $m_{\rm max}=22$ is applied isotropically across the sky. 
As theoretically expected, the multipoles $\ell \ge 1$ of $\eta$ are not biased.  Only the recovered monopole  $\mathcal{M}$ exhibits a systematic difference with respect to the input value. Since we subtract the locally estimated monopole rather than assuming a constant value or adopting a model for it and its redshift evolution, no bias leaks into the higher order multipoles.

The bias in the monopole, for the chosen $m_{\rm max}$ corresponds approximately to a spurious peculiar velocity ranging from $\sim30$ km/s for CF4a to $\sim1100$ km/s for CF4d. This bias is also roughly equivalent to an increase in the Hubble constant $H_0$ of $\sim 0.7$ km/s/Mpc, and an increase of the deceleration parameter $q_0$ by $\sim 0.3$. 

\begin{figure}
\begin{center}
\includegraphics[scale=0.35]{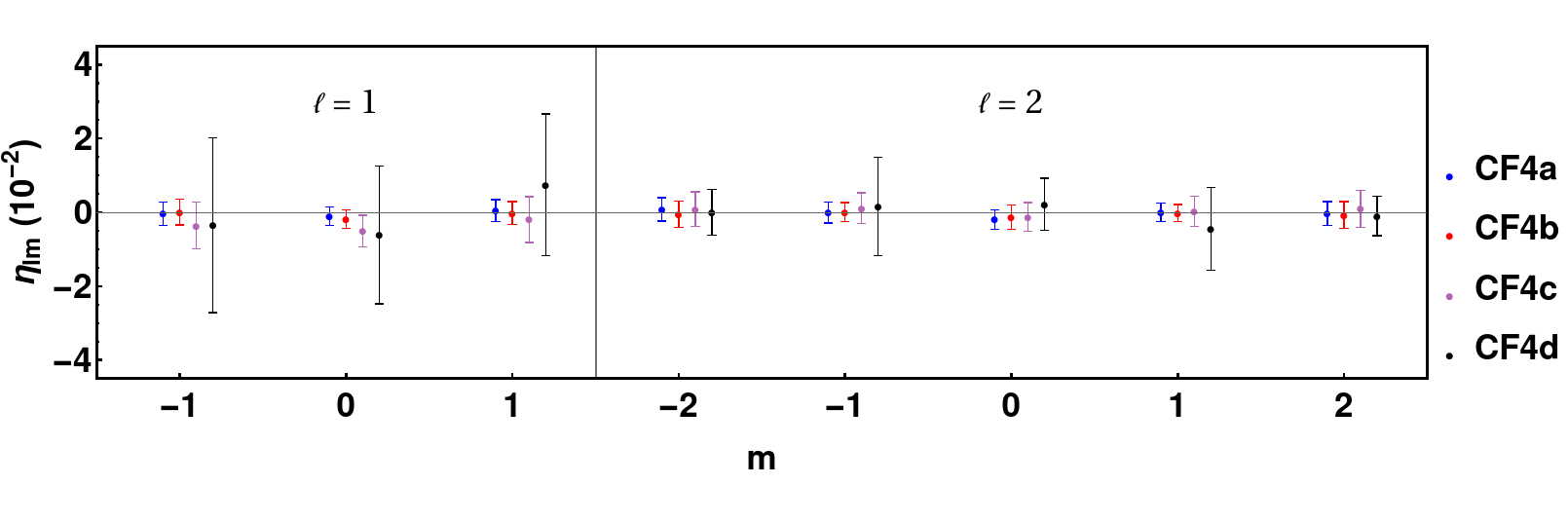}
\caption{The recovered multipoles for the CF4a, CF4b, CF4c, and CF4d mock catalogs, simulated with an anisotropic flux-limit :
$m_{\text{max}} = 22$ for $b > 25^\circ$, 
$m_{\text{max}} = 21$ for $b < -25^\circ$, 
and $m_{\text{max}} = 20$ for $-25^\circ < b < 25^\circ$.}
\label{Malm_sim2}
\centering
\end{center}
\end{figure}

In Figure~\ref{Malm_sim2}, we present the analysis of the 
$\ell \geq 1$ multipoles in simulations with an anisotropic 
flux limit. Specifically, we consider a case where the 
northern and southern polar caps have different magnitude 
limits ($m_{\text{max}} = 22$ for $b > 25^\circ$ and 
$m_{\text{max}} = 21$ for $b < -25^\circ$), and where the 
region around the Galactic plane is assigned a brighter 
magnitude cut ($m_{\text{max}} = 20$ for $-25^\circ < b < 25^\circ$). 
In this configuration, the bias becomes nonzero because the 
Malmquist effect acquires a directional dependence. However, 
the effect remains completely negligible compared to the 
signal observed in the data and largely subdominant 
with respect to the associated error bars (see 
Figure~\ref{alm_CF4}).

\section{$\eta$ field traced by galaxy groups}\label{app_group}

An additional advantage of the CF4 sample is that it enables tracing the expansion rate field using galaxy groups rather than individual galaxies.
Grouping CF4 galaxies allows us to average out local  contributions to the observed cosmological redshift. As a result, the measured redshift becomes less sensitive to small-scale gravitational fields and more representative of the large-scale structure of spacetime.

Although distance resolution improves—since the error on the average distance decreases with the square root of the number of galaxies within each structure—the overall statistical power in fixing the amplitude of the multipoles of the $\eta$ field remains largely unaffected, as the expansion rate fluctuation field is traced by a smaller number of grouped objects.  Indeed, for isotropic distribution of measurements and equal errors for the distance modulus $\sigma_\mu$, the error of the SH coefficients scales as  $\sigma_{\ell m }\approx {\sigma_\mu}\sqrt{{4\pi}/{N_g}}/5$.

In the catalog CF4-groups, there are 38057 structures, only 5689 containing more than one galaxy (these groups have on average 4.2 galaxies per group). As in the case of the individual galaxies, the groups with $(PGC=20679, 40498, 43296, 59762, 59927, 3097150)$ are removed from the analysis. 

\begin{figure}
\begin{center}
\includegraphics[scale=0.17]{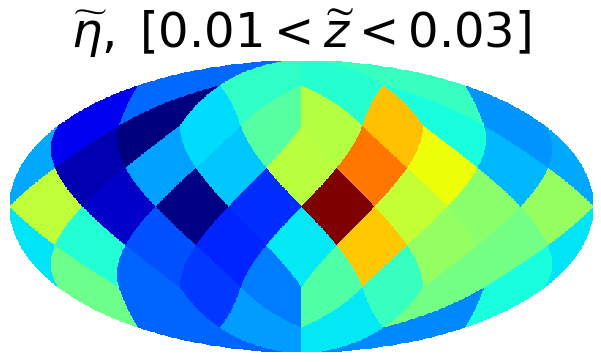}
\includegraphics[scale=0.17]{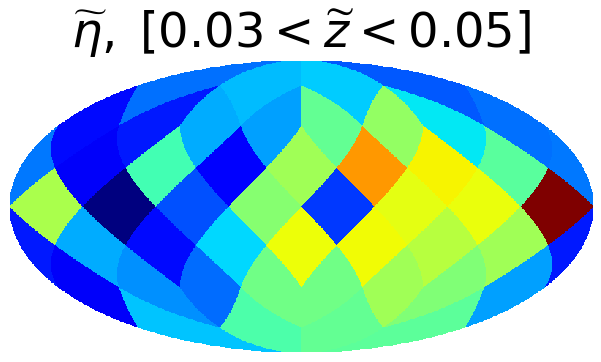}
\includegraphics[scale=0.17]{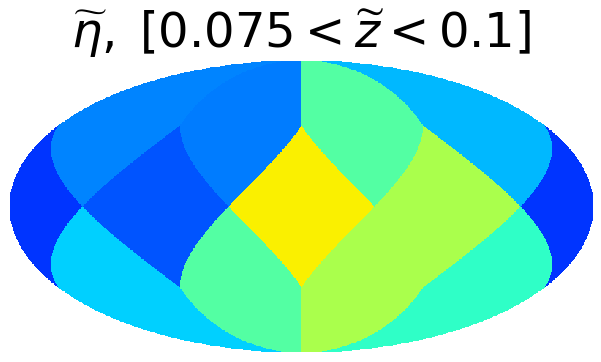}
\includegraphics[scale=0.17]{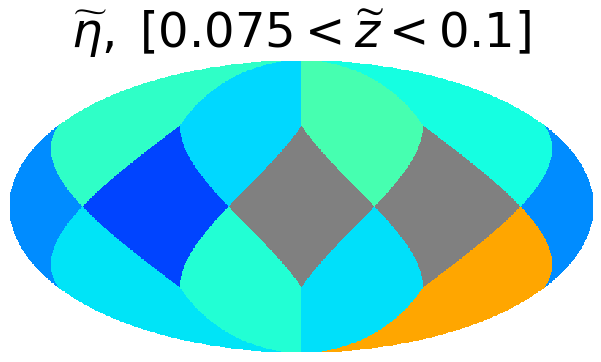}
\\
\includegraphics[scale=0.45]{figures/fig_99c1.png}
\\
\includegraphics[scale=0.17]{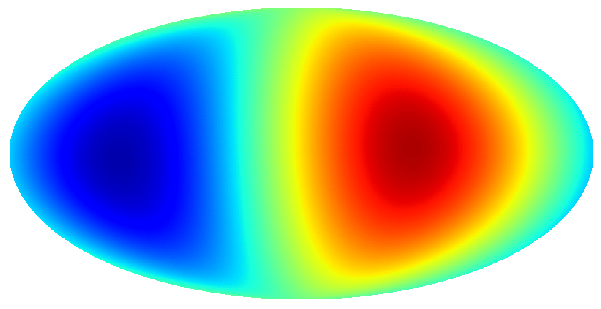}
\includegraphics[scale=0.17]{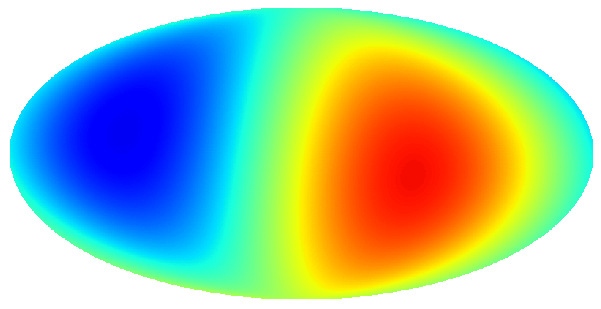}
\includegraphics[scale=0.17]{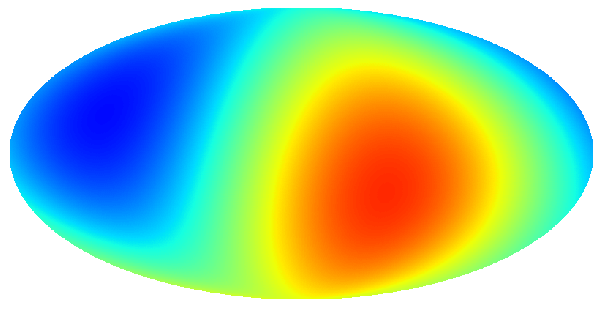}
\includegraphics[scale=0.17]{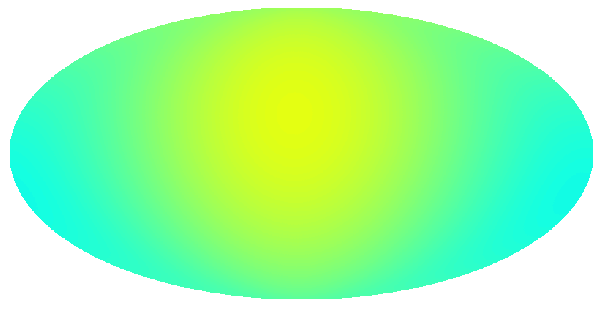}
\\
\includegraphics[scale=0.17]{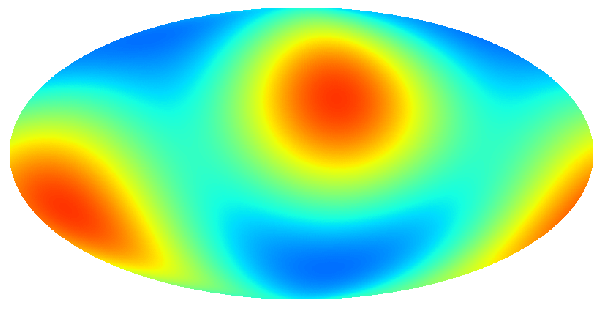}
\includegraphics[scale=0.17]{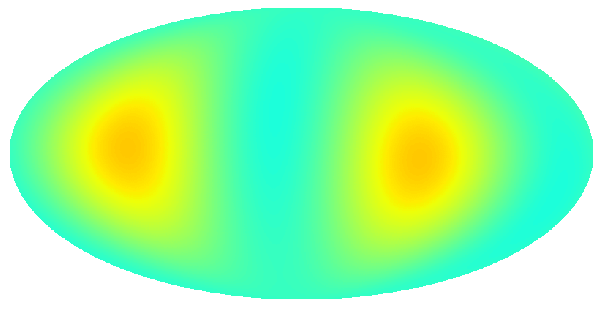}
\includegraphics[scale=0.17]{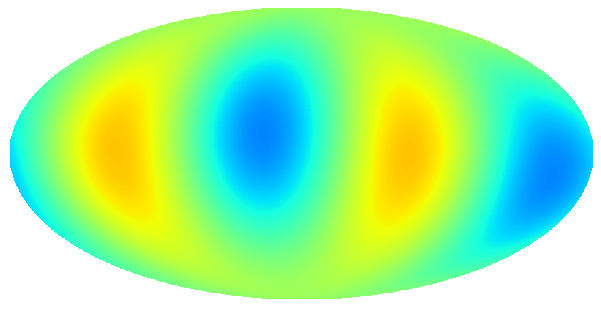}
\includegraphics[scale=0.17]{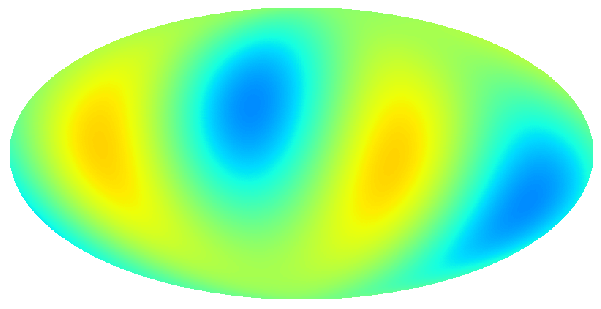}
\\
\includegraphics[scale=0.17]{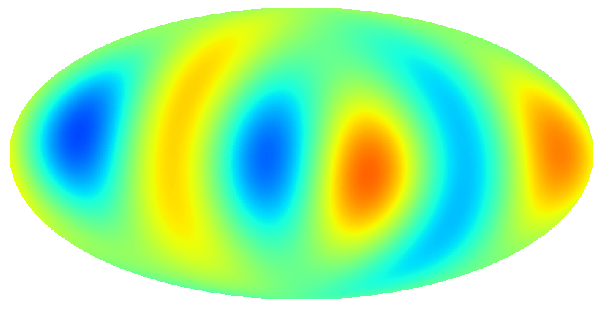}
\includegraphics[scale=0.17]{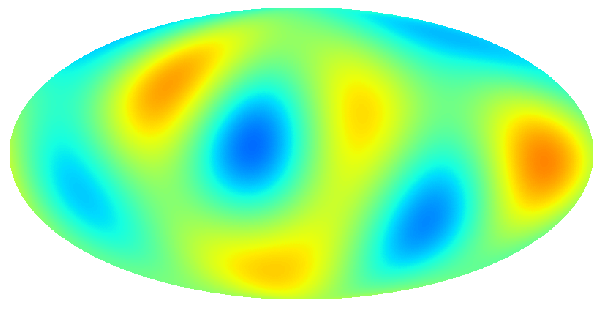}
\includegraphics[scale=0.17]{figures/fig_99emp.png}
\includegraphics[scale=0.17]{figures/fig_99emp.png}
\\
\includegraphics[scale=0.17]{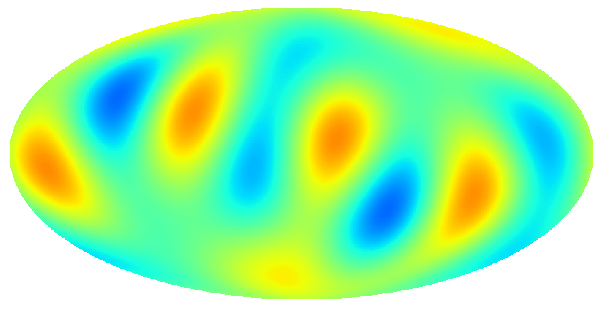}
\includegraphics[scale=0.17]{figures/fig_99emp.png}
\includegraphics[scale=0.17]{figures/fig_99emp.png}
\includegraphics[scale=0.17]{figures/fig_99emp.png}
\\
\includegraphics[scale=0.6]{figures/fig_99c2.png}
\caption{
From left to right, columns correspond to CF4-groups ($0.01 < \tilde z < 0.1$), CF4a-groups ($0.01 <\tilde z < 0.03$), CF4b-groups ($0.03 <\tilde z < 0.05$) and CF4c-groups ($0.05 <\tilde z < 0.1$).
{\it Top row:} Pixelized maps of the expansion fluctuation field $\tilde\eta$.
{\it Bottom 4 rows:} Representation of the multipolar decomposition of $\tilde\eta$. Rows correspond, from top to bottom, to the dipole, quadrupole, octupole and hexadecapole.}
\label{eta_cf4gr1}
\centering
\end{center}
\end{figure}

Figure~\ref{eta_cf4gr1} presents the HEALPix-pixelized maps of the expansion rate fluctuation field \(\tilde{\eta}\), traced by the CF4 groups sample across four redshift intervals, along with their corresponding multipole decompositions. A  comparison with the analogous multipole maps derived from individual galaxy distances (Figure~\ref{eta_l_map2}) demonstrates the effective quantitative equivalence of the two reconstructions.

\bibliographystyle{JHEP}
\bibliography{aabiblio}

\end{document}